%% file: IGFEM-TopOpt.tex
\documentclass[10pt,a4paper]{article}         

\usepackage{geometry}
\usepackage{graphicx}
\usepackage{moreverb}
\usepackage{paralist} 

\usepackage[usenames,dvipsnames,svgnames,table]{xcolor} 
\usepackage[utf8]{inputenc}
\usepackage[inline, shortlabels]{enumitem} 
\usepackage[export]{adjustbox} 
\usepackage[normalem]{ulem}
\usepackage{amsmath}
\usepackage{mathtools}  
\usepackage{amsfonts}  
\usepackage{tikz}
\usepackage[customcolors]{hf-tikz}
\hfsetfillcolor{black!10}
\hfsetbordercolor{black!0}
\usepackage{pgfplots}
\usetikzlibrary{plotmarks, calc, spy, pgfplots.polar, external}
\usepackage{siunitx}
\usepackage{amsmath,amsfonts,amssymb}
\usepackage{soul}
\usepackage{bm}
\usepackage{bbold} 
\usepackage{scalerel} 
\usepackage[]{standalone}
\usepackage{url}
\usepackage{subfig}

\graphicspath{{./figures/}}

\usepackage{pgfplots}
\newcommand{\bs}[1] {\bm{#1}}

\newcommand{\gv}[1]{\ensuremath{\mbox{\boldmath$ #1 $}}} 
\newcommand{\grad}[1]{\gv{\nabla} #1} 
\DeclareMathOperator*{\assembly}{\scalerel*{\mathbb{A}}{\sum}}

\newcommand\BibTeX{{\rmfamily B\kern-.05em \textsc{i\kern-.025em b}\kern-.08em
T\kern-.1667em\lower.7ex\hbox{E}\kern-.125emX}}

\newcommand{\SJVDB}[1]{#1}

\newcommand*{\standardDOF}{}
\DeclareRobustCommand*{\standardDOF}{%
  \ensuremath{\vcenter{\hbox{%
    \def\BoundingBoxCorrection{.55\pgflinewidth}%
    \tikz\path plot[
        mark=*,
        mark size=2,
        mark options={solid, fill=black},
      ] coordinates {(0, 0)}
      (current bounding box.south west)
      ++(-\BoundingBoxCorrection, -\BoundingBoxCorrection)
      (current bounding box.north east)
      ++(\BoundingBoxCorrection, \BoundingBoxCorrection)
    ;%
  }}}%
}

\newcommand*{\wrongDOF}{}
\DeclareRobustCommand*{\wrongDOF}{%
  \ensuremath{\vcenter{\hbox{%
    \def\BoundingBoxCorrection{.55\pgflinewidth}%
    \tikz\path plot[
        mark=*,
        mark size=2,
        mark options={solid, fill=red},
      ] coordinates {(0, 0)}
      (current bounding box.south west)
      ++(-\BoundingBoxCorrection, -\BoundingBoxCorrection)
      (current bounding box.north east)
      ++(\BoundingBoxCorrection, \BoundingBoxCorrection)
    ;%
  }}}%
}

\newcommand*{\enrichedDOF}{}
\DeclareRobustCommand*{\enrichedDOF}{%
  \ensuremath{\vcenter{\hbox{%
    \def\BoundingBoxCorrection{.55\pgflinewidth}%
    \tikz\path plot[
        mark=*,
        mark size=2,
        mark options={solid, fill=white},
      ] coordinates {(0, 0)}
      (current bounding box.south west)
      ++(-\BoundingBoxCorrection, -\BoundingBoxCorrection)
      (current bounding box.north east)
      ++(\BoundingBoxCorrection, \BoundingBoxCorrection)
    ;%
  }}}%
}

\DeclareMathOperator{\dx}{d}
\DeclareMathOperator{\Tr}{tr}
\DeclareMathOperator{\adj}{adj}

\DeclareMathOperator*{\argmin}{arg\min}

\begin{document}

\title{An \SJVDB{Interface}-enriched Generalized Finite Element Method for Levelset-based Topology Optimization
}

\author{S. J. van den Boom,
        J. Zhang,
        F. van Keulen, 
        A. M. Arag\'{o}n
}

\date{Department of Precision \& Microsystems Engineering (PME)\\
             Faculty of Mechanical, Maritime \& Materials Engineering (3ME)\\
             Delft University of Technology (TU Delft)\\
             Mekelweg 2, 2628 CD Delft, Netherlands \\
              a.m.aragon@tudelft.nl  \\
              \vspace{0.5cm}
              \today}

\maketitle

\begin{abstract}

During design optimization, a smooth description of the geometry is important, especially for problems that are sensitive to the way interfaces are resolved, \textit{e.g.}, wave propagation or fluid-structure interaction. A levelset description of the boundary, when combined with an enriched finite element formulation, offers a smoother description of the design than traditional density-based methods. However, existing \SJVDB{enriched} methods have drawbacks, including ill-conditioning and difficulties in prescribing essential boundary conditions.
In this work we introduce a new enriched topology optimization methodology that overcomes \SJVDB{the} aforementioned drawbacks; boundaries are resolved accurately by means of the Interface-enriched Generalized Finite Element Method (IGFEM), coupled to a levelset function constructed by radial basis functions. The enriched method used in this new approach to topology optimization has the same level of accuracy in the analysis as \SJVDB{standard the finite element method with} matching meshes, \SJVDB{but} without the need for remeshing.
 We derive the analytical sensitivities  and we discuss the behavior of the optimization process in detail. We establish that IGFEM-based levelset topology optimization generates correct topologies for well-known compliance minimization  problems.

\end{abstract}

\section{Introduction}

The use of enriched finite element methods in topology optimization approaches is not new; the eXtended/Generalized Finite Element Method (X/GFEM)~\cite{Duarte:1998,Moes:1999,Moes:2003,Belytschko:2009,Aragon:2010}, for example, has been explored in this context. 
\SJVDB{However, the Interface-enriched Generalized Finite Element Method (IGFEM) has been shown to have many advantages over X/GFEM~\cite{Soghrati:2012,vandenBoom:2018a}.}
In \SJVDB{this} work we extend IGFEM to be used in a levelset based topology optimization framework.

Topology optimization, first introduced by Bends{\o}e and Kikuchi~\cite{Bendsoe:1988}, has been widely used to obtain designs that are optimized for a certain functionality, \textit{e.g.}, minimum compliance. In the commonly-used density-based methods, a continuous design variable that represents a material density is assigned to each element in the discretization. The design is pushed towards a \textit{black and white} design by means of an interpolation function, \textit{e.g.}, the Solid Isotropic Material with Penalisation (SIMP)~\cite{Bendsoe:1989}, that disfavors intermediate density values, also referred to as gray values. A filter is then required to prevent checkerboard-like density patterns, and to impose a minimum feature size. However, due to the filter, gray values are introduced. Density based topology optimization is straightforward to implement and widely available in both research codes and commercial software. However, because the topology is described by a density field \SJVDB{on a (usually) structured mesh}, material interfaces not only contain gray values but also suffer from \textit{pixelization} or \textit{staircasing}---\SJVDB{staggered} boundaries \SJVDB{that follow} the finite element mesh. Although a postprocessing step can be performed to smoothen the final design, the analysis during optimization is still based on gray density fields and a staircased representation.  This may be detrimental to the approximate solution's accuracy, especially in cases that are sensitive to the boundary description, such as flow problems~\cite{Villanueva2017}. Furthermore, because the location of the material boundary is not well defined, it is difficult to track the evolving boundary during optimization, for example to impose contact \SJVDB{constraints}.

The aforementioned drawbacks could be alleviated by the use of geometry-fitted discretization methods, which have been widely used in shape optimization~\cite{Staten:2012}. In these methods, the location of the material interface is known throughout the optimization, and the analysis mesh is modified to completely eliminate the pixalization and gray values. Mesh-morphing methods such as the deformable simplex method \cite{Misztal:2012,Christiansen:2014,Christiansen:2015,Zhou:2018}, anisotropic elements~\cite{Jensen:2016}, and $r$-refinement~\cite{Yamasaki:2017}, have been demonstrated for topology optimization. Nevertheless, adapting the mesh in every design iteration remains a challenge. Not only is it an extra computational step, the changing discretization also introduces another complication in the optimization procedure because design variables need to be mapped to the new discretization~\cite{vanDijk:2013}. 

A more elegant option is to define the material interface independently from the FE discretization, \textit{e.g., implicitly} by means of the zero-contour of a levelset function. Although the required mapping between the geometry and the discretization mesh can be done with an Ersatz method~\cite{Allaire:2014}, this again introduces gray values and staircasing into the analysis. Similarly, NURBS-based topology optimization using the Finite Cell Method (FCM)~\cite{Gao:2019} provides a higher resolution boundary description, that is, however, still staircased. Alternatively, there are methods that allow for a one-to-one mapping of the topology to the analysis mesh, resulting in a non-pixalized boundary description. These methods combine the advantages of clearly defined material interfaces with the benefits of a fixed discretization mesh used in density-based methods. In the literature, levelset based topology optimization has been established based on CutFEM~\cite{Villanueva:2017} and the eXtended/Generalized Finite Element Method (X/GFEM)~\cite{Belytschko:2003,Villanueva:2014,Liu:2016b}.  

In enriched finite element methods such as X/GFEM, the standard finite element space is augmented by enrichment functions that account for \textit{a priori} knowledge of the discontinuity of \SJVDB{the field or its} gradient, due to the material interface. Although X/GFEM has been shown to be advantageous in many applications, \textit{e.g.}, fluid-structure interaction~\cite{Mayer:2010} and fracture mechanics~\cite{Fries:2010}, the method also has weaknesses: degrees of freedom (DOFs) corresponding to original mesh nodes do not automatically retain their physical meaning, and essential boundary conditions mostly have to be prescribed \textit{weakly}. Moreover, the X/GFEM may result in ill-conditioned matrices, in which case a Stable Generalized FEM (SGFEM) is needed~\cite{Babuvska:2012,Gupta:2013,Kergrene:2016}. \SJVDB{Furthermore, the approximation of stresses can be highly overestimated near material boundaries~\cite{VanMiegroet:2007,Noel:2017,Sharma:2018}.} Finally, as the enriched functions are associated with original mesh nodes, the accuracy of the approximation may degrade in blending elements---elements that do not have all nodes enriched\SJVDB{~\cite{Fries:2008}}.

The Interface-enriched Generalized Finite Element Method (IGFEM)~\cite{Soghrati:2012} was first introduced as a simplified generalized FEM to solve problems with weak discontinuities, \textit{i.e.}, where the gradient field is discontinuous. \SJVDB{The method} overcomes most problems of X/GFEM: 
In IGFEM, enriched nodes are placed along interfaces, and enrichment functions are non-zero only in \textit{cut elements}, \textit{i.e.}, elements that are intersected by a discontinuity. \SJVDB{Furthermore, enrichment functions are exactly zero at original mesh nodes.} Therefore,
 original mesh nodes retain their physical meaning and essential boundary conditions can be enforced directly on non-matching edges~\cite{vandenBoom:2018a,CubaRamos:2015,Aragon:2017}. 
It was shown that IGFEM is optimally convergent under mesh refinement for problems without singularities~\cite{Soghrati:2012,Soghrati:2012b}, and stable by means of scaling enrichment functions or a simple diagonal preconditioner~\cite{vandenBoom:2018a}. The method has been applied to the modeling of fibre-reinforced composites~\cite{Soghrati:2012b}, multi-scale damage evolution in heterogeneous adhesives~\cite{Aragon:2013}, microvascular materials with active cooling~\cite{Soghrati:2012,Soghrati:2012b,Soghrati:2012c,Soghrati:2013}, and the transverse failure of composite laminate~\cite{ZhangB:2019,Shakiba:2019}. Extensions of IGFEM are found in the Hierarchical Interface-enriched Finite Element Method (HIFEM)~\cite{Soghrati:2014}, that allows for intersecting discontinuities, and in the Discontinuity-Enriched Finite Element Method (DE-FEM)~\cite{Aragon:2017}, that provides a unified formulation for both weak and strong discontinuities. DE-FEM, which inherits the advantages of IGFEM over X/GFEM, has successfully been applied to problems in fracture mechanics~\cite{Aragon:2017,Zhang:2019} and fictitious domain or immersed boundary problems with strongly enforced essential boundary condition~\cite{vandenBoom:2018a}.

In the context of optimization, IGFEM has been explored for NURBS-based shape optimization~\cite{Najafi:2017}, the shape  optimization of microvascular channels~\cite{Tan:2017} and their combined shape and network topology optimization~\cite{Pejman:2019}, the optimization of microvascular panels for nanosatellites~\cite{Tan:2018b}, and optimal cooling of batteries~\cite{Tan:2018}. 
Nevertheless, IGFEM has not yet been used for continuum topology optimization. In this paper we show topology optimization based on a levelset function, parametrized with Radial Basis Functions (RBFs)~\cite{Wendland:1995,Wang:2006}, in combination with IGFEM. We demonstrate  the method on benchmark compliance problems. It should be noted that no \SJVDB{significant} performance improvement is expected for these cases, as they are not sensitive to the way the boundaries are discretized, but rather by the bulk. This paper should, however, be seen as the necessary proof of concept before considering more complex cases. The sensitivities are derived and the method is compared to density-based topology optimization and to the levelset-based Ersatz method.

\section{Formulation}

\subsection{IGFEM-based analysis}
\label{sec:formulation}

\begin{figure}
\centering
	 \def\svgwidth{230pt}
    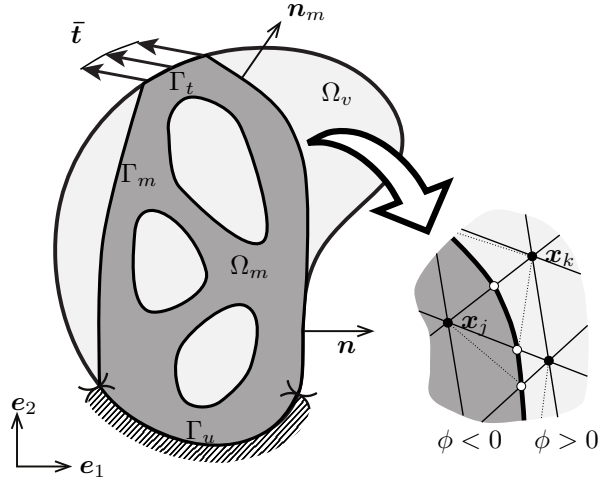
\caption{Mathematical representation of a topology optimization design domain $\Omega$. Essential and natural boundary conditions are prescribed on the part of the boundary denoted $\Gamma_\mathrm{u}$ and $\Gamma_\mathrm{t}$, respectively. The material domain is referred to as $\Omega_\mathrm{m}$, while the void region is denoted $\Omega_\mathrm{v}$. The insert shows the discretization with a material interface, defined by the zero-contour of the levelset function $\phi$, that is non-matching to the mesh. Original mesh nodes and enriched nodes are denoted with \standardDOF{} and \enrichedDOF{} symbols, respectively. }
 \label{fig:cracked_potato}
\end{figure}
In this work we focus on elastostatics and heat \SJVDB{conduction} problems on solid domains, as represented in Figure~\ref{fig:cracked_potato}. A design domain $\Omega \subset \mathbb{R}^d$ is referenced by a Cartesian coordinate system spanned by base vectors $\left\{ \bs e_i \right\}^d_{i=1}$. This domain is decomposed into a solid material domain and a void domain, denoted by $\Omega_\mathrm{m}$ and $\Omega_\mathrm{v}$, respectively, such that the domain closure is $\overline{\Omega} = \overline{\Omega}_\mathrm{m}  \cup \overline{\Omega}_\mathrm{v}$, and $\Omega_\mathrm{m}  \cap \Omega_\mathrm{v} = \emptyset$. The boundary of the design domain, $\partial \Omega \equiv \Gamma = \overline{\Omega} \setminus \Omega$, is subjected to essential (Dirichlet) boundary conditions on $\Gamma_\mathrm{u}$, and to natural (Neumann) boundary conditions on $\Gamma_\mathrm{t}$, such that $\overline{\Gamma} = \overline{\Gamma}_\mathrm{u}  \cup \overline{\Gamma}_\mathrm{t} $ and $\Gamma_\mathrm{u}  \cap \Gamma_\mathrm{t} = \emptyset$. The material boundary, $\Gamma_\mathrm{m} = \left(\overline{\Omega}_\mathrm{m} \cap \overline{\Omega}_\mathrm{v} \right) \setminus \Gamma$, is defined implicitly by a levelset function, $\phi \left( \bs x \right) = 0$, that is a function of the spatial coordinate $\bs x$.

\SJVDB{For any iteration in the elastostatic optimization procedure,} the boundary value problem is solved with prescribed displacements $\bar{\bs{u}} : \Gamma_{\mathrm{u}} \to \mathbb{R}^d$, prescribed tractions $\bar{\bs{t}}: \Gamma_\mathrm{t} \to \mathbb{R}^d$, and body forces $\bs b_i \equiv \left. \bs b \right |_{\Omega_i} : \Omega_i \to \mathbb{R}^d, \; \text{where} \; i = \mathrm{m}, \mathrm{v}$. We denote the field $\bs u_i$ as the restriction of $\bs u$ to domain $\Omega_i$, \textit{i.e.} $\bs u_i \equiv \left. \bs u \right |_{\Omega_i}$. Note that here the field is solved on both the material domain and the void domain. However, following the techniques described in~\cite{vandenBoom:2018a}, it is also possible to completely remove the void regions from the analysis.

We define a vector-valued function space, denoted $ \boldsymbol{\mathcal{V}}_0 \equiv \left[ \mathcal{H}_0^1 \left( \Omega \right) \right]^d $, where components of $\bs{v} \in \boldsymbol{\mathcal{V}}_0$ belong to the first-order Sobolev space that satisfies homogeneous essential boundary conditions on $\Gamma_{\mathrm{u}}$. In this work we only focus on problems with homogeneous Dirichlet boundary conditions. For problems with non-homogeneous essential boundary conditions, the reader is referred to~\cite{Aragon:2017}.
The weak form of the \SJVDB{elastostatics} boundary value problem can be written as: 
Find $\bs u \in \boldsymbol{\mathcal{V}}_0$ such that
\begin{equation} \label{eq:weak_abstract_form}
  B \left( \bs{u},  \bs{v}\right)  = L \left (  \bs{v} \right)  \qquad \forall \, \bs{v} \in  \boldsymbol{\mathcal{V}}_0,
\end{equation} where the bilinear and linear forms can be written as 
\begin{align}
&B \left( \bs{u},  \bs{v}\right)  = \sum_{i=\mathrm{m},\mathrm{v}} \;\int_{\Omega_i} \bs \sigma_i \left( \bs u_i \right)   : \bs \varepsilon_i \left( \bs v_i \right) \, \dx\Omega, \\
 \intertext{and}
&L \left (  \bs{v} \right) = \sum_{i=\mathrm{m},\mathrm{v}} \;\int_{\Omega_i} \bs v_i  \cdot  \bs b_i \, \dx\Omega +  \int_{\Gamma_{\mathrm{t}}} \bs v_i  \cdot   \bar{\bs t} \, \dx\Gamma,
\end{align} respectively, where the stress tensor $\bs \sigma_i \equiv \left. \bs \sigma \right |_{\Omega_i}$ follows Hooke's law for linear elastic materials, \SJVDB{$\bs \sigma_i = \bs C_i \bs \epsilon_i$, and $\bs C_i$ is the elasticity tensor}. Small strain theory is used for the strain tensor, \textit{i.e.}, $\bs \varepsilon \left( \bs u \right)= \frac{1}{2} \left( \grad \bs u + \grad \bs u^\intercal \right)$\SJVDB{, for the elastostatic boundary value problem}. 

For the heat conduction problem both the trial and the weight function are taken from the space $ \mathcal{V}_0 \equiv \mathcal{H}_0^1 \left( \Omega \right) $. For a prescribed temperature $u : \Gamma_u \to \mathbb{R}$, prescribed heat flux $q : \Gamma_q \to \mathbb{R}$, heat source $f_i : \Omega_i \to \mathbb{R}$, and conductivity tensor $\bs \kappa_i \equiv \left. \bs \kappa \right |_{\Omega_i} \to \mathbb{R}^d \times \mathbb{R}^d$, the bilinear and linear forms for \SJVDB{each iteration in} the heat \SJVDB{compliance problems} equation are written as 
\begin{align} \label{eq:weak_abstract_form_heat}
 &B \left( u,  v \right) = \sum_{i=\mathrm{m},\mathrm{v}} \; \int_{\Omega_i}  \nabla v_i \cdot \left(\bs \kappa_i \cdot \nabla u_i \right) \, \dx\Omega\\
 \intertext{and}
&L \left ( v \right) = \sum_{i=\mathrm{m},\mathrm{v}} \;\int_{\Omega_i} v_i \;    f_i \, \dx\Omega +  \int_{\Gamma_{\mathrm{t}}}  v_i    \; \bar{q} \, \dx\Gamma.
\end{align} 

It is worth noting that interface conditions that satisfy continuity of the field and its tractions (or fluxes) do not appear explicitly in Eq.~\eqref{eq:weak_abstract_form} (or Eq.~\eqref{eq:weak_abstract_form_heat}), because they drop out due to the weight \SJVDB{function $\bs v$ (or $v$)} being continuous along the interface.

The design domain is discretized without prior knowledge of the topology as $\Omega^h =  \bigcup_{i} \overline{e}_i  $, where $e_i$ is the $i$th \SJVDB{finite} element, resulting in a mesh that is non-matching to \SJVDB{material boundaries}. The levelset function, whose zero contour defines the interface between void and material, is then \SJVDB{evaluated} on the same mesh. This is done for efficiency, as the mapping needs to be computed only once, and results in discrete nodal levelset values. New \textit{enriched nodes} are placed at the intersection between element edges and the zero contour of the levelset. The locations of these enriched nodes, denoted $\bs x_n$, are found by linear interpolation between nodes \SJVDB{$\bs x_j$ and $\bs x_k$ of the original mesh:}
\begin{equation} \label{eq:discretizedLS}
\bs x_n = \bs x_j - \frac{\phi_j}{\phi_{k} - \phi_j}\left( \bs x_{k} - \bs x_j \right).
\end{equation}
Here $\bs x_j$ and $\bs x_{k}$ have levelset values of opposite sign.
Cut elements are then subdivided into \textit{integration elements}.

\begin{figure}
\centering
	 \def\svgwidth{180pt}
    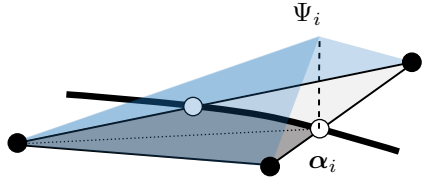
\caption{Schematic representation of enrichment function $\psi_i$ corresponding to the enriched DOFs $\bs \alpha_i$, where the enriched nodes are shown as \enrichedDOF{}. This enrichment function is constructed from the standard Lagrange shape function of the integration elements. }
 \label{fig:enrichment}
\end{figure}

Following a Bubnov-Galerkin procedure, the resulting finite dimensional problem is then solved by choosing trial and weight functions from the same enriched finite element space. The IGFEM  approximation can then be written as
\begin{equation} \label{eq:IGFEMapproximation}
\bs{u}^h(\bs{x}) = \underbrace{\sum_{i\in \iota_h} N_i \left( \bs{x} \right) \bs{U}_ i}_{\text{standard FEM}}+  \underbrace{\sum_{i\in \iota_w} \mathit{\psi}_{i} \left( \bs{x} \right) \bs{\alpha}_{i}}_{\text{enrichment}},  
\end{equation} \SJVDB{for elastostatics or \begin{equation} \label{eq:IGFEMapproximation}
u^h(\bs{x}) = \underbrace{\sum_{i\in \iota_h} N_i \left( \bs{x} \right) U_ i}_{\text{standard FEM}}+  \underbrace{\sum_{i\in \iota_w} \mathit{\psi}_{i} \left( \bs{x} \right) \alpha_{i}}_{\text{enrichment}},  
\end{equation} for heat conduction problems,} where the first term corresponds to the standard finite element approximation, with shape functions $N_i \left( \bs x\right)$ and corresponding standard degrees of freedom $\bs U_i$, and the second term refers to the enrichment, characterized by enrichment functions $\psi_i \left(\bs x\right)$ and associated enriched DOFs $\bs \alpha_i$. The index set containing all enriched nodes is denoted $\iota_w$. Enrichment functions $\psi_i$ can be conveniently constructed from Lagrange shape functions of integration elements, as illustrated in Figure~\ref{fig:enrichment}, while the underlying partition of unity shape functions are kept intact.

Subsequently, the local stiffness matrix $\bs k_e$ and force vector $\bs f_e$ are obtained numerically; elements that are not intersected follow the standard FEM procedures. For every integration element $\bs k_e$ and $\bs f_e$ are defined as \SJVDB{ 
\begin{equation}\label{eq:localassembly}
\begin{aligned}
\bs{k}_e &= \int_{e}{\begin{bmatrix} \bs \Delta \bs{N} \\ \bs \Delta\bs{\psi} \end{bmatrix}  }\bs{D}_e \begin{bmatrix} \bs \Delta\bs{N}^{\intercal} & \bs \Delta\bs{\psi}^\intercal \end{bmatrix}\dx\Omega, \; \text{and} \\ 
\bs{f}_e &= \int_{e} \begin{bmatrix} \bs{N} \\ \bs{\psi} \end{bmatrix} \bs{b}_e \; \dx\Omega + \int_{\delta e} \begin{bmatrix} \bs{N} \\ \bs{\psi} \end{bmatrix} \bar{\bs{t}} \; \dx\Gamma  ,
\end{aligned}
\end{equation} where $\bs D_e$ }is the constitutive matrix, the shape functions vector $\bs N$ and enrichment functions $\bs \psi$ are stacked together. The differential operator $\bs \Delta$ is defined as:
\begin{equation}
\renewcommand*{\arraystretch}{1.3}
\bs \Delta \equiv
\begin{bmatrix}
\frac{\partial}{\partial x} & 0 & \frac{\partial}{\partial y} \\
0 & \frac{\partial }{\partial y} & \frac{\partial}{\partial x} & \\
\end{bmatrix}^\top,  \; \renewcommand*{\arraystretch}{1.3}
\bs \Delta \equiv
\begin{bmatrix}
\frac{\partial}{\partial x} & 0 & 0 & \frac{\partial}{\partial y} &0 & \frac{\partial}{\partial z}\\
0 & \frac{\partial }{\partial y} & 0  & \frac{\partial}{\partial x} & \frac{\partial}{\partial z}  &0\\
0 & 0 & \frac{\partial}{\partial z} & 0 & \frac{\partial}{\partial y} & \frac{\partial}{\partial x}\\
\end{bmatrix}^\top, 
\end{equation} for elastostatics in 2-D and 3-D, respectively, and \begin{equation}
\renewcommand*{\arraystretch}{1.3}
\bs \Delta \equiv
\begin{bmatrix}
\frac{\partial}{\partial x} &  \frac{\partial}{\partial y} \\
\end{bmatrix}^\top,  \; \renewcommand*{\arraystretch}{1.3}
\bs \Delta \equiv
\begin{bmatrix}
\frac{\partial}{\partial x} & \frac{\partial}{\partial y}  & \frac{\partial}{\partial z}\\
\end{bmatrix}^\top, 
\end{equation} for the heat equation in 2-D and 3-D, respectively.

In this work, we are concerned with linear triangular elements, for which a single integration point in standard and integration elements is sufficient. The discrete system of linear equations $\bs K \bs U = \bs F$ is finally obtained through standard procedures, where
\begin{equation}
	\bs K = \assembly_{e} \bs k_e, \qquad \bs F = \assembly_{e} \bs f_e,
\end{equation}
and $\assembly$ denotes the standard finite element assembly operator.

For a more detailed description on IGFEM, the reader is referred to~\cite{Soghrati:2012}.

\subsection{Radial basis functions}

Although it is possible to directly use the levelset values $\phi_j$ on original nodes of the finite element mesh as design variables, we choose to use compactly supported radial basis functions for the levelset parametrization for a number of reasons~\cite{Wang:2006}:
\begin{enumerate}[\itshape i)]

\item RBFs give control over the complexity of the designs, and as such, they act similarly to a filter in density-based topology optimization;

\item By decoupling the finite element analysis mesh from the RBF grid, the design space can be restricted without deteriorating the finite element approximation. This can be used to mitigate approximation errors due to too coarse discretizations; and

\item By tuning the radius of support of RBFs, we can ensure that the influence of each design variable extends over multiple elements. This allows the optimizer to move the boundary further and therefore converge faster, while using fewer design variables.
\end{enumerate}

\begin{figure}
\centering
\includegraphics[scale=1.]{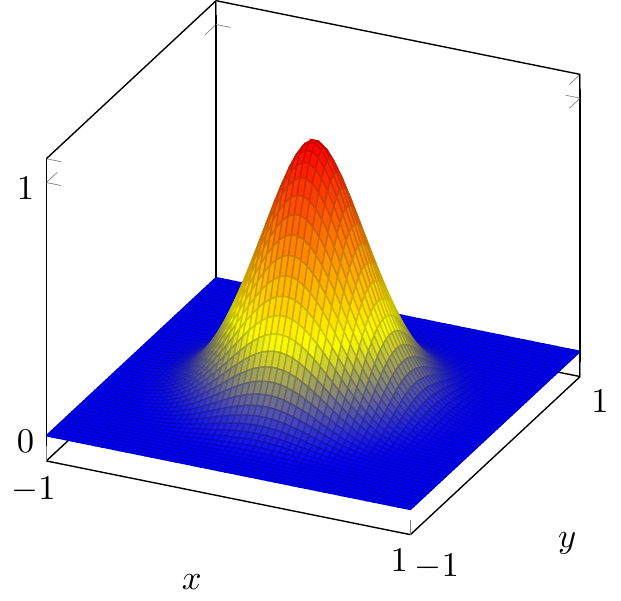}
\caption{Compactly supported RBF as given by Equation~\ref{eq:RBFs} with the coordinates $x_i = y_i = 0$ and the radius of influence $r_{\mathrm{s}}=1$.}
\label{fig:rbf}
\end{figure}

Figure~\ref{fig:rbf} illustrates a compactly supported RBF $\theta$~\cite{Wendland:1995} described by
\begin{equation}\label{eq:RBFs}
\theta_i \left( r_i\right) = \left(1-r_i\right)^4 \left( 4r_i + 1\right) ,
\end{equation} where the radius $r_i$ is defined as 
\begin{equation}\label{eq:radius}
r_i \left(\bs x, \bs x_i \right)= \frac{\sqrt{\left \| \bs x- \bs x_i \right \|}}{r_{\mathrm{s}}},
\end{equation} and $r_{\mathrm{s}}$ is the radius of support. In~\eqref{eq:radius} $\left\| \cdot \right\|$ denotes the Euclidian norm, and $\bs x_i$ the coordinate of the center of the RBF $\theta_i$. 

The scalar-valued levelset function $\phi\left(\bs x\right)$ is found as a summation of every non-zero RBF $\theta_i$, scaled with its corresponding design variable $s_i$:
\begin{equation}\label{eq:ls}
\phi \left( \bs x \right) = \bs \Theta \left( \bs x\right)^\intercal \bs s = \sum_{i \in \iota_s} \theta_i \left( \bs x\right) s_i,
\end{equation} \SJVDB{where $\iota_s$, the index set containing all design variables, and} $\bs s \in \bs{\mathcal{D}}$ is a vector of design variables with length \SJVDB{$\left| \iota_s \right|$, the cardinality of $\iota_s$}. Finally, evaluating this function at the original nodes of the finite element mesh results in the levelset vector
\begin{equation}
\bs \phi = \bs \theta^\intercal \bs s ,
\end{equation} where $\bs \theta$ is a matrix that needs to be computed only once.

\subsection{Optimization}

The optimization problem is chosen as a minimization of the compliance $C$ with respect to the design variables $\bs s$ that scale the radial basis functions. It needs to be emphasized that compliance minimization is merely a demonstrator problem, and the method is not limited to it. The minimization problem is subject to equilibrium and a volume constraint $V_c$. Furthermore, the design variables are bounded between $\bs s_{\text{min}}$ and  $\bs s_{\text{max}}$. This problem can be written as 
\begin{equation} \label{eq:opt}
\begin{aligned}
\bs s^\star =\argmin_{\bs{s} \in \bs{\mathcal{D}}} \; & &C =  \bs U^\intercal \bs K\left(\bs{s} \right) \bs U \\
\text{subject to} \; & & \bs K \bs U = \bs F\\
 & & V_{\Omega_\mathrm{m}} \leq V_c\\
 & & \bs s_{\text{min}} \leq \bs{s} \leq \bs s_{\text{max}}.
\end{aligned}
\end{equation} The Method of Moving Asymptotes (MMA)~\cite{Svanberg:1987} is employed to solve this optimization problem.

\subsubsection{Sensitivity analysis} \label{sec:sens}

The compliance minimization problem is self-adjoint~\cite{Bendsoe:2004}, resulting in the sensitivity of the compliance $C$ with respect to the design variables $\bs s$ as 
\begin{equation} \label{eq:dCdalpha}
\frac{\partial C}{\partial \bs s} = -\bs U^\intercal  \frac{\partial \bs K}{\partial \bs s} \bs U + 2 \bs U^\intercal \frac{\partial \bs F}{\partial \bs s} .
\end{equation} Applying the chain rule, the sensitivity of the compliance $C$ with respect to design variable $s_i$ can be written at the level of integration elements in terms of the nodal levelset values $\phi_j$: 
\begin{equation} \label{eq:dCdalpha2}
\begin{aligned}
\frac{\partial C}{\partial s_i} =   \sum_{j \in \iota_i }  \sum_{e \in \iota_j} \sum_{n \in \iota_e}  & \left( - \bs u_e^\intercal \frac{\partial \bs k_e}{\partial \bs x_n} \frac{\partial \bs x_n}{\partial \phi_j} \bs u_e \right.\\ &\left.+  2 \bs u_e^\intercal \frac{\partial \bs f_e}{\partial \bs x_n} \frac{\partial \bs x_n}{\partial \phi_j} \right)\frac{\partial \phi_j}{\partial s_i}.
\end{aligned}
\end{equation} In~\eqref{eq:dCdalpha2}, a summation is done over all the nodes in the index set $\iota_i$ which contains all the original mesh nodes that are in the support of the RBF corresponding to design variable $s_i$. Then, a summation is done over $\iota_j$, which refers to the index set of all integration elements $e$ in the support of original mesh node $j$, \textit{i.e.}, the region where the original shape function $N_j$ is nonzero. Lastly, a summation is done over the index set $\iota_e$, which contains all the enriched nodes $n$ in integration element $e$. The location of these enriched nodes is denoted  $\bs x_n$. Note that a number of terms can be identified in the sensitivity formulation: the derivatives of nodal levelset values with respect to the design variables, $\partial \phi_j / \partial s_i$, the design velocities $\partial \bs x_n / \partial \phi_j$, and the sensitivity of the element stiffness matrix and force vector with respect to the location of the $n$th enriched node, $\partial \bs k_e / \partial \bs x_n$ and $\partial \bs f_e / \partial \bs x_n$, respectively.

First, the sensitivity of the nodal levelset values with respect to the design variables is simply computed by taking the derivative of~\eqref{eq:ls} with respect to $\bs s$ as
\begin{equation} \label{eq:dphidalpha}
\frac{\partial \bs \phi}{\partial \bs s} = \bs \theta^\intercal .
\end{equation}
The design velocities $\partial \bs x_n / \partial \phi_j$ also remain straightforward as they are computed by taking the derivative of Eq.~\eqref{eq:discretizedLS} as
\begin{equation} \label{eq:dxdphi}
\frac{\partial \bs x_n}{\partial \phi_j} = - \frac{\phi_{k}}{\left(\phi_j - \phi_{k}\right) ^2} \left( \bs x_j - \bs x_{k}\right) .
\end{equation} 
Note that the enriched nodes remain on the element edges of the finite element mesh, and thus the direction of the design velocity is known \textit{a priori}. 

More involved is the sensitivity of the $e$th integration element stiffness matrix $\bs k_e$  with respect to the location of enriched node $n$, which for a single integration point can be written as:
\begin{equation} \label{eq:dKdx}
\frac{\partial \bs k_e}{\partial \bs x_n} = \frac{\partial j_e}{\partial \bs x_n} \bs B_e^\intercal \bs D_e \bs B_e + j_e \frac{\partial \bs B_e^\intercal}{\partial \bs x_n} \bs D_e \bs B_e + j_e \bs B_e^\intercal \bs D_e \frac{\partial \bs B_e}{\partial \bs x_n}  ,
\end{equation}\SJVDB{where $\bs B_e = \left[ \Delta \bs N_e \bs J_p^{-1} \;\; \Delta \bs \psi_e \bs J_e^{-1}\right]$ and $\bs J_p$ and $\bs J_e$ are the Jacobian of the parent and integration element, respectively.} Recall that the material within each integration element remains constant, and therefore $\partial \bs D_e / \partial \bs x_n = \bs 0$. The first term in~\eqref{eq:dKdx} contains the sensitivity of the Jacobian determinant, and represents the effect of the changing integration element area; the second and third term contain the sensitivity of the element $\bs B_e$ matrix, and represents the effect of the changing shape and enrichment functions. The latter is computed as
\begin{equation} \label{eq:dBdx}
\frac{\partial \bs B_e}{\partial \bs x_n} = \left[ \bs 0  \;\;
\;\; \bs \Delta \bs \psi_e\frac{\partial \bs J_e^{-1}}{\partial \bs x_n} \right] .
\end{equation} Observe that only the enriched part of the formulation has an influence, as for linear elements the background shape function derivatives are constant throughout the element ($\partial \Delta \bs N_e / \partial \bs n = \bs 0$). The Jacobian of the parent element is not influenced by the enriched node location either ($\partial \bs J_p / \partial \bs x_n = \bs 0$). The enrichment functions have a constant value at the integration point of the integration element ($\partial \Delta \bs \psi_e / \partial \bs n = \bs 0$). Appendix~\ref{appendix} describes how to compute the derivative of the Jacobian inverse and determinant, $\partial \bs J_e^{-1} / \partial \bs x_n$ and $\partial j_e / \partial \bs x_n$, respectively, by straightforward differentiation.

Finally, the sensitivity of the design-dependent force vector $\bs f_e$ is evaluated. Due to the IGFEM discretization, enriched nodes \SJVDB{whose support is subjected to a line or body load} contribute to the force vector, implying that the derivatives of the force vector are nonzero for cases with line loads or body forces. Similarly to the sensitivity of the element stiffness matrix, the sensitivity of the element force vector consists of two terms, one relating to the Jacobian derivative, and another containing the function derivatives:
\begin{equation} \label{eq:dfdx}
\frac{\partial \bs f_e}{\partial \bs x_n} = \frac{\partial j_e}{\partial \bs x_n} \left[\bs N_e^\intercal \;\;
\;\; \bs \psi^\intercal \right] \bs b_e + j_e \left[\frac{\partial \bs N_e^\intercal}{\partial \bs x_n} \;\;
\;\; \bs 0^\intercal \right] \bs b_e .
\end{equation} Here, the element right hand side \SJVDB{$\bs f_e$} is computed as a function of body sources, but line loads \SJVDB{and tractions} can be handled completely analogously. In the second term, only the original shape functions have a contribution. This is because  enriched functions always have the same value on the Gauss point of an integration element. However, as the location of the Gauss points with respect to the background element changes, $\partial \bs N_e / \partial \bs x_n$ is nonzero, and can be evaluated as 
\begin{equation} \label{eq:dNdx}
\frac{\partial \bs N_e}{\partial \bs x_n} =  \bs \Delta \bs N_e \bs A \frac{\partial \bs x_e}{\partial \bs x_n} \bs N_e ,
\end{equation} where $\bs A$ is the inverse isoparametric mapping that maps global coordinates to the local \SJVDB{master} coordinate system of the parent element.

Although the sensitivity analysis seems involved, every partial derivative is relatively \SJVDB{straightforward} and cheap to compute.

\section{Numerical examples}
The enriched method outlined above is demonstrated on a number of classical compliance optimization problems. The results generated by this approach are compared to those generated by open source optimization codes, and the influence of the design discretization is investigated. A 3-D compliance optimization case and a heat sink problem are also considered. 

In this section, no units are specified; therefore, any consistent unit system can be \SJVDB{assumed}. For the MMA optimizer \cite{Svanberg:1987}, the following settings are used unless otherwise specified:
\begin{itemize}
\item The lower and upper bounds on the design variables $s_i$ are given by $ s_{\text{min},i} = -1$ and $s_{\text{max},i} = 1$ for $i = 1,\ldots,I$;
\item The move limit used by MMA is set to $0.01$;
\item A value of $10$ is used for the coefficient that controls how aggressively the constraints are enforced.
\end{itemize}

\subsection{Cantilever beam}
\label{sec:cantilever}
\begin{figure}
\centering
\includegraphics[scale=1.]{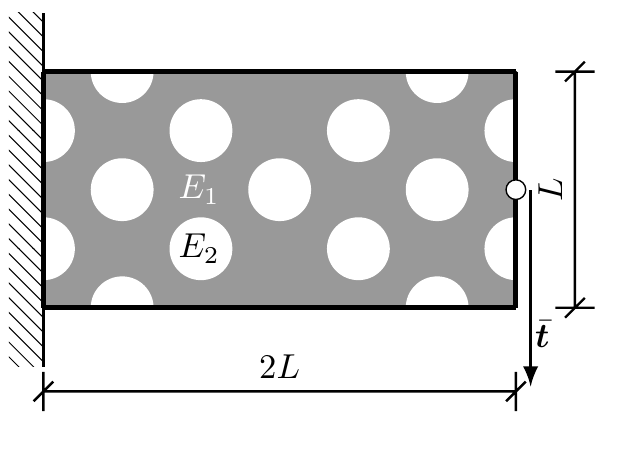}
\caption{Problem description and initial design for the cantilever beam example in \S\ref{sec:cantilever}. The domain is clamped on the left and a downward force is applied in the middle of the right side.}
\label{fig:cantilever}
\end{figure}

\begin{figure*}
\centering
\def\svgwidth{\textwidth}
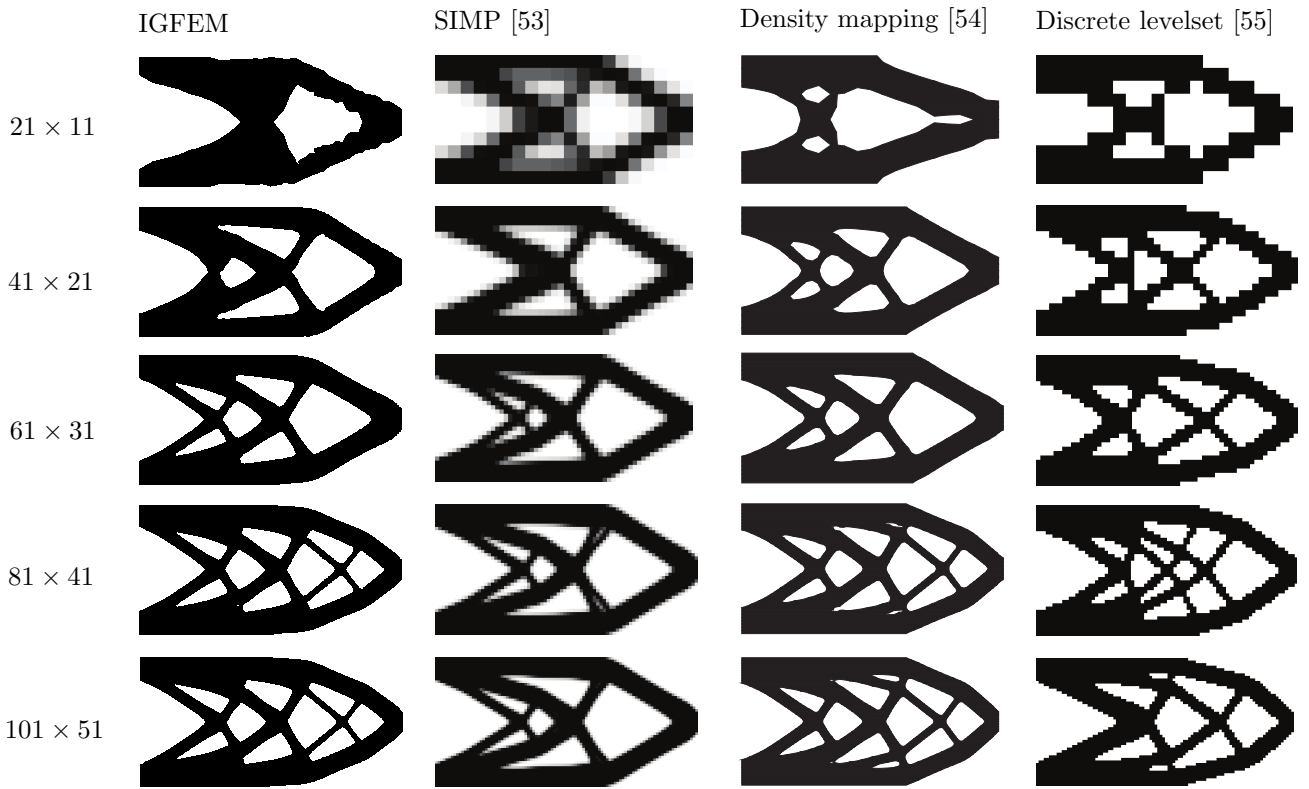
\caption{Final designs for a cantilever beam obtained by \SJVDB{the proposed method} and the other methods considered in this study. In the columns, the different methods are shown: \SJVDB{our IGFEM-based method}, SIMP, a levelset method with density mapping, and a discrete levelset method. The rows show final designs obtained on meshes defined on grids of $21 \time 11$, $41 \times 21$, $61 \times 31$, $81 \times 41$, and $101 \times 51$ nodes, respectively. }
\label{fig:cantilever_beam}
\end{figure*}

First, \SJVDB{our approach to enriched} levelset-based topology optimization is compared to the following open source codes: \begin{inparaenum}[\it i)]\item the 99-line SIMP-based code by Sigmund~\cite{sigmund:2001}; \item an 88-line code for parameterized levelset optimization using radial-basis functions and density mapping, proposed by Wei \textit{et al.}~\cite{wei:2018}; and \item a code for discrete levelset topology optimization with topological derivatives by Challis~\cite{challis:2010}.\end{inparaenum}

The optimization problem for this comparison is the widely-used cantilever beam problem, as illustrated in Figure~\ref{fig:cantilever}. It consists of a $2L \times L $ rectangular domain that is clamped on the left and subjected to a downward point load \SJVDB{$\bar{\bs t}$} in the middle of the right side.  We set $L$ equal to $1$,  the volume constraint to 55\% of the design domain volume, and use \SJVDB{$\left| \bar{\bs t}\right| = 1$.} The material domain $\Omega_\mathrm{m}$ is assigned a Young's modulus $E_1 = 1$, whereas the void domain $\Omega_\mathrm{v}$ has Young's modulus $E_2 = 10^{-6}$. Both domains have a Poisson ratio $\nu_1 = \nu_2 = 0.3$. Note that it is also possible to give the void regions a stiffness of exactly zero by removing DOFs~\cite{vandenBoom:2018a}. However, this would entail extra overhead, and to ensure a fair comparison with the other models, in this work it is chosen to use a small value for the void stiffness. 

Figure~\ref{fig:cantilever} shows the initial design that is used for the IGFEM-based optimization, which is the same as that used in the paper describing the 88-line code \cite{wei:2018}. The other two codes do not require an initial design, as they are able to nucleate holes. The optimization problem is solved on meshes defined on rectangular grids of $21 \times 11$, $41 \times 21$, $61 \times 31$, $81 \times 41$, and $101 \times 51$ nodes. \SJVDB{Our proposed method} makes use of triangular meshes, whereas the other methods use quadrilateral meshes. The RBF mesh used in the IGFEM\SJVDB{-based} solutions is the same as the analysis mesh, and a radius of influence of $r_s = \sqrt{2} \cdot a $ is used, where $a$ is the distance between two RBFs.

\begin{figure*}
\centering
\subfloat[][]{\begin{tikzpicture}[scale =1]
\pgfplotsset{
    scale only axis,
    xmin=0, xmax=80
}

\begin{axis}[
  axis y line*=left,
  ymin=20, ymax=250,
  xlabel=Iteration,
  ylabel= Compliance $C$,
  ylabel style={xshift=0cm, yshift=-0.25cm},
  width = 6cm,
            legend cell align=right,
            legend plot pos=right,
legend style={at={(0.28, 0.90)}, anchor=north west,  draw=none, fill=none, text height=1.5ex, text depth=0.5ex}
]
\addplot[smooth,black]
  coordinates{
( 0 , 102.361064 )
( 1 , 86.272391 )
( 2 , 123.194061 )
( 3 , 96.628964 )
( 4 , 79.618163 )
( 5 , 68.954579 )
( 6 , 64.111391 )
( 7 , 60.92604 )
( 8 , 59.125404 )
( 9 , 57.655222 )
( 10 , 56.99311 )
( 11 , 56.300322 )
( 12 , 55.918934 )
( 13 , 55.56875 )
( 14 , 55.471107 )
( 15 , 55.275404 )
( 16 , 55.260435 )
( 17 , 55.166647 )
( 18 , 55.178865 )
( 19 , 55.113856 )
( 20 , 55.026369 )
( 21 , 54.998629 )
( 22 , 54.930329 )
( 23 , 54.901562 )
( 24 , 54.916421 )
( 25 , 54.882898 )
( 26 , 54.905511 )
( 27 , 54.889493 )
( 28 , 54.940786 )
( 29 , 57.409329 )
( 30 , 56.948135 )
( 31 , 55.947793 )
( 32 , 55.49371 )
( 33 , 55.268216 )
( 34 , 55.170517 )
( 35 , 55.116222 )
( 36 , 55.053193 )
( 37 , 55.037891 )
( 38 , 54.998002 )
( 39 , 54.979347 )
( 40 , 54.966521 )
( 41 , 54.987446 )
( 42 , 54.95496 )
( 43 , 57.459499 )
( 44 , 56.545337 )
( 45 , 55.560879 )
( 46 , 55.267632 )
( 47 , 55.178421 )
( 48 , 55.147462 )
( 49 , 55.113562 )
( 50 , 55.095756 )
( 51 , 55.080802 )
( 52 , 55.06582 )
( 53 , 55.044179 )
( 54 , 55.037123 )
( 55 , 55.07177 )
( 56 , 55.041783 )
( 57 , 55.023687 )
( 58 , 55.022698 )
( 59 , 55.015906 )
( 60 , 55.021625 )
( 61 , 55.025231 )
( 62 , 55.025587 )
( 63 , 55.008731 )
( 64 , 55.027843 )
( 65 , 55.030408 )
( 66 , 54.998931 )
( 67 , 55.020512 )
( 68 , 54.996597 )
( 69 , 55.005362 )
( 70 , 54.997656 )
( 71 , 54.988949 )
( 72 , 54.979313 )
( 73 , 55.003612 )
( 74 , 54.983738 )
( 75 , 54.997131 )
( 76 , 54.996617 )
( 77 , 54.987423 )
( 78 , 54.978938 )
( 79 , 54.989711 )
( 80 , 54.979812 )
( 81 , 54.986357 )
( 82 , 54.982096 )
( 83 , 54.973052 )
( 84 , 54.977155 )
( 85 , 54.981328 )
( 86 , 54.970517 )
( 87 , 54.97278 )
( 88 , 54.983823 )
( 89 , 54.962663 )
( 90 , 54.992414 )
( 91 , 54.988679 )
( 92 , 54.961455 )
( 93 , 54.98307 )
( 94 , 54.981513 )
( 95 , 54.985268 )
( 96 , 54.980395 )
( 97 , 54.971449 )
( 98 , 54.948832 )
( 99 , 54.975724 )
( 100 , 54.971302 )
( 101 , 54.989021 )
( 102 , 54.983278 )
( 103 , 54.969516 )
( 104 , 54.923269 )
( 105 , 54.968512 )
( 106 , 54.966544 )
( 107 , 55.004201 )
( 108 , 54.999859 )
( 109 , 54.997833 )
( 110 , 54.971251 )
( 111 , 54.990284 )
( 112 , 55.015894 )
( 113 , 54.975011 )
( 114 , 54.992659 )
( 115 , 54.98772 )
( 116 , 54.988145 )
( 117 , 54.917527 )
( 118 , 55.023253 )
( 119 , 54.986241 )
( 120 , 55.027192 )
( 121 , 54.973371 )
( 122 , 54.926846 )
( 123 , 54.958745 )
( 124 , 54.967909 )
( 125 , 54.971419 )
( 126 , 54.959487 )
( 127 , 54.960981 )
( 128 , 54.952369 )
( 129 , 54.946037 )
( 130 , 54.985504 )
( 131 , 54.956964 )
( 132 , 55.05767 )
( 133 , 54.984676 )
( 134 , 54.966621 )
( 135 , 54.959478 )
( 136 , 54.979621 )
( 137 , 54.98566 )
( 138 , 54.959705 )
( 139 , 54.959211 )
( 140 , 54.970467 )
( 141 , 54.969998 )
( 142 , 54.976987 )
( 143 , 54.962041 )
( 144 , 54.957573 )
( 145 , 54.959517 )
( 146 , 54.980816 )
( 147 , 54.972117 )
( 148 , 54.971965 )
( 149 , 54.976553 )
( 150 , 54.964414 )
( 151 , 54.964913 )
( 152 , 54.952801 )
( 153 , 54.991188 )
( 154 , 54.97106 )
( 155 , 54.967855 )
( 156 , 54.948816 )
( 157 , 54.977732 )
( 158 , 54.956239 )
( 159 , 54.993531 )
( 160 , 54.969375 )
( 161 , 54.969767 )
( 162 , 54.963064 )
( 163 , 54.966039 )
( 164 , 54.966883 )
( 165 , 54.98417 )
( 166 , 54.981845 )
( 167 , 54.973873 )
( 168 , 54.971658 )
( 169 , 54.954995 )
( 170 , 54.951116 )
( 171 , 55.030203 )
( 172 , 55.00908 )
( 173 , 54.982179 )
( 174 , 54.965896 )
( 175 , 54.972787 )
( 176 , 54.987313 )
( 177 , 54.985498 )
( 178 , 54.989999 )
( 179 , 54.969674 )
( 180 , 54.963076 )
( 181 , 54.954456 )
( 182 , 54.953935 )
( 183 , 54.956673 )
( 184 , 54.985414 )
( 185 , 54.967336 )
( 186 , 54.961976 )
( 187 , 54.950358 )
( 188 , 54.954963 )
( 189 , 54.976614 )
( 190 , 54.977507 )
( 191 , 54.956215 )
( 192 , 54.965836 )
( 193 , 54.958678 )
( 194 , 54.954965 )
( 195 , 55.005318 )
( 196 , 54.989444 )
( 197 , 54.978934 )
( 198 , 54.96005 )
( 199 , 54.951276 )
( 200 , 54.950295 )
}; \label{plot_0}
\addlegendentry{IGFEM}
\addplot[smooth,blue]
  coordinates{
( 1,237.6724)
( 2,154.9407)
( 3,114.5690)
( 4, 98.8262)
( 5, 90.9340)
( 6, 84.0245)
( 7, 79.2354)
( 8, 75.9516)
( 9, 73.8634)
(10, 72.2894)
(11, 70.9350)
(12, 69.6216)
(13, 68.2531)
(14, 67.0775)
(15, 65.9829)
(16, 64.9452)
(17, 63.7923)
(18, 62.6161)
(19, 61.5415)
(20, 60.8276)
(21, 60.4902)
(22, 60.3184)
(23, 60.2417)
(24, 60.2060)
(25, 60.1689)
(26, 60.1497)
(27, 60.1246)
(28, 60.1059)
(29, 60.0941)
(30, 60.0892)
(31, 60.0917)
(32, 60.0804)
(33, 60.0804)
(34, 60.0807)
(35, 60.0850)
(36, 60.0913)
(37, 60.0882)
(38, 60.0945)
(39, 60.0899)
(40, 60.0961)
(41, 60.0937)
(42, 60.0927)
(43, 60.0901)
(44, 60.0895)
(45, 60.0899)
(46, 60.0908)
(47, 60.0910)
(48, 60.0904)
(49, 60.0892)
(50, 60.0985)
(51, 60.0882)
(52, 60.0825)
(53, 60.0773)
(54, 60.0836)
(55, 60.0691)
(56, 60.0718)
(57, 60.0676)
(58, 60.0642)
(59, 60.0598)
(60, 60.0690)
(61, 60.0706)
(62, 60.0715)
(63, 60.0711)
(64, 60.0702)
(65, 60.0686)
(66, 60.0785)
(67, 60.0686)
(68, 60.0767)
(69, 60.0767)
(70, 60.0792)
(71, 60.0814)
(72, 60.0835)
(73, 60.0842)
(74, 60.0964)
(75, 60.1009)
(76, 60.0924)
(77, 60.1007)
(78, 60.1020)
(79, 60.1037)
(80, 60.1048)
(81, 60.1064)
(82, 60.1074)
(83, 60.1081)
(84, 60.1084)
(85, 60.1205)
(86, 60.1129)
(87, 60.1243)
(88, 60.1145)
}; \label{plot_zero}
\addlegendentry{SIMP~\cite{sigmund:2001}}
\addplot[smooth,red]
  coordinates{
(1, 103.497570)
(2, 61.194241)
(3, 56.035854)
(4, 51.809971)
(5, 49.484049)
(6, 49.166037)
(7, 48.427271)
(8, 48.124486)
(9, 47.891308)
(10, 47.936747)
(11, 48.060963)
(12, 48.233575)
(13, 48.468596)
(14, 48.671649)
(15, 48.878869)
(16, 49.104775)
(17, 49.323779)
(18, 49.553504)
(19, 49.791375)
(20, 50.034204)
(21, 50.283937)
(22, 50.550978)
(23, 50.813775)
(24, 51.091896)
(25, 51.359383)
(26, 51.636397)
(27, 51.942850)
(28, 52.180764)
(29, 52.491595)
(30, 52.793017)
(31, 53.083710)
(32, 53.352429)
(33, 53.599588)
(34, 53.832583)
(35, 54.050738)
(36, 54.225820)
(37, 54.411935)
(38, 54.596026)
(39, 54.786992)
(40, 54.960591)
(41, 55.095716)
(42, 55.251324)
(43, 55.387168)
(44, 55.502704)
(45, 55.569138)
(46, 55.638135)
(47, 55.687144)
(48, 55.725431)
(49, 55.703683)
(50, 55.680977)
(51, 55.625379)
(52, 55.586082)
(53, 55.563617)
(54, 55.512819)
(55, 55.443275)
(56, 55.379365)
(57, 55.307871)
(58, 55.246577)
(59, 55.247811)
(60, 55.246379)
(61, 55.231993)
(62, 55.238854)
(63, 55.255916)
(64, 55.295393)
(65, 55.341446)
(66, 55.379968)
(67, 55.399002)
(68, 55.407372)
(69, 55.398909)
(70, 55.385751)
(71, 55.380680)
(72, 55.361098)
(73, 55.349684)
(74, 55.339207)
(75, 55.326234)
(76, 55.297191)
(77, 55.301252)
(78, 55.298935)
(79, 55.322708)
(80, 55.340817)
}; \label{plot_one}
\addlegendentry{Wei \textit{et al.}~\cite{wei:2018}}
\addplot[smooth, teal]
  coordinates{
(1, 39.5427)
(2, 39.5982)
(3, 39.6413)
(4, 39.7170)
(5, 39.7546)
(6, 39.8368)
(7, 39.9105)
(8, 39.9979)
(9, 40.0154)
(10, 40.1595)
(11, 40.2265)
(12, 40.3352)
(13, 40.4239)
(14, 40.6406)
(15, 40.7101)
(16, 40.9317)
(17, 41.1147)
(18, 41.4579)
(19, 41.6509)
(20, 41.9556)
(21, 44.3685)
(22, 45.7735)
(23, 47.4076)
(24, 47.8157)
(25, 49.0802)
(26, 51.4009)
(27, 51.9141)
(28, 54.7136)
(29, 55.7107)
(30, 56.4038)
(31, 57.2025)
(32, 57.6724)
(33, 58.1969)
(34, 58.4518)
(35, 58.7442)
(36, 58.8424)
(37, 59.1597)
(38, 59.1597)
(39, 59.1597)
(40, 59.1597)
(41, 59.1597)
(42, 59.1597)
(43, 59.1597)
(44, 59.0083)
(45, 58.7326)
(46, 58.7326)
(47, 58.7326)
(48, 58.7326)
(49, 58.7326)
(50, 58.7326)
(51, 58.7326)
(52, 58.5178)
(53, 58.3383)
(54, 58.2295)
(55, 57.6837)
(56, 57.6837)
(57, 57.6837)
(58, 57.6837)
(59, 57.5748)
(60, 57.5748)
}; \label{plot_two}
\addlegendentry{Challis~\cite{challis:2010}}
\end{axis}

\begin{axis}[
  axis y line*=right,
  axis x line=none,
  ymin=0, ymax=1,
  ylabel=Volume fraction $V_{\Omega_\mathrm{m}} / V_\Omega$,
  ylabel style={xshift=0cm, yshift=-8.25cm},
  yticklabel pos=right,
  width = 6cm,
  legend cell align=right,
  legend plot pos=right,
  legend style={at={(0.82, 0.90)}, anchor=north west, draw=none, fill=none, text height=1.5ex, text depth=.5ex}
]

\addplot[smooth, black, dashed]
  coordinates{
( 0 , 0.700828 )
( 1 , 0.657799 )
( 2 , 0.553095 )
( 3 , 0.544747 )
( 4 , 0.545738 )
( 5 , 0.549564 )
( 6 , 0.544974 )
( 7 , 0.545746 )
( 8 , 0.544798 )
( 9 , 0.546887 )
( 10 , 0.545203 )
( 11 , 0.546401 )
( 12 , 0.547382 )
( 13 , 0.548187 )
( 14 , 0.548364 )
( 15 , 0.548781 )
( 16 , 0.548943 )
( 17 , 0.549115 )
( 18 , 0.548696 )
( 19 , 0.548658 )
( 20 , 0.549151 )
( 21 , 0.549211 )
( 22 , 0.549884 )
( 23 , 0.549839 )
( 24 , 0.549612 )
( 25 , 0.549742 )
( 26 , 0.549258 )
( 27 , 0.549461 )
( 28 , 0.549201 )
( 29 , 0.549038 )
( 30 , 0.542734 )
( 31 , 0.548314 )
( 32 , 0.549162 )
( 33 , 0.549472 )
( 34 , 0.549565 )
( 35 , 0.549531 )
( 36 , 0.549691 )
( 37 , 0.549758 )
( 38 , 0.549688 )
( 39 , 0.549761 )
( 40 , 0.549831 )
( 41 , 0.549431 )
( 42 , 0.549603 )
( 43 , 0.549661 )
( 44 , 0.544795 )
( 45 , 0.548453 )
( 46 , 0.550204 )
( 47 , 0.549903 )
( 48 , 0.549608 )
( 49 , 0.549618 )
( 50 , 0.549652 )
( 51 , 0.549704 )
( 52 , 0.549755 )
( 53 , 0.549894 )
( 54 , 0.549903 )
( 55 , 0.549561 )
( 56 , 0.549758 )
( 57 , 0.549917 )
( 58 , 0.549851 )
( 59 , 0.549849 )
( 60 , 0.549823 )
( 61 , 0.549669 )
( 62 , 0.549708 )
( 63 , 0.549896 )
( 64 , 0.5496 )
( 65 , 0.549545 )
( 66 , 0.549877 )
( 67 , 0.549692 )
( 68 , 0.549828 )
( 69 , 0.549762 )
( 70 , 0.549773 )
( 71 , 0.549917 )
( 72 , 0.550025 )
( 73 , 0.549648 )
( 74 , 0.549893 )
( 75 , 0.549779 )
( 76 , 0.549715 )
( 77 , 0.549801 )
( 78 , 0.549914 )
( 79 , 0.549754 )
( 80 , 0.549876 )
( 81 , 0.549809 )
( 82 , 0.549835 )
( 83 , 0.549937 )
( 84 , 0.549857 )
( 85 , 0.549848 )
( 86 , 0.549943 )
( 87 , 0.549911 )
( 88 , 0.549791 )
( 89 , 0.550005 )
( 90 , 0.549715 )
( 91 , 0.549707 )
( 92 , 0.54995 )
( 93 , 0.549806 )
( 94 , 0.549837 )
( 95 , 0.54976 )
( 96 , 0.549752 )
( 97 , 0.549855 )
( 98 , 0.550191 )
( 99 , 0.549842 )
( 100 , 0.549803 )
( 101 , 0.549667 )
( 102 , 0.549715 )
( 103 , 0.54994 )
( 104 , 0.550474 )
( 105 , 0.54986 )
( 106 , 0.549867 )
( 107 , 0.549723 )
( 108 , 0.54954 )
( 109 , 0.549554 )
( 110 , 0.549842 )
( 111 , 0.550038 )
( 112 , 0.549283 )
( 113 , 0.549764 )
( 114 , 0.549768 )
( 115 , 0.549533 )
( 116 , 0.549815 )
( 117 , 0.550644 )
( 118 , 0.549876 )
( 119 , 0.54969 )
( 120 , 0.54969 )
( 121 , 0.5499 )
( 122 , 0.550421 )
( 123 , 0.54993 )
( 124 , 0.549905 )
( 125 , 0.549753 )
( 126 , 0.549893 )
( 127 , 0.549892 )
( 128 , 0.549963 )
( 129 , 0.550069 )
( 130 , 0.549627 )
( 131 , 0.549927 )
( 132 , 0.549488 )
( 133 , 0.549814 )
( 134 , 0.549865 )
( 135 , 0.549921 )
( 136 , 0.549723 )
( 137 , 0.549721 )
( 138 , 0.549899 )
( 139 , 0.549907 )
( 140 , 0.549844 )
( 141 , 0.549873 )
( 142 , 0.54976 )
( 143 , 0.54992 )
( 144 , 0.549928 )
( 145 , 0.549865 )
( 146 , 0.5498 )
( 147 , 0.549881 )
( 148 , 0.549863 )
( 149 , 0.549801 )
( 150 , 0.54988 )
( 151 , 0.549848 )
( 152 , 0.54996 )
( 153 , 0.549689 )
( 154 , 0.549891 )
( 155 , 0.549865 )
( 156 , 0.550055 )
( 157 , 0.549714 )
( 158 , 0.54994 )
( 159 , 0.54968 )
( 160 , 0.549916 )
( 161 , 0.549844 )
( 162 , 0.54989 )
( 163 , 0.549848 )
( 164 , 0.549786 )
( 165 , 0.549795 )
( 166 , 0.549785 )
( 167 , 0.549833 )
( 168 , 0.549834 )
( 169 , 0.54995 )
( 170 , 0.549953 )
( 171 , 0.549709 )
( 172 , 0.549757 )
( 173 , 0.549785 )
( 174 , 0.54985 )
( 175 , 0.549776 )
( 176 , 0.549725 )
( 177 , 0.549771 )
( 178 , 0.549713 )
( 179 , 0.549863 )
( 180 , 0.549893 )
( 181 , 0.549918 )
( 182 , 0.549939 )
( 183 , 0.55001 )
( 184 , 0.54964 )
( 185 , 0.549866 )
( 186 , 0.549872 )
( 187 , 0.54994 )
( 188 , 0.549947 )
( 189 , 0.549727 )
( 190 , 0.549682 )
( 191 , 0.549943 )
( 192 , 0.549852 )
( 193 , 0.549942 )
( 194 , 0.549949 )
( 195 , 0.549785 )
( 196 , 0.549648 )
( 197 , 0.549722 )
( 198 , 0.549896 )
( 199 , 0.549955 )
( 200 , 0.549931 )
}; \addlegendentry{ }
\addplot[smooth, blue, dashed]
  coordinates{
( 1, 0.550)
( 2, 0.550)
( 3, 0.550)
( 4, 0.550)
( 5, 0.550)
( 6, 0.550)
( 7, 0.550)
( 8, 0.550)
( 9, 0.550)
(10, 0.550)
(11, 0.550)
(12, 0.550)
(13, 0.550)
(14, 0.550)
(15, 0.550)
(16, 0.550)
(17, 0.550)
(18, 0.550)
(19, 0.550)
(20, 0.550)
(21, 0.550)
(22, 0.550)
(23, 0.550)
(24, 0.550)
(25, 0.550)
(26, 0.550)
(27, 0.550)
(28, 0.550)
(29, 0.550)
(30, 0.550)
(31, 0.550)
(32, 0.550)
(33, 0.550)
(34, 0.550)
(35, 0.550)
(36, 0.550)
(37, 0.550)
(38, 0.550)
(39, 0.550)
(40, 0.550)
(41, 0.550)
(42, 0.550)
(43, 0.550)
(44, 0.550)
(45, 0.550)
(46, 0.550)
(47, 0.550)
(48, 0.550)
(49, 0.550)
(50, 0.550)
(51, 0.550)
(52, 0.550)
(53, 0.550)
(54, 0.550)
(55, 0.550)
(56, 0.550)
(57, 0.550)
(58, 0.550)
(59, 0.550)
(60, 0.550)
(61, 0.550)
(62, 0.550)
(63, 0.550)
(64, 0.550)
(65, 0.550)
(66, 0.550)
(67, 0.550)
(68, 0.550)
(69, 0.550)
(70, 0.550)
(71, 0.550)
(72, 0.550)
(73, 0.550)
(74, 0.550)
(75, 0.550)
(76, 0.550)
(77, 0.550)
(78, 0.550)
(79, 0.550)
(80, 0.550)
(81, 0.550)
(82, 0.550)
(83, 0.550)
(84, 0.550)
(85, 0.550)
(86, 0.550)
(87, 0.550)
(88, 0.550)
}; \addlegendentry{ }
\addplot[smooth,red, dashed]
  coordinates{
(1, 0.700277)
(2, 0.778851)
(3, 0.739987)
(4, 0.709932)
(5, 0.714689)
(6, 0.695281)
(7, 0.697934)
(8, 0.693328)
(9, 0.693046)
(10, 0.688108)
(11, 0.682915)
(12, 0.677400)
(13, 0.671844)
(14, 0.667717)
(15, 0.663039)
(16, 0.657891)
(17, 0.652945)
(18, 0.648080)
(19, 0.643187)
(20, 0.638312)
(21, 0.633331)
(22, 0.628360)
(23, 0.623585)
(24, 0.618566)
(25, 0.613734)
(26, 0.608904)
(27, 0.603741)
(28, 0.599690)
(29, 0.594669)
(30, 0.590005)
(31, 0.585344)
(32, 0.581114)
(33, 0.577302)
(34, 0.573754)
(35, 0.570438)
(36, 0.567750)
(37, 0.565001)
(38, 0.562273)
(39, 0.559544)
(40, 0.557057)
(41, 0.555059)
(42, 0.552814)
(43, 0.550852)
(44, 0.549176)
(45, 0.548151)
(46, 0.547135)
(47, 0.546397)
(48, 0.545780)
(49, 0.545936)
(50, 0.546180)
(51, 0.546825)
(52, 0.547246)
(53, 0.547496)
(54, 0.548088)
(55, 0.548982)
(56, 0.549781)
(57, 0.550693)
(58, 0.551502)
(59, 0.551476)
(60, 0.551459)
(61, 0.551625)
(62, 0.551534)
(63, 0.551277)
(64, 0.550746)
(65, 0.550111)
(66, 0.549589)
(67, 0.549312)
(68, 0.549169)
(69, 0.549274)
(70, 0.549418)
(71, 0.549463)
(72, 0.549675)
(73, 0.549811)
(74, 0.549919)
(75, 0.550076)
(76, 0.550431)
(77, 0.550380)
(78, 0.550370)
(79, 0.550043)
(80, 0.549778)
}; \addlegendentry{}
\addplot[smooth,teal, dashed]
  coordinates{
(1, 1.000)
(2, 0.970)
(3, 0.958)
(4, 0.946)
(5, 0.942)
(6, 0.933)
(7, 0.928)
(8, 0.922)
(9, 0.921)
(10, 0.912)
(11, 0.908)
(12, 0.902)
(13, 0.898)
(14, 0.889)
(15, 0.886)
(16, 0.878)
(17, 0.871)
(18, 0.859)
(19, 0.853)
(20, 0.844)
(21, 0.801)
(22, 0.761)
(23, 0.728)
(24, 0.716)
(25, 0.684)
(26, 0.660)
(27, 0.646)
(28, 0.620)
(29, 0.602)
(30, 0.584)
(31, 0.568)
(32, 0.560)
(33, 0.552)
(34, 0.547)
(35, 0.542)
(36, 0.541)
(37, 0.538)
(38, 0.538)
(39, 0.538)
(40, 0.538)
(41, 0.538)
(42, 0.538)
(43, 0.538)
(44, 0.539)
(45, 0.540)
(46, 0.540)
(47, 0.540)
(48, 0.540)
(49, 0.540)
(50, 0.540)
(51, 0.540)
(52, 0.542)
(53, 0.543)
(54, 0.544)
(55, 0.549)
(56, 0.549)
(57, 0.549)
(58, 0.549)
(59, 0.550)
(60, 0.550)
}; \addlegendentry{}
\end{axis}
\node[draw=none] at (4.1,4.8) {$C$};
\node[draw=none] at (5.4, 4.8) {$V_{\Omega_\mathrm{m}} / V_\Omega$};
\end{tikzpicture}\label{fig:comp_conva}}\;\;
\subfloat[][]{\begin{tikzpicture}
\begin{semilogxaxis}[
  width = 7.6cm,
  ymin=50, ymax=70,
xmin=400, xmax=13000,
  xlabel=DOFs,
  ylabel=Compliance $C$,
legend style={at={(0.99, 0.99)}, anchor=north east, draw=none, fill=none}
]
\addplot[black, mark=*]
  coordinates{
(752, 56.99831)
(2472, 55.424173)
(4964, 54.950295)
(8720, 54.979812)
(12944, 55.190879)
}; \label{plot_0}
\addlegendentry{IGFEM}
\addplot[blue, mark=square*]
  coordinates{
( 462, 68.5859)
( 1722, 61.1179)
( 3782, 60.1145)
( 6642, 59.1095)
( 10302, 58.6506)
}; \label{plot_zero}
\addlegendentry{SIMP \cite{sigmund:2001}}
\addplot[red, mark = triangle*]
  coordinates{
( 462, 57.575547)
( 1722, 55.819577)
( 3782, 55.331784)
( 6642, 55.614283)
( 10302, 55.832185)
}; \label{plot_one}
\addlegendentry{Wei \textit{et al.} \cite{wei:2018}}
\addplot[teal, mark=diamond*]
  coordinates{
( 462, 61.4341)
( 1722, 58.5638 )
( 3782 ,57.5748)
( 6642, 57.1234)
( 10302, 57.0177)
}; \label{plot_two}
\addlegendentry{Challis \cite{challis:2010}}
\end{semilogxaxis}
\end{tikzpicture}\label{fig:comp_convb}}
\caption{\SJVDB{Results of the cantilever beam problem; (a) shows the compliance and volume ratio as a function of the number of iterations, showing the convergence of the different methods considered in the study, (b) illustrates the final compliance of the cantilever beam as a function of the number of DOFs, optimized using the different methods considered in \S\ref{sec:cantilever}.}}\label{fig:comp_conv}
\end{figure*}
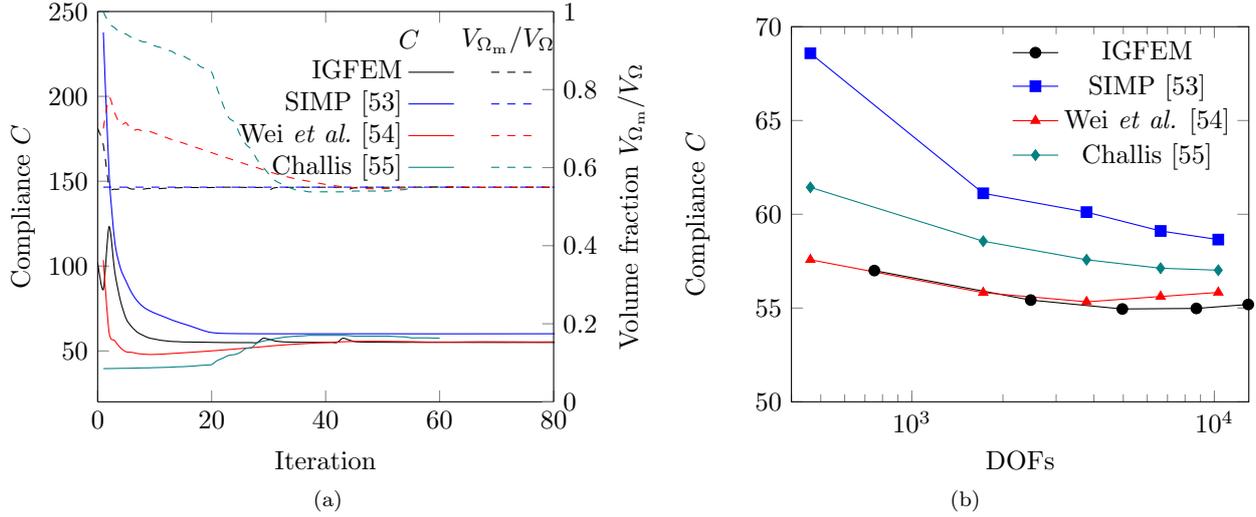

The results for each code are illustrated in Figure~\ref{fig:cantilever_beam}. For all methods, the design becomes more detailed when the mesh size is increased. Furthermore, the topologies obtained by each method are roughly the same. It is observed that the resulting designs are similar to those given by the code of Wei~\textit{et al.}, especially for the finer meshes. Indeed, the \SJVDB{our proposed method yields} results \SJVDB{that} have clearly defined (black and white) non-\SJVDB{staircased} boundaries. Figure~\ref{fig:comp_conva} shows the convergence behavior of the different codes for the finest mesh. It is observed that \SJVDB{our method} leads to the lowest objective function value, which again is similar to that obtained by Wei~\textit{et alli}'s code, while initially converging faster in the volume fraction.

Figure~\ref{fig:comp_convb} shows the final compliance as a function of the number of DOFs. Initially, the different methods all find lower compliance values as the mesh is refined, but the method by Wei \textit{et al.} and \SJVDB{our method} find slightly higher values for the finest mesh sizes. This may be explained by the optimizer converging to a local optimum. For each mesh size, \SJVDB{the proposed method} finds the lowest compliance value at the cost of adding some enriched DOFs.

\subsection{MBB beam}
\label{sec:example_rbf}
\begin{figure}
\centering
\includegraphics[width=0.5\columnwidth]{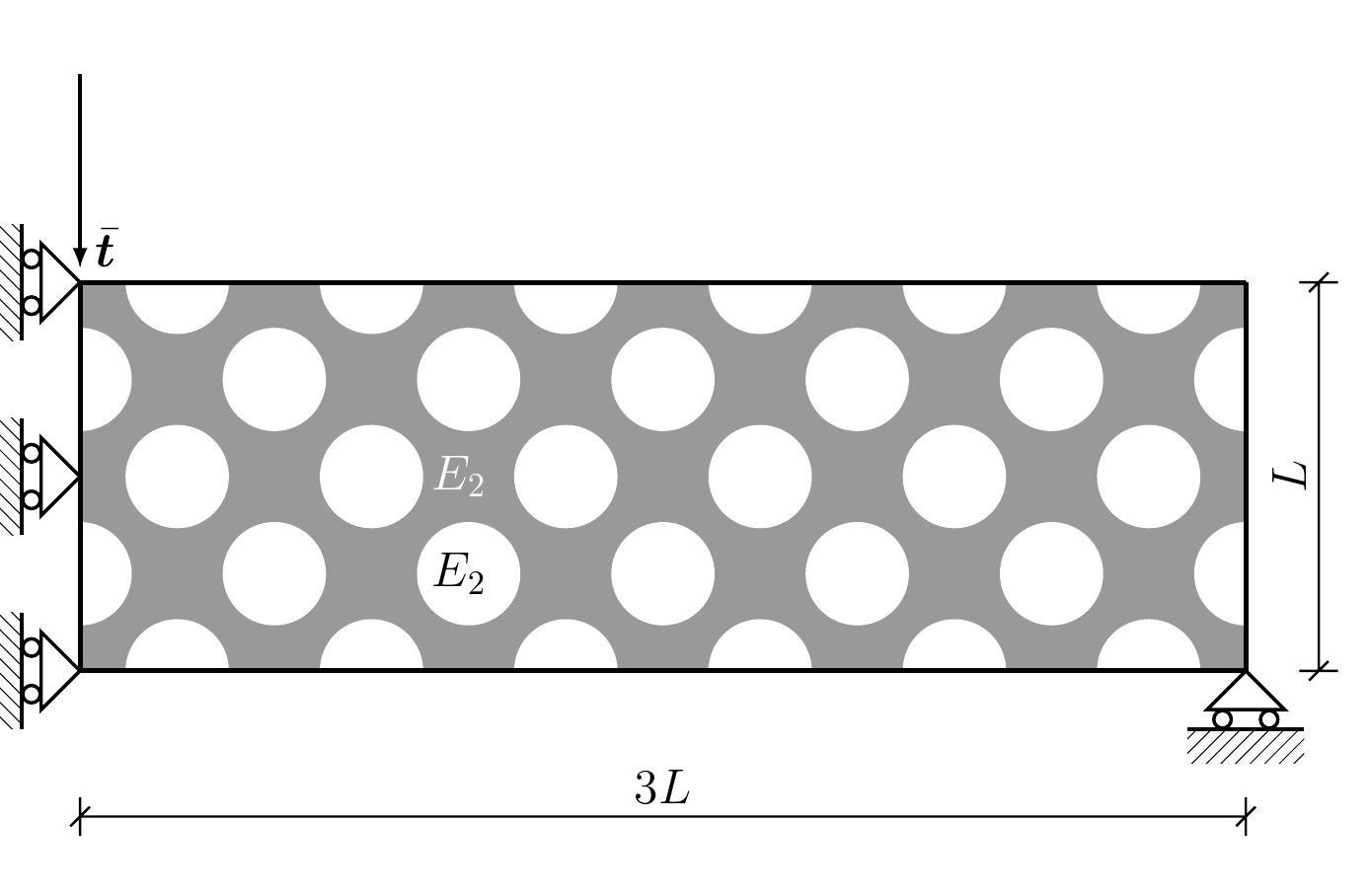}
\caption{Problem description and initial design for the MBB beam example in \S\ref{sec:example_rbf}. Symmetry conditions are applied on the left of the domain, and the bottom-right corner is simply supported. A downward force is applied at the top-left side on the domain, in the middle of the beam.}
\label{fig:mbb}
\end{figure}

The influence of the number of radial basis functions is investigated on the well-known MBB beam\footnote{The original Messerschmitt-Bölkow-Blohm (MBB) beam problem, as introduced by Olhoff~\cite{Olhoff:1991}, also specified that the upper and lower surfaces have to remain planar, in addition to a maximum allowable deflection and maximum stress. Over the years a more free interpretation of the problem formulation has become commonplace.}, which is illustrated in Figure~\ref{fig:mbb}.  The optimization problem consists of a $3L \times L$ domain with symmetry conditions on the left. On the bottom right corner, the domain is simply supported, and a downward force $\bar{\bs t}$ is applied on the top left corner. As in the previous example, the volume constraint is set to $55\%$ of the volume of the total design domain. The initial design is also indicated in the figure, and the same material properties as in the previous example \SJVDB{are used}.

This optimization problem is solved on a triangular analysis mesh defined on a grid of size $151 \times 51$, using  a discretization of the design space consisting of $61 \times 21$, $91 \times 31$, $121 \times 41$ and $151 \times 51$ radial basis functions, so that only for the finest design space discretization, both resolutions match, and an RBF is assigned to every node in the analysis mesh. The support radius $r_s$ is changed together with the design grid so that the overlap of RBFs is the same in each case: $r_s = \sqrt{2} \cdot a$, where $a$ is again the distance between two RBFs.

\begin{figure}
\centering
\includegraphics[width =0.4\columnwidth]{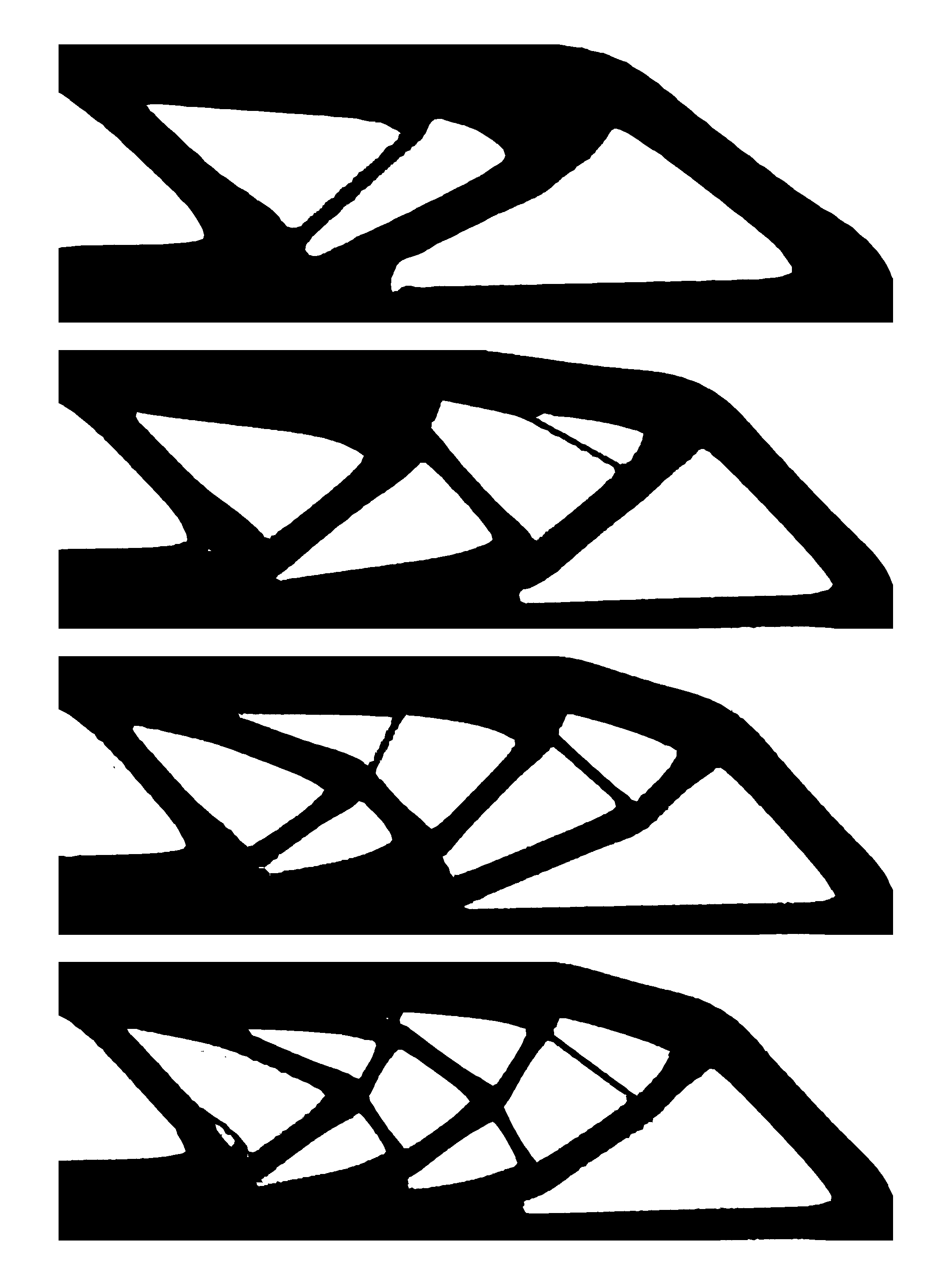}
\caption{Influence of the RBF mesh on the final design. Using symmetry conditions, only half of the MBB-beam is considered in the optimization. For each optimization, a structured mesh consisting of $150 \times 50 \times 2$ triangular finite elements is used. From top to bottom, \SJVDB{final} designs \SJVDB{are shown} obtained with design meshes consisting of $21 \times 61$, $31 \times 91$, $41 \times 121$ and $51 \times 151$ RBFs.  }
\label{fig:MBB_beam}
\end{figure}

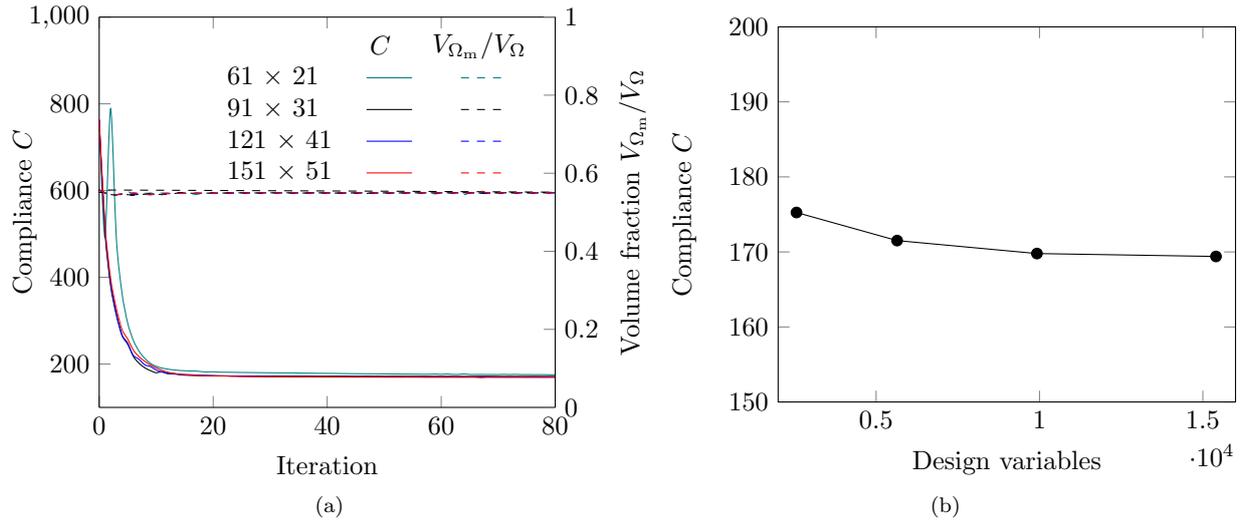
\begin{figure*}
\centering
\subfloat[][]{\begin{tikzpicture}[scale=1]
\pgfplotsset{
    scale only axis,
    xmin=0, xmax=80
}

\begin{axis}[
  axis y line*=left,
  ymin=100, ymax=1000,
  xlabel=Iteration,
  ylabel=Compliance $C$,
  ylabel style={xshift=0cm, yshift=-0.25cm},
width =6cm,
            legend cell align=right,
            legend plot pos=right,
legend style={at={(0.25, 0.90)}, anchor=north west, draw=none, fill=none, text width= 5em, text height=1.5ex, text depth=0.5ex}
]
\addplot[smooth, teal]
  coordinates{
( 0 , 7.629335659877399394e+02 )
( 1 , 4.920842237418647187e+02 )
( 2 , 7.887621711998199316e+02 )
( 3 , 4.914473553356127695e+02 )
( 4 , 3.639555312472429591e+02 )
( 5 , 2.944080189791612270e+02 )
( 6 , 2.542073486979249139e+02 )
( 7 , 2.292965782730486524e+02 )
( 8 , 2.131190603847764180e+02 )
( 9 , 2.014709614370544557e+02 )
( 10 , 1.948283350271732957e+02 )
( 11 , 1.908366119866145709e+02 )
( 12 , 1.882700555941414962e+02 )
( 13 , 1.866888305863675726e+02 )
( 14 , 1.853082838684628371e+02 )
( 15 , 1.845909571946301639e+02 )
( 16 , 1.839390277318362337e+02 )
( 17 , 1.839445833772059586e+02 )
( 18 , 1.823628188157359773e+02 )
( 19 , 1.818536200631339455e+02 )
( 20 , 1.822306384825062082e+02 )
( 21 , 1.812514785782688307e+02 )
( 22 , 1.811376172610760307e+02 )
( 23 , 1.810367328066465973e+02 )
( 24 , 1.808309479289417254e+02 )
( 25 , 1.805826244608352340e+02 )
( 26 , 1.804633649251399277e+02 )
( 27 , 1.803059736496487631e+02 )
( 28 , 1.801447226085265356e+02 )
( 29 , 1.800079107295756842e+02 )
( 30 , 1.799101822065087504e+02 )
( 31 , 1.797961195049011280e+02 )
( 32 , 1.796738216835292121e+02 )
( 33 , 1.796532334865329688e+02 )
( 34 , 1.796701181047112357e+02 )
( 35 , 1.791770980242893927e+02 )
( 36 , 1.790421443421185188e+02 )
( 37 , 1.790491793746774647e+02 )
( 38 , 1.787165880068504862e+02 )
( 39 , 1.786432923852169949e+02 )
( 40 , 1.784555766527619483e+02 )
( 41 , 1.784270012937150796e+02 )
( 42 , 1.782132077018093241e+02 )
( 43 , 1.780014933445868337e+02 )
( 44 , 1.778541719963830019e+02 )
( 45 , 1.777126375958143001e+02 )
( 46 , 1.776448592110386073e+02 )
( 47 , 1.775357505594438976e+02 )
( 48 , 1.775540575866997131e+02 )
( 49 , 1.773525399833264657e+02 )
( 50 , 1.772911043459346843e+02 )
( 51 , 1.770590043319361087e+02 )
( 52 , 1.770582984974141993e+02 )
( 53 , 1.768491600165693853e+02 )
( 54 , 1.768299647411498654e+02 )
( 55 , 1.766980557870491566e+02 )
( 56 , 1.765772986280642272e+02 )
( 57 , 1.764060029525616642e+02 )
( 58 , 1.765352267206652641e+02 )
( 59 , 1.769532109202898198e+02 )
( 60 , 1.762120146836510912e+02 )
( 61 , 1.770706200966353094e+02 )
( 62 , 1.762053047531002221e+02 )
( 63 , 1.760879041975008477e+02 )
( 64 , 1.771477616430409512e+02 )
( 65 , 1.760276690569529308e+02 )
( 66 , 1.759368926711501331e+02 )
( 67 , 1.760653041743943845e+02 )
( 68 , 1.758828227073086339e+02 )
( 69 , 1.758849605867048354e+02 )
( 70 , 1.757792246155685234e+02 )
( 71 , 1.762800215689908327e+02 )
( 72 , 1.757789591170237316e+02 )
( 73 , 1.762547512944163373e+02 )
( 74 , 1.756906278013152303e+02 )
( 75 , 1.757796445645540757e+02 )
( 76 , 1.756884507460031841e+02 )
( 77 , 1.756090899379230450e+02 )
( 78 , 1.756023419616949184e+02 )
( 79 , 1.752198717003498700e+02 )
( 80 , 1.752558892338277872e+02 )
( 81 , 1.754842601007713938e+02 )
( 82 , 1.754041836335491951e+02 )
( 83 , 1.752496487088476442e+02 )
( 84 , 1.754409039738010563e+02 )
( 85 , 1.753618479129228263e+02 )
( 86 , 1.753909248284006708e+02 )
( 87 , 1.751607204309712245e+02 )
( 88 , 1.753017373063298692e+02 )
( 89 , 1.752885288694073154e+02 )
( 90 , 1.751942130225931180e+02 )
( 91 , 1.754526294945275993e+02 )
( 92 , 1.755057097663502361e+02 )
( 93 , 1.753026231145972247e+02 )
( 94 , 1.753107518010745878e+02 )
( 95 , 1.752309750855416723e+02 )
( 96 , 1.753983511129112571e+02 )
( 97 , 1.753490038276170537e+02 )
( 98 , 1.752457523420187044e+02 )
( 99 , 1.752893391572612813e+02 )
( 100 , 1.752519442569460750e+02 )
}; \label{plot_two}
\addlegendentry{61 $\times$ 21}
\addplot[smooth, black]
  coordinates{
( 0 , 7.629335659877399394e+02 )
( 1 , 5.072228673826271006e+02 )
( 2 , 3.813126298112738937e+02 )
( 3 , 3.138370010994379982e+02 )
( 4 , 2.671765301067835026e+02 )
( 5 , 2.480290832122250038e+02 )
( 6 , 2.174377275619677050e+02 )
( 7 , 2.016414563845169141e+02 )
( 8 , 1.914309936894846089e+02 )
( 9 , 1.846878884541726222e+02 )
( 10 , 1.799793483046354368e+02 )
( 11 , 1.823530003032090860e+02 )
( 12 , 1.777915037572381607e+02 )
( 13 , 1.781463374241066902e+02 )
( 14 , 1.756456865296579792e+02 )
( 15 , 1.742904280846430822e+02 )
( 16 , 1.738834410635516008e+02 )
( 17 , 1.735536115946550524e+02 )
( 18 , 1.731616168158078892e+02 )
( 19 , 1.729378897037051672e+02 )
( 20 , 1.727747062464152066e+02 )
( 21 , 1.726532659811371104e+02 )
( 22 , 1.725090722698648165e+02 )
( 23 , 1.724985422614663833e+02 )
( 24 , 1.723414173319665679e+02 )
( 25 , 1.722146492754986582e+02 )
( 26 , 1.723104589713735777e+02 )
( 27 , 1.722032645333160588e+02 )
( 28 , 1.722722804019586533e+02 )
( 29 , 1.720900750385043239e+02 )
( 30 , 1.721092101137952852e+02 )
( 31 , 1.720351824144352122e+02 )
( 32 , 1.720659924432078469e+02 )
( 33 , 1.719969878895050215e+02 )
( 34 , 1.719996552793595868e+02 )
( 35 , 1.719828556781154418e+02 )
( 36 , 1.721052245820041549e+02 )
( 37 , 1.719878149171328801e+02 )
( 38 , 1.719140487389281589e+02 )
( 39 , 1.719236725379066684e+02 )
( 40 , 1.719573353194597303e+02 )
( 41 , 1.720756669395062772e+02 )
( 42 , 1.720671304399229200e+02 )
( 43 , 1.720434716308521388e+02 )
( 44 , 1.719616349184128978e+02 )
( 45 , 1.718648394032589408e+02 )
( 46 , 1.719017826561335767e+02 )
( 47 , 1.721014309823238193e+02 )
( 48 , 1.718636302914747773e+02 )
( 49 , 1.719867722653046940e+02 )
( 50 , 1.717927024407615590e+02 )
( 51 , 1.718476316518799649e+02 )
( 52 , 1.719216976376235380e+02 )
( 53 , 1.717524356235233540e+02 )
( 54 , 1.717661871038689867e+02 )
( 55 , 1.717102311564466390e+02 )
( 56 , 1.717308587369815882e+02 )
( 57 , 1.717483479494573260e+02 )
( 58 , 1.716840732263179632e+02 )
( 59 , 1.716842886005734954e+02 )
( 60 , 1.716781200089698700e+02 )
( 61 , 1.716976489140028832e+02 )
( 62 , 1.717030173301667730e+02 )
( 63 , 1.715972692805761710e+02 )
( 64 , 1.715799722145171700e+02 )
( 65 , 1.715590311910104049e+02 )
( 66 , 1.716542557259232069e+02 )
( 67 , 1.718288922894376469e+02 )
( 68 , 1.717874366102183785e+02 )
( 69 , 1.717461701547860855e+02 )
( 70 , 1.715907484842109909e+02 )
( 71 , 1.715358570976474368e+02 )
( 72 , 1.715752313949529935e+02 )
( 73 , 1.715445160937293281e+02 )
( 74 , 1.715265199080427863e+02 )
( 75 , 1.715701131070875931e+02 )
( 76 , 1.716540391027595831e+02 )
( 77 , 1.716174310416349158e+02 )
( 78 , 1.715296560233847742e+02 )
( 79 , 1.715109261946181221e+02 )
( 80 , 1.715123431191224199e+02 )
( 81 , 1.715099442776659373e+02 )
( 82 , 1.715000758662745284e+02 )
( 83 , 1.715096690209549308e+02 )
( 84 , 1.714970078494466463e+02 )
( 85 , 1.715065823700001033e+02 )
( 86 , 1.714965647855403859e+02 )
( 87 , 1.714873150947521481e+02 )
( 88 , 1.714930048008477570e+02 )
( 89 , 1.715164126323110736e+02 )
( 90 , 1.715426520740108742e+02 )
( 91 , 1.715605608132075304e+02 )
( 92 , 1.715006169834944387e+02 )
( 93 , 1.714865362574740288e+02 )
( 94 , 1.714835836216220457e+02 )
( 95 , 1.714749755777631322e+02 )
( 96 , 1.714983802812779743e+02 )
( 97 , 1.715051534973017340e+02 )
( 98 , 1.717692822357607270e+02 )
( 99 , 1.715171082864987682e+02 )
( 100 , 1.715583638070345671e+02 )
}; \label{plot_three}
\addlegendentry{91 $\times$ 31}
\addplot[smooth, blue]
  coordinates{
( 0 , 7.629335659877399394e+02 )
( 1 , 5.156664543548077972e+02 )
( 2 , 3.870266561577852826e+02 )
( 3 , 3.161228069826696014e+02 )
( 4 , 2.661204199163392445e+02 )
( 5 , 2.455447905998385636e+02 )
( 6 , 2.198956109756234127e+02 )
( 7 , 2.108762364124238218e+02 )
( 8 , 1.982516207769454866e+02 )
( 9 , 1.942202167337142669e+02 )
( 10 , 1.862064146157583480e+02 )
( 11 , 1.816004344134398139e+02 )
( 12 , 1.789888977071495049e+02 )
( 13 , 1.772633410928343096e+02 )
( 14 , 1.762068529337555560e+02 )
( 15 , 1.754829271087714062e+02 )
( 16 , 1.754741915132759118e+02 )
( 17 , 1.743928178563587892e+02 )
( 18 , 1.741015486611713641e+02 )
( 19 , 1.736265643214437091e+02 )
( 20 , 1.732738410296103666e+02 )
( 21 , 1.729655080433239505e+02 )
( 22 , 1.727945305378128751e+02 )
( 23 , 1.724434625765241265e+02 )
( 24 , 1.724344479087514230e+02 )
( 25 , 1.720535585828675664e+02 )
( 26 , 1.718661591427335225e+02 )
( 27 , 1.717157108361710414e+02 )
( 28 , 1.714763016615947322e+02 )
( 29 , 1.713470200327842292e+02 )
( 30 , 1.712899354952743636e+02 )
( 31 , 1.711519561587289502e+02 )
( 32 , 1.710586839864013768e+02 )
( 33 , 1.709837694776493322e+02 )
( 34 , 1.709898145018624405e+02 )
( 35 , 1.710605749817807180e+02 )
( 36 , 1.710270684953871694e+02 )
( 37 , 1.708138879341993004e+02 )
( 38 , 1.706962507975819676e+02 )
( 39 , 1.706313494673644016e+02 )
( 40 , 1.707115444502148023e+02 )
( 41 , 1.705940298830045379e+02 )
( 42 , 1.705739533635419605e+02 )
( 43 , 1.705314745829032290e+02 )
( 44 , 1.705071511174901389e+02 )
( 45 , 1.704505853382085547e+02 )
( 46 , 1.704233004402272229e+02 )
( 47 , 1.703810638745427752e+02 )
( 48 , 1.703160301997549766e+02 )
( 49 , 1.703608073154124725e+02 )
( 50 , 1.702830383897315301e+02 )
( 51 , 1.702077842790920954e+02 )
( 52 , 1.702399500745303271e+02 )
( 53 , 1.701450035595667600e+02 )
( 54 , 1.701399549648201344e+02 )
( 55 , 1.700964216305030732e+02 )
( 56 , 1.701225312256196105e+02 )
( 57 , 1.700835680053221211e+02 )
( 58 , 1.700179479727287344e+02 )
( 59 , 1.700604879578801842e+02 )
( 60 , 1.700039458790129459e+02 )
( 61 , 1.699753697615468866e+02 )
( 62 , 1.699996901501554873e+02 )
( 63 , 1.699957132797324277e+02 )
( 64 , 1.699762098581413454e+02 )
( 65 , 1.700035514185572936e+02 )
( 66 , 1.699483454165348633e+02 )
( 67 , 1.698952936702209229e+02 )
( 68 , 1.698859335235575543e+02 )
( 69 , 1.698618415647669337e+02 )
( 70 , 1.698774384678197862e+02 )
( 71 , 1.697944708868069483e+02 )
( 72 , 1.697578009203776332e+02 )
( 73 , 1.699415133757758554e+02 )
( 74 , 1.697979382311602876e+02 )
( 75 , 1.697087439759819745e+02 )
( 76 , 1.698188302417618729e+02 )
( 77 , 1.697459369614245190e+02 )
( 78 , 1.697167745912012151e+02 )
( 79 , 1.697675206296557633e+02 )
( 80 , 1.697847356629327749e+02 )
( 81 , 1.697131698282711625e+02 )
( 82 , 1.697151927725163887e+02 )
( 83 , 1.697068378463336842e+02 )
( 84 , 1.697192176157797405e+02 )
( 85 , 1.696767337106353750e+02 )
( 86 , 1.696574542692638374e+02 )
( 87 , 1.695223287631983169e+02 )
( 88 , 1.697177544294368943e+02 )
( 89 , 1.696723994856842808e+02 )
( 90 , 1.696913810816398325e+02 )
( 91 , 1.696923816326849703e+02 )
( 92 , 1.697201671464547132e+02 )
( 93 , 1.696592031113576979e+02 )
( 94 , 1.696775151671006654e+02 )
( 95 , 1.696643509819288056e+02 )
( 96 , 1.696354139588718226e+02 )
( 97 , 1.696635523539345627e+02 )
( 98 , 1.696386399464781505e+02 )
( 99 , 1.696148423056467323e+02 )
( 100 , 1.696121296473910718e+02 )
}; \label{plot_four}
\addlegendentry{121 $\times$ 41}
\addplot[smooth, red]
  coordinates{
( 0 , 7.629335659877399394e+02 )
( 1 , 5.238113610718230575e+02 )
( 2 , 3.979762789409114134e+02 )
( 3 , 3.287966569948117694e+02 )
( 4 , 2.795975593957634260e+02 )
( 5 , 2.576656628094294206e+02 )
( 6 , 2.318850328285153068e+02 )
( 7 , 2.168733558266120554e+02 )
( 8 , 2.063187205951231249e+02 )
( 9 , 1.978659301080818125e+02 )
( 10 , 1.912498094089533538e+02 )
( 11 , 1.856760426322593105e+02 )
( 12 , 1.813961594962842412e+02 )
( 13 , 1.786393663780471570e+02 )
( 14 , 1.770279820157728352e+02 )
( 15 , 1.756828244695473131e+02 )
( 16 , 1.744489041000210250e+02 )
( 17 , 1.738890325401068253e+02 )
( 18 , 1.733051585445429623e+02 )
( 19 , 1.729158426397393384e+02 )
( 20 , 1.726708573591139952e+02 )
( 21 , 1.724013187649494228e+02 )
( 22 , 1.725554355919049101e+02 )
( 23 , 1.717600955816349426e+02 )
( 24 , 1.715737044617111451e+02 )
( 25 , 1.711882917326217353e+02 )
( 26 , 1.709373430354605432e+02 )
( 27 , 1.707302423889984766e+02 )
( 28 , 1.706261296760847301e+02 )
( 29 , 1.705170185205471114e+02 )
( 30 , 1.703631539934406760e+02 )
( 31 , 1.703164502489941299e+02 )
( 32 , 1.702706290841900056e+02 )
( 33 , 1.702080714230607725e+02 )
( 34 , 1.701783936489113103e+02 )
( 35 , 1.701225190181596076e+02 )
( 36 , 1.701354543646540094e+02 )
( 37 , 1.702381350903669386e+02 )
( 38 , 1.700605892566581474e+02 )
( 39 , 1.700272161757281708e+02 )
( 40 , 1.700496881136973855e+02 )
( 41 , 1.699849370799865937e+02 )
( 42 , 1.699345868010455547e+02 )
( 43 , 1.699300911573811277e+02 )
( 44 , 1.698514044673567298e+02 )
( 45 , 1.698559145413241254e+02 )
( 46 , 1.698852379627213338e+02 )
( 47 , 1.698151866454854257e+02 )
( 48 , 1.697759958654806383e+02 )
( 49 , 1.697518754244077570e+02 )
( 50 , 1.696486974616026657e+02 )
( 51 , 1.697027058818367209e+02 )
( 52 , 1.696633904287199925e+02 )
( 53 , 1.696588816958169161e+02 )
( 54 , 1.697556432762797840e+02 )
( 55 , 1.696974872452190368e+02 )
( 56 , 1.697046284474972140e+02 )
( 57 , 1.695578272547401752e+02 )
( 58 , 1.695241495116706574e+02 )
( 59 , 1.695768002022506948e+02 )
( 60 , 1.695696280078109908e+02 )
( 61 , 1.695459144658155992e+02 )
( 62 , 1.696094099621459463e+02 )
( 63 , 1.694994927529139090e+02 )
( 64 , 1.695592623298780097e+02 )
( 65 , 1.696286015553883146e+02 )
( 66 , 1.695533854887779341e+02 )
( 67 , 1.686126054634153206e+02 )
( 68 , 1.694695754243790020e+02 )
( 69 , 1.694646942048975973e+02 )
( 70 , 1.694287260670592161e+02 )
( 71 , 1.694314182404994256e+02 )
( 72 , 1.694086488494436082e+02 )
( 73 , 1.693918595627991408e+02 )
( 74 , 1.694238856790792056e+02 )
( 75 , 1.693867740688870072e+02 )
( 76 , 1.693654224718235071e+02 )
( 77 , 1.693971899501471228e+02 )
( 78 , 1.693558291040995414e+02 )
( 79 , 1.694444030179483605e+02 )
( 80 , 1.693984583540296001e+02 )
( 81 , 1.693137613466271034e+02 )
( 82 , 1.692801639400231863e+02 )
( 83 , 1.692929706109675863e+02 )
( 84 , 1.692867137802942068e+02 )
( 85 , 1.692985849166990988e+02 )
( 86 , 1.693193003372394401e+02 )
( 87 , 1.692497644333242874e+02 )
( 88 , 1.692428889354223429e+02 )
( 89 , 1.692709713348288858e+02 )
( 90 , 1.692496650231002491e+02 )
( 91 , 1.692751026309677798e+02 )
( 92 , 1.693219121089659893e+02 )
( 93 , 1.693615504443023383e+02 )
( 94 , 1.692873491130596051e+02 )
( 95 , 1.692134468256778064e+02 )
( 96 , 1.692406506119378946e+02 )
( 97 , 1.692273204205581862e+02 )
( 98 , 1.692072750232252076e+02 )
( 99 , 1.692435235291646904e+02 )
( 100 , 1.693350005132546983e+02 )
}; \label{plot_five}
\addlegendentry{151 $\times$ 51}
\end{axis}
\begin{axis}[
  axis y line*=right,
  axis x line=none,
  ymin=0, ymax=1,
  ylabel=Volume fraction $V_{\Omega_\mathrm{m}} / V_\Omega$,
  ylabel style={xshift=0cm, yshift=-8.25cm},
  yticklabel pos=right,
width = 6cm, 
            legend cell align=right,
            legend plot pos=right,
legend style={at={(0.75, 0.90)}, anchor=north west,  draw=none, fill=none, text height=1.5ex, text depth=0.5ex}
]
\addplot[smooth,teal, dashed]
  coordinates{
( 0 , 0.5570922734073306 )
( 1 , 0.5483281254130352 )
( 2 , 0.5406405929756655 )
( 3 , 0.5453962356106099 )
( 4 , 0.5454017367703594 )
( 5 , 0.5520033716964654 )
( 6 , 0.548148111339068 )
( 7 , 0.5491351486719251 )
( 8 , 0.5484589463290185 )
( 9 , 0.5492228519572695 )
( 10 , 0.5491479546433915 )
( 11 , 0.549202330063348 )
( 12 , 0.5492069141557523 )
( 13 , 0.5487248192923018 )
( 14 , 0.5488847748220937 )
( 15 , 0.5487586045771946 )
( 16 , 0.5482751897612417 )
( 17 , 0.5455418290874092 )
( 18 , 0.5484202327031799 )
( 19 , 0.5488604371269658 )
( 20 , 0.5481366164940839 )
( 21 , 0.5495124149089599 )
( 22 , 0.549080433285052 )
( 23 , 0.549297661402626 )
( 24 , 0.5492106882334785 )
( 25 , 0.5496923991045961 )
( 26 , 0.5496097103321355 )
( 27 , 0.549769751389689 )
( 28 , 0.5497460710550375 )
( 29 , 0.5498800037191066 )
( 30 , 0.5498209928497018 )
( 31 , 0.5497175146025856 )
( 32 , 0.5496960388590864 )
( 33 , 0.5495845471801779 )
( 34 , 0.5485608227869279 )
( 35 , 0.5498223864063192 )
( 36 , 0.5497178695989021 )
( 37 , 0.5490195239837005 )
( 38 , 0.5496949126831859 )
( 39 , 0.5494788939345022 )
( 40 , 0.5496710947473262 )
( 41 , 0.5492860199310431 )
( 42 , 0.5495529029523674 )
( 43 , 0.5497827194977611 )
( 44 , 0.5498529592819388 )
( 45 , 0.5498905926888392 )
( 46 , 0.5497154284800846 )
( 47 , 0.5496797679058368 )
( 48 , 0.549567468406091 )
( 49 , 0.549403796907552 )
( 50 , 0.5495480600909594 )
( 51 , 0.5496633231262776 )
( 52 , 0.5495152678384192 )
( 53 , 0.5496561539210358 )
( 54 , 0.5496051428777189 )
( 55 , 0.5494701035272541 )
( 56 , 0.5496668167085473 )
( 57 , 0.5497906207519706 )
( 58 , 0.549489982617662 )
( 59 , 0.5469524935796326 )
( 60 , 0.5497786611112944 )
( 61 , 0.5491310700320435 )
( 62 , 0.5494541707806235 )
( 63 , 0.5498017053402985 )
( 64 , 0.5450720036788567 )
( 65 , 0.5496130663289899 )
( 66 , 0.549823555992156 )
( 67 , 0.5490396366402599 )
( 68 , 0.5496925070038685 )
( 69 , 0.5494597054132309 )
( 70 , 0.5497907231521393 )
( 71 , 0.5477261822883043 )
( 72 , 0.5494975539418436 )
( 73 , 0.5489268527931419 )
( 74 , 0.549787806971334 )
( 75 , 0.549099446685498 )
( 76 , 0.5496516474060138 )
( 77 , 0.5496002639579419 )
( 78 , 0.5497214808257601 )
( 79 , 0.551311915685746 )
( 80 , 0.5511115029311296 )
( 81 , 0.5497027950277256 )
( 82 , 0.550021378404765 )
( 83 , 0.5506561187795526 )
( 84 , 0.5496666854342763 )
( 85 , 0.5499529662712141 )
( 86 , 0.5499208905759551 )
( 87 , 0.5506915062307206 )
( 88 , 0.5499352267190154 )
( 89 , 0.5499897422990999 )
( 90 , 0.5503345410308039 )
( 91 , 0.5495941266356527 )
( 92 , 0.5489106042325879 )
( 93 , 0.5498765987372675 )
( 94 , 0.5496363031849394 )
( 95 , 0.5499487262885194 )
( 96 , 0.5496320508505189 )
( 97 , 0.5493729772044399 )
( 98 , 0.5498850352922167 )
( 99 , 0.5495923084035165 )
( 100 , 0.5497397489664746 )
}; \label{plot_vol_two}
\addlegendentry{}
\addplot[smooth,black, dashed]
  coordinates{
( 0 , 0.5570922734073306 )
( 1 , 0.5501099793002615 )
( 2 , 0.5453886336958479 )
( 3 , 0.5447252182902292 )
( 4 , 0.5471050391721551 )
( 5 , 0.5449758858558561 )
( 6 , 0.54447727473539 )
( 7 , 0.545875839273229 )
( 8 , 0.5460165816873697 )
( 9 , 0.5467940038856738 )
( 10 , 0.5468526922404184 )
( 11 , 0.5449649810478704 )
( 12 , 0.5477772867573951 )
( 13 , 0.5473164528904039 )
( 14 , 0.5499738237596435 )
( 15 , 0.5491600134652816 )
( 16 , 0.5488667996767442 )
( 17 , 0.5484957597675353 )
( 18 , 0.548893030072011 )
( 19 , 0.5489947558521107 )
( 20 , 0.5490568948480254 )
( 21 , 0.5493139384680548 )
( 22 , 0.549280826250251 )
( 23 , 0.5493386764991012 )
( 24 , 0.5495139269238086 )
( 25 , 0.549870151650528 )
( 26 , 0.5492766249760968 )
( 27 , 0.5497530402825819 )
( 28 , 0.5492208895562367 )
( 29 , 0.5498703532032355 )
( 30 , 0.5497993436957923 )
( 31 , 0.5498997083046298 )
( 32 , 0.5497682026197387 )
( 33 , 0.5498826117632336 )
( 34 , 0.5498221998392799 )
( 35 , 0.5498145916955017 )
( 36 , 0.5491760206001165 )
( 37 , 0.5496911868359925 )
( 38 , 0.5497880320523548 )
( 39 , 0.5497449464137163 )
( 40 , 0.5496652203788825 )
( 41 , 0.5491165521985676 )
( 42 , 0.5493372208063291 )
( 43 , 0.5489764712167514 )
( 44 , 0.5495598077377627 )
( 45 , 0.5495634196884646 )
( 46 , 0.5497366489501443 )
( 47 , 0.5484350541227598 )
( 48 , 0.5497007110298652 )
( 49 , 0.5488634364500118 )
( 50 , 0.5497190962258143 )
( 51 , 0.5495852554503232 )
( 52 , 0.5490784758682304 )
( 53 , 0.5497912063941818 )
( 54 , 0.5496739007497756 )
( 55 , 0.5499351345699189 )
( 56 , 0.5498443746196007 )
( 57 , 0.5496986360344003 )
( 58 , 0.5499369534475369 )
( 59 , 0.5499181801859109 )
( 60 , 0.5499374243468191 )
( 61 , 0.5498651336455948 )
( 62 , 0.5497876452484362 )
( 63 , 0.5498894715431425 )
( 64 , 0.5498939461420635 )
( 65 , 0.5499516549164093 )
( 66 , 0.5498243499439376 )
( 67 , 0.5489858731773183 )
( 68 , 0.5491387757988926 )
( 69 , 0.5491465260641245 )
( 70 , 0.5497106101338772 )
( 71 , 0.5499531977632526 )
( 72 , 0.5498455304371899 )
( 73 , 0.5498939192471025 )
( 74 , 0.5499603265175298 )
( 75 , 0.5497507365579891 )
( 76 , 0.5498414466780346 )
( 77 , 0.5495292969793376 )
( 78 , 0.5498909899263633 )
( 79 , 0.5499644099323795 )
( 80 , 0.5499593382221183 )
( 81 , 0.5499501474865631 )
( 82 , 0.5499717971233358 )
( 83 , 0.5499383366944145 )
( 84 , 0.5499749901611712 )
( 85 , 0.5499242432690375 )
( 86 , 0.5499609506903386 )
( 87 , 0.5499817546308855 )
( 88 , 0.5499594637650564 )
( 89 , 0.5498604515745439 )
( 90 , 0.5497660938738022 )
( 91 , 0.5496945917108347 )
( 92 , 0.5498839864425732 )
( 93 , 0.5499191136283474 )
( 94 , 0.5499344301832945 )
( 95 , 0.549947211038925 )
( 96 , 0.549824114245697 )
( 97 , 0.549835695232413 )
( 98 , 0.5485620699451907 )
( 99 , 0.5496850489872994 )
( 100 , 0.5499015308974771 )( 0 , 0.5570922734073306 )
( 1 , 0.5501099793002615 )
( 2 , 0.5453886336958479 )
( 3 , 0.5447252182902292 )
( 4 , 0.5471050391721551 )
( 5 , 0.5449758858558561 )
( 6 , 0.54447727473539 )
( 7 , 0.545875839273229 )
( 8 , 0.5460165816873697 )
( 9 , 0.5467940038856738 )
( 10 , 0.5468526922404184 )
( 11 , 0.5449649810478704 )
( 12 , 0.5477772867573951 )
( 13 , 0.5473164528904039 )
( 14 , 0.5499738237596435 )
( 15 , 0.5491600134652816 )
( 16 , 0.5488667996767442 )
( 17 , 0.5484957597675353 )
( 18 , 0.548893030072011 )
( 19 , 0.5489947558521107 )
( 20 , 0.5490568948480254 )
( 21 , 0.5493139384680548 )
( 22 , 0.549280826250251 )
( 23 , 0.5493386764991012 )
( 24 , 0.5495139269238086 )
( 25 , 0.549870151650528 )
( 26 , 0.5492766249760968 )
( 27 , 0.5497530402825819 )
( 28 , 0.5492208895562367 )
( 29 , 0.5498703532032355 )
( 30 , 0.5497993436957923 )
( 31 , 0.5498997083046298 )
( 32 , 0.5497682026197387 )
( 33 , 0.5498826117632336 )
( 34 , 0.5498221998392799 )
( 35 , 0.5498145916955017 )
( 36 , 0.5491760206001165 )
( 37 , 0.5496911868359925 )
( 38 , 0.5497880320523548 )
( 39 , 0.5497449464137163 )
( 40 , 0.5496652203788825 )
( 41 , 0.5491165521985676 )
( 42 , 0.5493372208063291 )
( 43 , 0.5489764712167514 )
( 44 , 0.5495598077377627 )
( 45 , 0.5495634196884646 )
( 46 , 0.5497366489501443 )
( 47 , 0.5484350541227598 )
( 48 , 0.5497007110298652 )
( 49 , 0.5488634364500118 )
( 50 , 0.5497190962258143 )
( 51 , 0.5495852554503232 )
( 52 , 0.5490784758682304 )
( 53 , 0.5497912063941818 )
( 54 , 0.5496739007497756 )
( 55 , 0.5499351345699189 )
( 56 , 0.5498443746196007 )
( 57 , 0.5496986360344003 )
( 58 , 0.5499369534475369 )
( 59 , 0.5499181801859109 )
( 60 , 0.5499374243468191 )
( 61 , 0.5498651336455948 )
( 62 , 0.5497876452484362 )
( 63 , 0.5498894715431425 )
( 64 , 0.5498939461420635 )
( 65 , 0.5499516549164093 )
( 66 , 0.5498243499439376 )
( 67 , 0.5489858731773183 )
( 68 , 0.5491387757988926 )
( 69 , 0.5491465260641245 )
( 70 , 0.5497106101338772 )
( 71 , 0.5499531977632526 )
( 72 , 0.5498455304371899 )
( 73 , 0.5498939192471025 )
( 74 , 0.5499603265175298 )
( 75 , 0.5497507365579891 )
( 76 , 0.5498414466780346 )
( 77 , 0.5495292969793376 )
( 78 , 0.5498909899263633 )
( 79 , 0.5499644099323795 )
( 80 , 0.5499593382221183 )
( 81 , 0.5499501474865631 )
( 82 , 0.5499717971233358 )
( 83 , 0.5499383366944145 )
( 84 , 0.5499749901611712 )
( 85 , 0.5499242432690375 )
( 86 , 0.5499609506903386 )
( 87 , 0.5499817546308855 )
( 88 , 0.5499594637650564 )
( 89 , 0.5498604515745439 )
( 90 , 0.5497660938738022 )
( 91 , 0.5496945917108347 )
( 92 , 0.5498839864425732 )
( 93 , 0.5499191136283474 )
( 94 , 0.5499344301832945 )
( 95 , 0.549947211038925 )
( 96 , 0.549824114245697 )
( 97 , 0.549835695232413 )
( 98 , 0.5485620699451907 )
( 99 , 0.5496850489872994 )
( 100 , 0.5499015308974771 )
}; \label{plot_vol_three}
\addlegendentry{}
\addplot[smooth,blue, dashed]
  coordinates{
( 0 , 0.5570922734073306 )
( 1 , 0.5495883577150318 )
( 2 , 0.5463175174819178 )
( 3 , 0.5449123032248439 )
( 4 , 0.5463443650561727 )
( 5 , 0.5450208225205805 )
( 6 , 0.5470369824919067 )
( 7 , 0.5463410188877306 )
( 8 , 0.5460814056962768 )
( 9 , 0.544532556866957 )
( 10 , 0.5468656293393693 )
( 11 , 0.5480165552374567 )
( 12 , 0.548532564348031 )
( 13 , 0.5486122737678351 )
( 14 , 0.5487181904599171 )
( 15 , 0.5487991521541271 )
( 16 , 0.546511602222198 )
( 17 , 0.5491852033045358 )
( 18 , 0.5486154261802193 )
( 19 , 0.5490646766551947 )
( 20 , 0.5488056814476525 )
( 21 , 0.5489628654137702 )
( 22 , 0.5486377118163708 )
( 23 , 0.5494049590414881 )
( 24 , 0.5491631342542365 )
( 25 , 0.5496383819326986 )
( 26 , 0.5496205059355594 )
( 27 , 0.549662142099456 )
( 28 , 0.5496145587936261 )
( 29 , 0.5497695422595581 )
( 30 , 0.5494962567998662 )
( 31 , 0.5496808932969286 )
( 32 , 0.5497043232291263 )
( 33 , 0.5497527299337083 )
( 34 , 0.5495269432404982 )
( 35 , 0.5488889320670026 )
( 36 , 0.5488407175318563 )
( 37 , 0.5495507011310958 )
( 38 , 0.5497730707975159 )
( 39 , 0.5497882003192022 )
( 40 , 0.549523807545634 )
( 41 , 0.5495222717273598 )
( 42 , 0.549901004470063 )
( 43 , 0.5496034152360869 )
( 44 , 0.5494328578194276 )
( 45 , 0.5495311701874829 )
( 46 , 0.5495281317846985 )
( 47 , 0.5496208055854324 )
( 48 , 0.5497323292071066 )
( 49 , 0.5495506110888438 )
( 50 , 0.5495752320835096 )
( 51 , 0.5497435309168679 )
( 52 , 0.5495459636966012 )
( 53 , 0.5497514774659192 )
( 54 , 0.5497095131512895 )
( 55 , 0.5498524929022145 )
( 56 , 0.549643219957236 )
( 57 , 0.5498441826940886 )
( 58 , 0.5500641892844649 )
( 59 , 0.5497633145663274 )
( 60 , 0.5499600311489702 )
( 61 , 0.5500428535536411 )
( 62 , 0.5498778670189866 )
( 63 , 0.5498496012567686 )
( 64 , 0.5498306150777278 )
( 65 , 0.5496281397012894 )
( 66 , 0.5498257779513831 )
( 67 , 0.5500302341650587 )
( 68 , 0.5497884605285032 )
( 69 , 0.5497195015962084 )
( 70 , 0.5496877463526754 )
( 71 , 0.5503948922596562 )
( 72 , 0.5500989368026269 )
( 73 , 0.5491791148337012 )
( 74 , 0.550180207715372 )
( 75 , 0.5503167370844573 )
( 76 , 0.5496886876586998 )
( 77 , 0.550046972354368 )
( 78 , 0.5502698722836586 )
( 79 , 0.5497026144484767 )
( 80 , 0.5498067720404233 )
( 81 , 0.5498998561373282 )
( 82 , 0.5498962721454509 )
( 83 , 0.5498834610019382 )
( 84 , 0.5497949484986017 )
( 85 , 0.5499228943905847 )
( 86 , 0.5500228172918586 )
( 87 , 0.5507128546789632 )
( 88 , 0.5499308257522256 )
( 89 , 0.5500588379057526 )
( 90 , 0.5499908626978096 )
( 91 , 0.5498845641795975 )
( 92 , 0.5498751015374664 )
( 93 , 0.5501233702956989 )
( 94 , 0.5499298144017796 )
( 95 , 0.5499800916781599 )
( 96 , 0.5501453948687541 )
( 97 , 0.54990240407736 )
( 98 , 0.549995510637487 )
( 99 , 0.5500869197353784 )
( 100 , 0.5499629006181158 )
}; \label{plot_vol_four}
\addlegendentry{}
\addplot[smooth,red, dashed]
  coordinates{
( 0 , 0.5570922734073306 )
( 1 , 0.5501554644515905 )
( 2 , 0.5454115399156914 )
( 3 , 0.5451731673160025 )
( 4 , 0.5474742389496288 )
( 5 , 0.5449999674820033 )
( 6 , 0.5502511232156893 )
( 7 , 0.5475407398478904 )
( 8 , 0.546215525676433 )
( 9 , 0.5459303366825903 )
( 10 , 0.5459089523379246 )
( 11 , 0.5464162211805741 )
( 12 , 0.5479837666854335 )
( 13 , 0.5488370076607426 )
( 14 , 0.5484322898255704 )
( 15 , 0.5484011281604848 )
( 16 , 0.5487811534638445 )
( 17 , 0.5483036214817412 )
( 18 , 0.54887518012981 )
( 19 , 0.5491361740845099 )
( 20 , 0.5489353776489515 )
( 21 , 0.5493460950541925 )
( 22 , 0.5492267287398483 )
( 23 , 0.5494950673384843 )
( 24 , 0.5492636518941232 )
( 25 , 0.5494621975957132 )
( 26 , 0.5497087104333284 )
( 27 , 0.5499502927655896 )
( 28 , 0.5498320642828418 )
( 29 , 0.5499628500649013 )
( 30 , 0.5499116409099107 )
( 31 , 0.5498766722887142 )
( 32 , 0.5498653422532211 )
( 33 , 0.5499314844227923 )
( 34 , 0.54988743342977 )
( 35 , 0.5498484940399153 )
( 36 , 0.549702099280607 )
( 37 , 0.5490431025512625 )
( 38 , 0.5497180743783809 )
( 39 , 0.5497756502593927 )
( 40 , 0.5498268003653168 )
( 41 , 0.5497610282875393 )
( 42 , 0.5498430000322784 )
( 43 , 0.5498088105466461 )
( 44 , 0.5500885338689273 )
( 45 , 0.5501029808739795 )
( 46 , 0.549687049010167 )
( 47 , 0.549977730121579 )
( 48 , 0.5496783189409028 )
( 49 , 0.5497831165345948 )
( 50 , 0.5501554985507991 )
( 51 , 0.5498137852548413 )
( 52 , 0.5499174968599053 )
( 53 , 0.549920080050146 )
( 54 , 0.5494874899775823 )
( 55 , 0.5495726273257069 )
( 56 , 0.5495955570600966 )
( 57 , 0.5502091496723222 )
( 58 , 0.5502144337885346 )
( 59 , 0.5498302315045219 )
( 60 , 0.5497546156746188 )
( 61 , 0.5498186569347799 )
( 62 , 0.5494336868177092 )
( 63 , 0.5498691203364402 )
( 64 , 0.5498024822296704 )
( 65 , 0.5496310910631481 )
( 66 , 0.5497608914636568 )
( 67 , 0.5543922121352937 )
( 68 , 0.5498738862783 )
( 69 , 0.5497970101884326 )
( 70 , 0.5499902096246809 )
( 71 , 0.5498464361112738 )
( 72 , 0.5499060763699206 )
( 73 , 0.5499303424542517 )
( 74 , 0.5497419616179313 )
( 75 , 0.549910268273517 )
( 76 , 0.5498702498885739 )
( 77 , 0.5497818963606769 )
( 78 , 0.54984871734504 )
( 79 , 0.5498733513636638 )
( 80 , 0.5496216060037978 )
( 81 , 0.5499620809266429 )
( 82 , 0.5500936646450391 )
( 83 , 0.5499940639666211 )
( 84 , 0.5499471316970751 )
( 85 , 0.5499269755491253 )
( 86 , 0.5497876167657655 )
( 87 , 0.5500939070635327 )
( 88 , 0.5500916046901094 )
( 89 , 0.5499228562707129 )
( 90 , 0.5499776644188386 )
( 91 , 0.5498952752894247 )
( 92 , 0.5496776075965882 )
( 93 , 0.5493943442238192 )
( 94 , 0.5497204189212799 )
( 95 , 0.5500286267031772 )
( 96 , 0.5498595079740086 )
( 97 , 0.5498732509247627 )
( 98 , 0.5499168130374962 )
( 99 , 0.5498139846852543 )
( 100 , 0.5492713362525296 )
}; \label{plot_vol_five}
\addlegendentry{}
\end{axis}
\node[draw=none] at (3.7,4.8) {$C$};
\node[draw=none] at (5.0, 4.8) {$V_{\Omega_\mathrm{m}} / V_\Omega$};
\end{tikzpicture}\label{fig:MBB_convergence-a}}
\subfloat[][]{
\begin{tikzpicture}
\begin{axis}[
  width = 7.6cm,
  xmin=2000, xmax=16000,
  ymin=150, ymax=200,
  xlabel= Design variables,
  ylabel=Compliance $C$,
legend style={at={(0.98, 0.99)}, anchor=north east, draw=none, fill=none}
]
\addplot[black, mark = *]
  coordinates{
( 2562, 1.752558892338277872e+02)
( 5642, 1.715123431191224199e+02)
( 9922, 1.697847356629327749e+02 )
( 15402, 1.693984583540296001e+02 )
};
\end{axis}
\end{tikzpicture}\label{fig:MBB_convergence-b}}
\caption{\SJVDB{Subfigure (a) shows the convergence of the compliance $C$ and volume fraction $V_{\Omega_\mathrm{m}} / V_\Omega$ of the MBB beam using different discretizations of the design space; (b) shows the final compliance of the MBB beam as a function of the number of design variables. The compliance value no longer decreases at the finest design mesh.}}\label{fig:MBB_convergence}
\end{figure*}

Figure~\ref{fig:MBB_beam} shows the optimized designs. As expected, the level of detail in the design can be controlled by the RBF discretization. However, it is noted that in the finest RBF mesh, artifacts appear on the design boundary. This behavior will be further analysed in \S\ref{sec:zigzag}. In Figure~\ref{fig:MBB_convergence-a} the convergence behavior of the different RBF meshes is shown. Although the coarsest RBF mesh shows some initial oscillations, the overall convergence behavior is similar in all cases.  Moreover, as shown in Figure~\ref{fig:MBB_convergence-b}, the compliance no longer significantly improves for the finest RBF discretization.  

\subsection{3-D cantilever beam}
\begin{figure}
\centering
	 \def\svgwidth{180pt}
	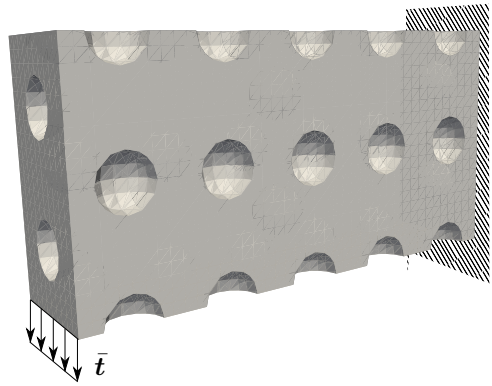
\caption{Initial design of the 3-D example with a schematic illustration of the boundary conditions: the right side is fixed and a vertical downward line load is applied on the bottom-left edge.}
\label{fig:3D_initial}
\end{figure}

To show that the method is not restricted to 2-D, a 3-D cantilever beam example is also considered. The material properties are the same as those of previous examples. The size of this cantilever beam is $2L \times L \times 0.5L$, and a structured mesh with $40 \times 20 \times 10 \times 6$ tetrahedral elements is used to discretize the model. The design space is discretized using a grid of $21 \times 11 \times 6$ RBFs, with $r_s = \sqrt{2} \cdot a$. Figure~\ref{fig:3D_initial} shows the initial design, along with the boundary conditions; the right surface is clamped, and a distributed line load with \SJVDB{$\left| \bar{\bs t}\right| = 0.2$ per unit length} is applied on the bottom-left edge. The move limit for MMA in this example is set to $0.001$ to prevent the optimizer from moving the boundaries too fast, as only a small number of RBFs is used with a large $r_s$ compared to the analysis mesh.
 The objective function is again the structural compliance, and the volume constraint is set to $40\%$ of the total design domain. 

\begin{figure*}
\centering
\subfloat[][]{\includegraphics[width =0.45\columnwidth]{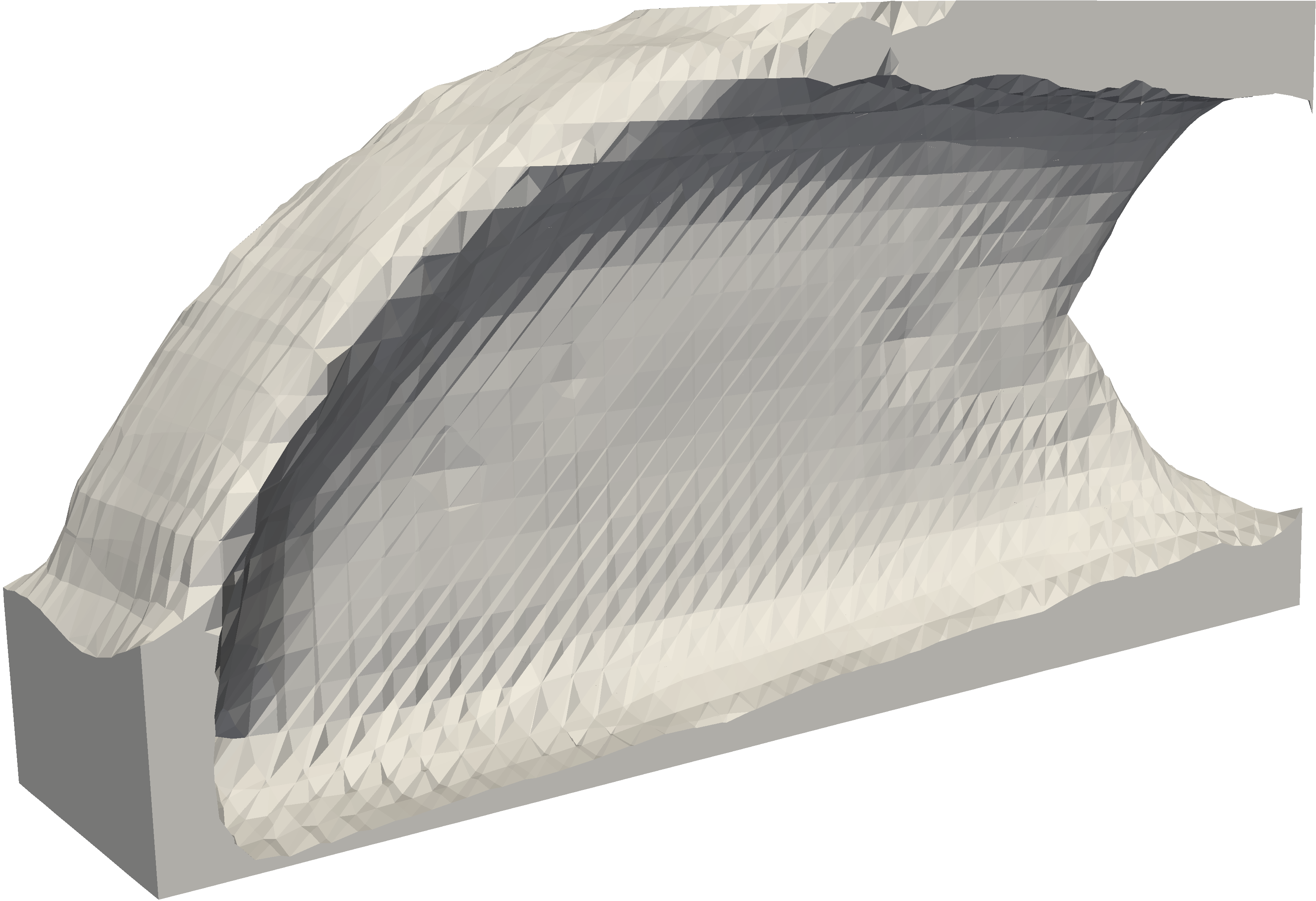}\label{fig:3D_final}}\;\;
\subfloat[][]{\begin{tikzpicture}[scale=1]      
\pgfplotsset{
    scale only axis,
    xmin=0, xmax=300
}

\begin{axis}[
  axis y line*=left,
  ymin=1, ymax=5,
  xlabel=Iteration,
  ylabel=Compliance $C$,
  ylabel style={xshift=0cm, yshift=-0.25cm},
width =6cm,
legend style={at={(0.60, 0.99)}, anchor=north west, draw=none, fill=none}
]
\addplot[smooth,black]
  coordinates{
( 0 , 0.942957 )
( 1 , 0.980314 )
( 2 , 1.006101 )
( 3 , 1.034394 )
( 4 , 1.065096 )
( 5 , 1.099761 )
( 6 , 1.140378 )
( 7 , 1.186569 )
( 8 , 1.239213 )
( 9 , 1.30058 )
( 10 , 1.373293 )
( 11 , 1.464613 )
( 12 , 1.588851 )
( 13 , 1.771996 )
( 14 , 2.014148 )
( 15 , 2.304653 )
( 16 , 2.667443 )
( 17 , 3.124528 )
( 18 , 3.606068 )
( 19 , 3.954941 )
( 20 , 3.792348 )
( 21 , 3.572702 )
( 22 , 3.405338 )
( 23 , 3.229052 )
( 24 , 3.089965 )
( 25 , 2.919222 )
( 26 , 2.749051 )
( 27 , 2.663479 )
( 28 , 2.548864 )
( 29 , 2.499412 )
( 30 , 2.412916 )
( 31 , 2.368288 )
( 32 , 2.306209 )
( 33 , 2.272144 )
( 34 , 2.226933 )
( 35 , 2.187271 )
( 36 , 2.14808 )
( 37 , 2.113115 )
( 38 , 2.08488 )
( 39 , 2.049643 )
( 40 , 2.019415 )
( 41 , 1.993444 )
( 42 , 1.965164 )
( 43 , 1.9373 )
( 44 , 1.913735 )
( 45 , 1.890248 )
( 46 , 1.865337 )
( 47 , 1.847041 )
( 48 , 1.827233 )
( 49 , 1.806241 )
( 50 , 1.791401 )
( 51 , 1.786595 )
( 52 , 1.767086 )
( 53 , 1.749495 )
( 54 , 1.730571 )
( 55 , 1.707645 )
( 56 , 1.693676 )
( 57 , 1.679928 )
( 58 , 1.668561 )
( 59 , 1.655914 )
( 60 , 1.646515 )
( 61 , 1.633303 )
( 62 , 1.622996 )
( 63 , 1.614049 )
( 64 , 1.60781 )
( 65 , 1.602332 )
( 66 , 1.595819 )
( 67 , 1.591469 )
( 68 , 1.586879 )
( 69 , 1.582127 )
( 70 , 1.576519 )
( 71 , 1.57312 )
( 72 , 1.569223 )
( 73 , 1.565121 )
( 74 , 1.562626 )
( 75 , 1.557587 )
( 76 , 1.552454 )
( 77 , 1.54788 )
( 78 , 1.543324 )
( 79 , 1.538319 )
( 80 , 1.533679 )
( 81 , 1.528107 )
( 82 , 1.523376 )
( 83 , 1.518803 )
( 84 , 1.51709 )
( 85 , 1.511677 )
( 86 , 1.511306 )
( 87 , 1.507156 )
( 88 , 1.505215 )
( 89 , 1.502331 )
( 90 , 1.500298 )
( 91 , 1.498521 )
( 92 , 1.496346 )
( 93 , 1.494174 )
( 94 , 1.491947 )
( 95 , 1.490312 )
( 96 , 1.488549 )
( 97 , 1.487457 )
( 98 , 1.485397 )
( 99 , 1.484107 )
( 100 , 1.483047 )
( 101 , 1.482242 )
( 102 , 1.480139 )
( 103 , 1.479509 )
( 104 , 1.47808 )
( 105 , 1.476708 )
( 106 , 1.476282 )
( 107 , 1.474184 )
( 108 , 1.473135 )
( 109 , 1.471243 )
( 110 , 1.470264 )
( 111 , 1.469242 )
( 112 , 1.467822 )
( 113 , 1.46648 )
( 114 , 1.466072 )
( 115 , 1.464334 )
( 116 , 1.463353 )
( 117 , 1.462323 )
( 118 , 1.461243 )
( 119 , 1.460307 )
( 120 , 1.458871 )
( 121 , 1.458266 )
( 122 , 1.456918 )
( 123 , 1.455887 )
( 124 , 1.455345 )
( 125 , 1.45348 )
( 126 , 1.452132 )
( 127 , 1.451079 )
( 128 , 1.450253 )
( 129 , 1.448618 )
( 130 , 1.448093 )
( 131 , 1.446616 )
( 132 , 1.446416 )
( 133 , 1.445376 )
( 134 , 1.444597 )
( 135 , 1.443004 )
( 136 , 1.441797 )
( 137 , 1.440921 )
( 138 , 1.439421 )
( 139 , 1.438453 )
( 140 , 1.4364 )
( 141 , 1.43525 )
( 142 , 1.434088 )
( 143 , 1.43326 )
( 144 , 1.432206 )
( 145 , 1.431509 )
( 146 , 1.430623 )
( 147 , 1.430001 )
( 148 , 1.429378 )
( 149 , 1.429036 )
( 150 , 1.428238 )
( 151 , 1.427671 )
( 152 , 1.427023 )
( 153 , 1.426893 )
( 154 , 1.426064 )
( 155 , 1.425572 )
( 156 , 1.424919 )
( 157 , 1.424709 )
( 158 , 1.424637 )
( 159 , 1.424223 )
( 160 , 1.423714 )
( 161 , 1.423425 )
( 162 , 1.423476 )
( 163 , 1.423017 )
( 164 , 1.423009 )
( 165 , 1.422427 )
( 166 , 1.422472 )
( 167 , 1.42202 )
( 168 , 1.421476 )
( 169 , 1.421528 )
( 170 , 1.421436 )
( 171 , 1.421315 )
( 172 , 1.420781 )
( 173 , 1.420572 )
( 174 , 1.420205 )
( 175 , 1.419931 )
( 176 , 1.419839 )
( 177 , 1.420165 )
( 178 , 1.419906 )
( 179 , 1.420221 )
( 180 , 1.419447 )
( 181 , 1.419647 )
( 182 , 1.419373 )
( 183 , 1.419608 )
( 184 , 1.419064 )
( 185 , 1.419273 )
( 186 , 1.418855 )
( 187 , 1.419162 )
( 188 , 1.418419 )
( 189 , 1.418358 )
( 190 , 1.41809 )
( 191 , 1.418794 )
( 192 , 1.41802 )
( 193 , 1.418396 )
( 194 , 1.418545 )
( 195 , 1.418214 )
( 196 , 1.418051 )
( 197 , 1.417745 )
( 198 , 1.417685 )
( 199 , 1.417654 )
( 200 , 1.417604 )
( 201 , 1.417157 )
( 202 , 1.417049 )
( 203 , 1.417281 )
( 204 , 1.417254 )
( 205 , 1.417445 )
( 206 , 1.4178 )
( 207 , 1.417283 )
( 208 , 1.417023 )
( 209 , 1.41718 )
( 210 , 1.416804 )
( 211 , 1.417075 )
( 212 , 1.416972 )
( 213 , 1.417056 )
( 214 , 1.416371 )
( 215 , 1.416247 )
( 216 , 1.416224 )
( 217 , 1.416347 )
( 218 , 1.415742 )
( 219 , 1.416523 )
( 220 , 1.415541 )
( 221 , 1.416089 )
( 222 , 1.415878 )
( 223 , 1.416837 )
( 224 , 1.416443 )
( 225 , 1.416323 )
( 226 , 1.415643 )
( 227 , 1.416188 )
( 228 , 1.416064 )
( 229 , 1.416041 )
( 230 , 1.415825 )
( 231 , 1.41571 )
( 232 , 1.415862 )
( 233 , 1.41529 )
( 234 , 1.414913 )
( 235 , 1.415162 )
( 236 , 1.414948 )
( 237 , 1.414909 )
( 238 , 1.415886 )
( 239 , 1.415283 )
( 240 , 1.415675 )
( 241 , 1.414861 )
( 242 , 1.41577 )
( 243 , 1.414601 )
( 244 , 1.4135 )
( 245 , 1.413605 )
( 246 , 1.413658 )
( 247 , 1.413203 )
( 248 , 1.41332 )
( 249 , 1.412872 )
( 250 , 1.413327 )
( 251 , 1.414515 )
( 252 , 1.413177 )
( 253 , 1.412834 )
( 254 , 1.412659 )
( 255 , 1.413004 )
( 256 , 1.412833 )
( 257 , 1.412624 )
( 258 , 1.41328 )
( 259 , 1.413055 )
( 260 , 1.412666 )
( 261 , 1.412933 )
( 262 , 1.412504 )
( 263 , 1.41227 )
( 264 , 1.412381 )
( 265 , 1.412732 )
( 266 , 1.413072 )
( 267 , 1.412462 )
( 268 , 1.412301 )
( 269 , 1.412421 )
( 270 , 1.412616 )
( 271 , 1.412468 )
( 272 , 1.412735 )
( 273 , 1.412299 )
( 274 , 1.412467 )
( 275 , 1.412194 )
( 276 , 1.412256 )
( 277 , 1.412055 )
( 278 , 1.411866 )
( 279 , 1.411855 )
( 280 , 1.41237 )
( 281 , 1.411942 )
( 282 , 1.412016 )
( 283 , 1.411568 )
( 284 , 1.411965 )
( 285 , 1.412852 )
( 286 , 1.411995 )
( 287 , 1.411716 )
( 288 , 1.411762 )
( 289 , 1.41166 )
( 290 , 1.41186 )
( 291 , 1.412031 )
( 292 , 1.41196 )
( 293 , 1.411972 )
( 294 , 1.41173 )
( 295 , 1.411547 )
( 296 , 1.411838 )
( 297 , 1.412243 )
( 298 , 1.411755 )
( 299 , 1.411327 )
( 300 , 1.411824 )
}; \label{plot_0}
\addlegendentry{$C$}
\end{axis}
\begin{axis}[
  axis y line*=right,
  axis x line=none,
  ymin=0, ymax=1,
  ylabel=Volume fraction $V_{\Omega_\mathrm{m}} / V_\Omega$,
  ylabel style={xshift=0cm, yshift=-8.25cm},
  yticklabel pos=right,
width =6cm,
  legend style={at={(0.60, 0.93)}, anchor=north west, draw=none, fill=none}
]
\addplot[smooth, black, dashed]
  coordinates{
( 0 ,  0.9346 )
( 1 ,  0.91439 )
( 2 ,  0.90134 )
( 3 ,  0.88732 )
( 4 ,  0.87235 )
( 5 ,  0.85636 )
( 6 ,  0.83908 )
( 7 ,  0.82032 )
( 8 ,  0.79992 )
( 9 ,  0.77766 )
( 10 , 0.75335 )
( 11 ,  0.72636 )
( 12 ,  0.69548 )
( 13 ,  0.65982 )
( 14 ,  0.62231 )
( 15 ,  0.58406 )
( 16 ,  0.54603 )
( 17 ,  0.50745 )
( 18 ,  0.47017 )
( 19 ,  0.44187 )
( 20 ,  0.43029 )
( 21 ,  0.42154 )
( 22 ,  0.41298 )
( 23 ,  0.4065 )
( 24 ,  0.39998 )
( 25 ,  0.4 )
( 26 ,  0.40003 )
( 27 ,  0.40011 )
( 28 ,  0.40032 )
( 29 ,  0.39998 )
( 30 ,  0.40025 )
( 31 ,  0.39976 )
( 32 ,  0.40032 )
( 33 ,  0.40006 )
( 34 ,  0.40006 )
( 35 ,  0.39978 )
( 36 ,  0.39988 )
( 37 ,  0.39981 )
( 38 ,  0.39983 )
( 39 ,  0.39984 )
( 40 ,  0.39994 )
( 41 ,  0.39969 )
( 42 ,  0.39989 )
( 43 ,  0.39983 )
( 44 ,  0.39983 )
( 45 ,  0.39971 )
( 46 ,  0.39997 )
( 47 ,  0.39997 )
( 48 ,  0.39988 )
( 49 ,  0.39974 )
( 50 ,  0.39966 )
( 51 ,  0.39979 )
( 52 ,  0.39968 )
( 53 ,  0.39996 )
( 54 ,  0.39978 )
( 55 ,  0.4001 )
( 56 ,  0.39992 )
( 57 ,  0.39994 )
( 58 ,  0.39978 )
( 59 ,  0.39978 )
( 60 ,  0.39978 )
( 61 ,  0.39995 )
( 62 ,  0.39993 )
( 63 ,  0.39996 )
( 64 ,  0.39974 )
( 65 ,  0.3998 )
( 66 ,  0.39977 )
( 67 ,  0.39979 )
( 68 ,  0.39959 )
( 69 ,  0.39974 )
( 70 ,  0.39973 )
( 71 ,  0.3998 )
( 72 ,  0.39956 )
( 73 ,  0.39993 )
( 74 ,  0.39943 )
( 75 ,  0.39981 )
( 76 ,  0.39976 )
( 77 ,  0.39989 )
( 78 ,  0.39985 )
( 79 ,  0.39993 )
( 80 ,  0.3999 )
( 81 ,  0.39998 )
( 82 ,  0.3999 )
( 83 ,  0.39991 )
( 84 ,  0.39964 )
( 85 ,  0.39974 )
( 86 ,  0.39917 )
( 87 ,  0.39945 )
( 88 ,  0.39932 )
( 89 ,  0.39946 )
( 90 ,  0.39952 )
( 91 ,  0.39951 )
( 92 ,  0.39961 )
( 93 ,  0.39967 )
( 94 ,  0.39974 )
( 95 ,  0.39972 )
( 96 ,  0.39982 )
( 97 ,  0.39972 )
( 98 ,  0.3998 )
( 99 ,  0.39976 )
( 101 ,  0.3996 )
( 102 ,  0.39976 )
( 103 ,  0.39967 )
( 104 ,  0.39972 )
( 105 ,  0.39975 )
( 106 ,  0.39977 )
( 107 ,  0.39975 )
( 108 ,  0.3999 )
( 109 ,  0.3999 )
( 110 ,  0.39993 )
( 111 ,  0.39993 )
( 112 ,  0.39994 )
( 113 ,  0.39993 )
( 114 ,  0.39991 )
( 115 ,  0.3999 )
( 116 ,  0.39994 )
( 117 ,  0.3999 )
( 118 ,  0.39985 )
( 119 ,  0.39993 )
( 120 ,  0.39993 )
( 121 ,  0.39989 )
( 122 ,  0.39988 )
( 123 ,  0.39992 )
( 124 ,  0.39995 )
( 125 ,  0.39993 )
( 126 ,  0.39998 )
( 127 ,  0.39994 )
( 128 ,  0.40002 )
( 129 ,  0.4 )
( 130 ,  0.40002 )
( 131 ,  0.39996 )
( 132 ,  0.39993 )
( 133 ,  0.39984 )
( 134 ,  0.3998 )
( 135 ,  0.39991 )
( 136 ,  0.39994 )
( 137 ,  0.39988 )
( 138 ,  0.3999 )
( 139 ,  0.39985 )
( 140 ,  0.39988 )
( 141 ,  0.39995 )
( 142 ,  0.39993 )
( 143 ,  0.39994 )
( 144 ,  0.39995 )
( 145 ,  0.39996 )
( 146 ,  0.39997 )
( 147 ,  0.39996 )
( 148 ,  0.39995 )
( 149 ,  0.39988 )
( 150 ,  0.39993 )
( 151 ,  0.39996 )
( 152 ,  0.39994 )
( 153 ,  0.39992 )
( 154 ,  0.39996 )
( 155 ,  0.39996 )
( 156 ,  0.39999 )
( 157 ,  0.39999 )
( 158 ,  0.39988 )
( 159 ,  0.39992 )
( 160 ,  0.39994 )
( 161 ,  0.39991 )
( 162 ,  0.39985 )
( 163 ,  0.39985 )
( 164 ,  0.39982 )
( 165 ,  0.39985 )
( 166 ,  0.39984 )
( 167 ,  0.39987 )
( 168 ,  0.39991 )
( 169 ,  0.39987 )
( 170 ,  0.39986 )
( 171 ,  0.39983 )
( 172 ,  0.39989 )
( 173 ,  0.39992 )
( 174 ,  0.39992 )
( 175 ,  0.39992 )
( 176 ,  0.3999 )
( 177 ,  0.39981 )
( 178 ,  0.39985 )
( 179 ,  0.3998 )
( 180 ,  0.39983 )
( 181 ,  0.39988 )
( 182 ,  0.39994 )
( 183 ,  0.39985 )
( 184 ,  0.39991 )
( 185 ,  0.39986 )
( 186 ,  0.3999 )
( 187 ,  0.39988 )
( 188 ,  0.39992 )
( 189 ,  0.39996 )
( 190 ,  0.39996 )
( 191 ,  0.39993 )
( 192 ,  0.39994 )
( 193 ,  0.39992 )
( 194 ,  0.39982 )
( 195 ,  0.39987 )
( 196 ,  0.3999 )
( 197 ,  0.39994 )
( 198 ,  0.39989 )
( 199 ,  0.39992 )
( 200 ,  0.39989 )
( 201 ,  0.39995 )
( 202 ,  0.39997 )
( 203 ,  0.39993 )
( 204 ,  0.39992 )
( 205 ,  0.39991 )
( 206 ,  0.39979 )
( 207 ,  0.39979 )
( 208 ,  0.39992 )
( 209 ,  0.39989 )
( 210 ,  0.39984 )
( 211 ,  0.39994 )
( 212 ,  0.39978 )
( 213 ,  0.39992 )
( 214 ,  0.39989 )
( 215 ,  0.39994 )
( 216 ,  0.39995 )
( 217 ,  0.39994 )
( 218 ,  0.39994 )
( 219 ,  0.39992 )
( 220 ,  0.39994 )
( 221 ,  0.39996 )
( 222 ,  0.39988 )
( 223 ,  0.39986 )
( 224 ,  0.39979 )
( 225 ,  0.3998 )
( 226 ,  0.39991 )
( 227 ,  0.39993 )
( 228 ,  0.3998 )
( 229 ,  0.39986 )
( 230 ,  0.39986 )
( 231 ,  0.39992 )
( 232 ,  0.39986 )
( 233 ,  0.39997 )
( 234 ,  0.39996 )
( 235 ,  0.40008 )
( 236 ,  0.39998 )
( 237 ,  0.39996 )
( 238 ,  0.39986 )
( 239 ,  0.3999 )
( 240 ,  0.39994 )
( 241 ,  0.39996 )
( 242 ,  0.39995 )
( 243 ,  0.40001 )
( 244 ,  0.40004 )
( 245 ,  0.39994 )
( 246 ,  0.39987 )
( 247 ,  0.39992 )
( 248 ,  0.39995 )
( 249 ,  0.39997 )
( 250 ,  0.39991 )
( 251 ,  0.39974 )
( 252 ,  0.39986 )
( 253 ,  0.39995 )
( 254 ,  0.40008 )
( 255 ,  0.39988 )
( 256 ,  0.39986 )
( 257 ,  0.39992 )
( 258 ,  0.39983 )
( 259 ,  0.3998 )
( 260 ,  0.39993 )
( 261 ,  0.39988 )
( 262 ,  0.3999 )
( 263 ,  0.39994 )
( 264 ,  0.39999 )
( 265 ,  0.39992 )
( 266 ,  0.39972 )
( 267 ,  0.39992 )
( 268 ,  0.39996 )
( 269 ,  0.39992 )
( 270 ,  0.39992 )
( 271 ,  0.39985 )
( 272 ,  0.39989 )
( 273 ,  0.39993 )
( 274 ,  0.3998 )
( 275 ,  0.39991 )
( 276 ,  0.39994 )
( 277 ,  0.39996 )
( 278 ,  0.39996 )
( 279 ,  0.39996 )
( 280 ,  0.3999 )
( 281 ,  0.39996 )
( 282 ,  0.39997 )
( 283 ,  0.40008 )
( 284 ,  0.39992 )
( 285 ,  0.39976 )
( 286 ,  0.39986 )
( 287 ,  0.39996 )
( 288 ,  0.39998 )
( 289 ,  0.39998 )
( 290 ,  0.39991 )
( 291 ,  0.3999 )
( 292 ,  0.39997 )
( 293 ,  0.39989 )
( 294 ,  0.39996 )
( 295 ,  0.39997 )
( 296 ,  0.3999 )
( 297 ,  0.39994 )
( 298 ,  0.39996 )
( 299 ,  0.40017 )
}; \addlegendentry{$V_{\Omega_\mathrm{m} / V_\Omega}$}
\end{axis}

\end{tikzpicture}\label{fig:3D_convergence}}
\caption{\SJVDB{Optimized design for the 3-D cantilever beam optimization example (a), and the convergence of the compliance $C$ and volume fraction $V_{\Omega_\mathrm{m}} / V_\Omega$ (b).}}
\end{figure*}
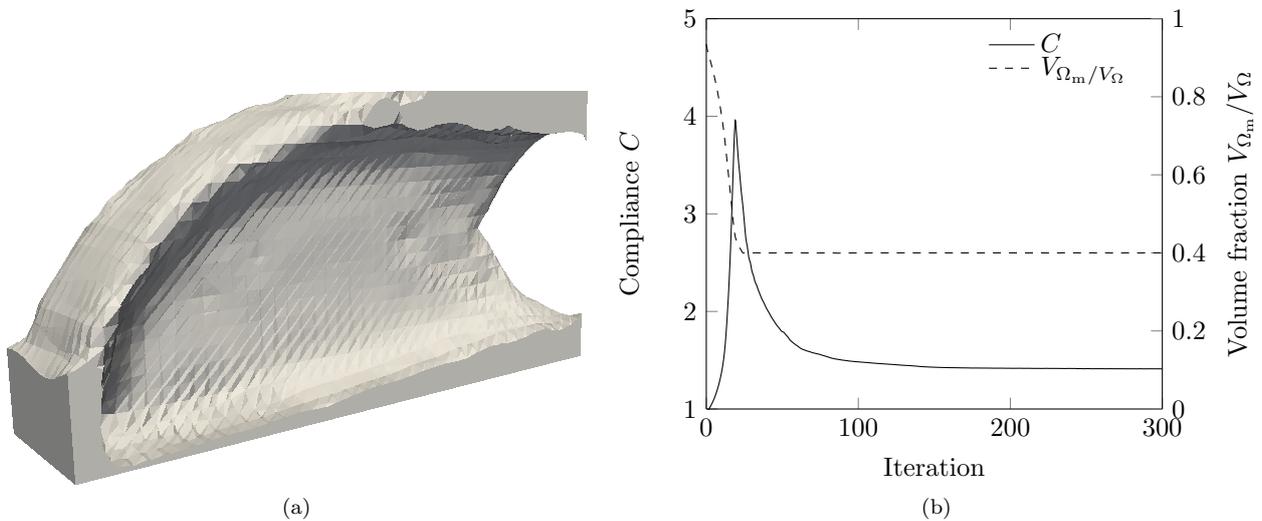

Figure \ref{fig:3D_final} displays the optimized design; the corresponding convergence plot is shown in Figure \ref{fig:3D_convergence}, where it can be seen that the volume satisfied the constraint, and the objective function converges smoothly.

\subsection{Heat sink}
\begin{figure}
\centering
\includegraphics[scale=1.]{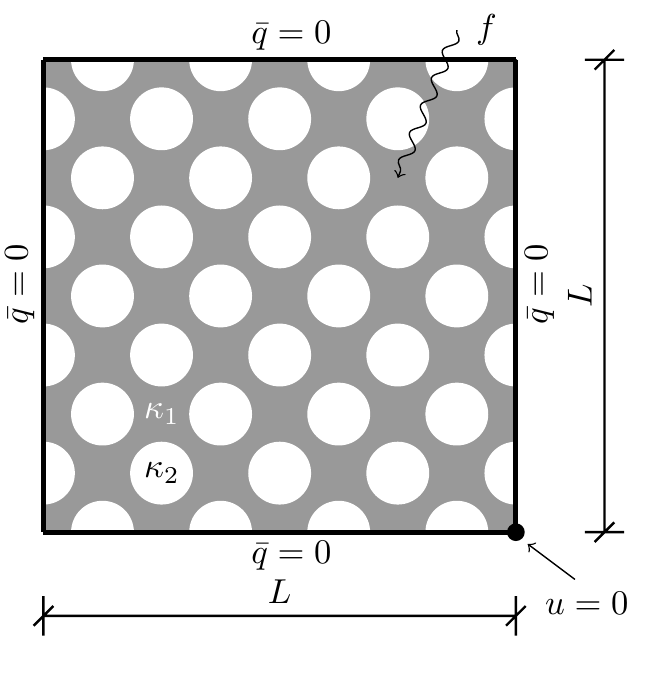}
\caption{Problem description and initial design for the heat sink. A fixed temperature is applied to the bottom right corner, and a uniform heat source is applied throughout the entire square domain.}
\label{fig:heat}
\end{figure}

Lastly, we consider a heat compliance minimization \SJVDB{problem}, illustrated in Figure~\ref{fig:heat}. In this two-material problem, a highly conductive material ($\kappa_1 = 1$) is distributed within an $L \times L$ square domain with a lower conductivity ($\kappa_2 = 0.01$). The bottom-right corner of the domain has a heat sink, with $u = 0$, whereas the domain edges are \SJVDB{adiabatic boundaries, \textit{i.e.},} $\bar q = 0$. The entire domain is subjected to uniform heat source $f = 1$. The problem is solved on a $41 \times 41$ node analysis mesh, using $31 \times 31 $ RBFs with $r_s = \sqrt{2} \cdot a$.

As this problem considers a case with a body load, the load vector also contains enriched degrees of freedom that depend on the locations of the enriched nodes. Therefore, the right hand side is design dependent,  \textit{i.e.}, $\partial \bs F / \partial \bs s \neq \bs 0$, even though the body load is constant throughout the entire domain.

\begin{figure*}
\centering
\subfloat[][]{\centering \includegraphics[width =0.4\columnwidth]{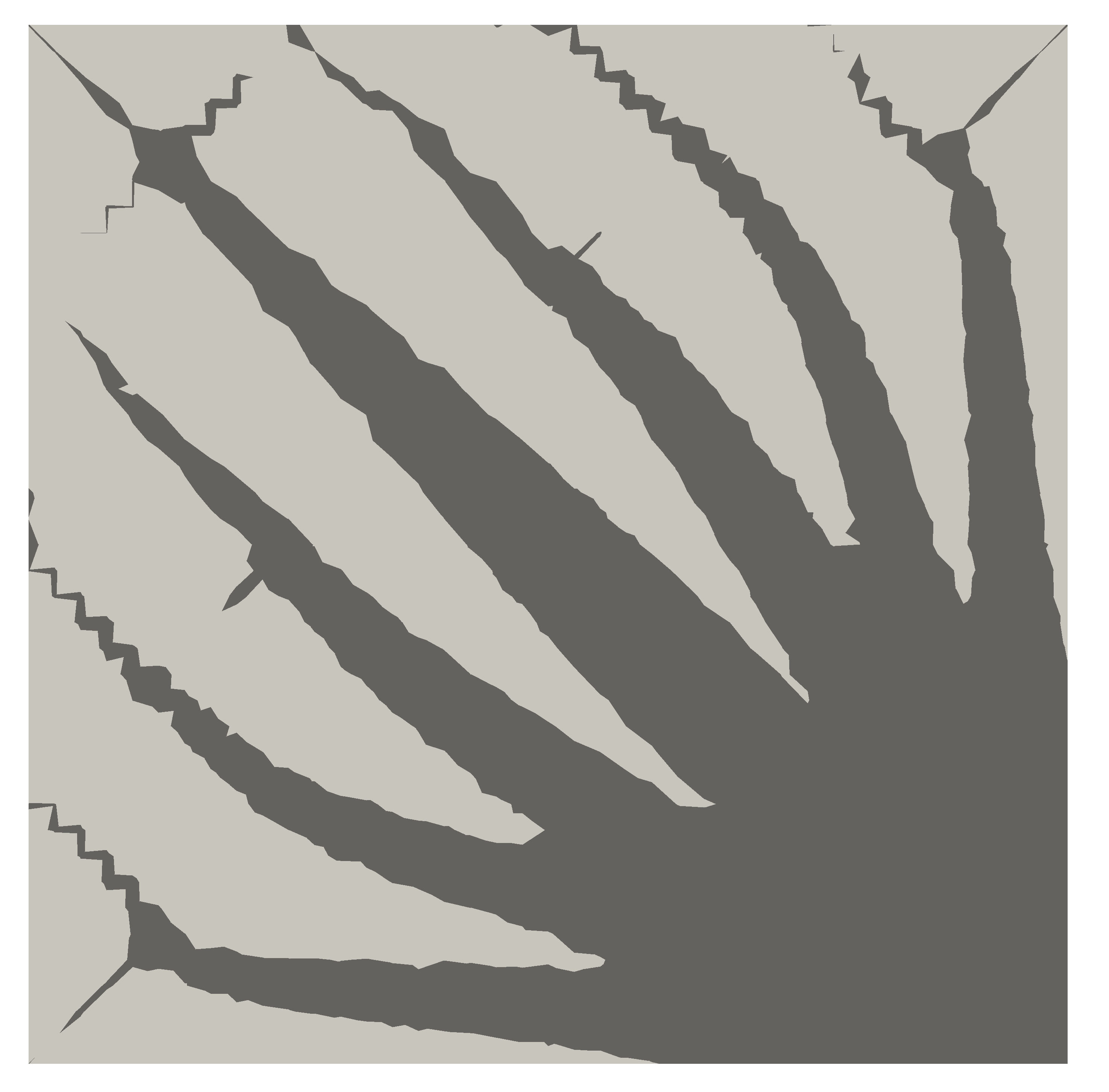}}\;\;
\subfloat[][]{\begin{tikzpicture}[scale=1]
\pgfplotsset{
    scale only axis,
    xmin=0, xmax=100
}

\begin{axis}[
  axis y line*=left,
  ymin=3, ymax=6,
  xlabel=Iteration,
  ylabel=Compliance $C$,
  ylabel style={xshift=0cm, yshift=-0.25cm},
width = 6cm,
legend style={at={(0.6, 0.99)}, anchor=north west, draw=none, fill=none}
]
\addplot[smooth,black]
  coordinates{
( 0 , 4.815195 )
( 1 , 3.959437 )
( 2 , 4.21228 )
( 3 , 3.504908 )
( 4 , 3.709745 )
( 5 , 3.404798 )
( 6 , 3.638394 )
( 7 , 3.430535 )
( 8 , 3.548886 )
( 9 , 3.433155 )
( 10 , 3.543752 )
( 11 , 3.386401 )
( 12 , 3.350254 )
( 13 , 3.314243 )
( 14 , 3.301933 )
( 15 , 3.324661 )
( 16 , 3.296741 )
( 17 , 3.28063 )
( 18 , 3.279174 )
( 19 , 3.278583 )
( 20 , 3.277241 )
( 21 , 3.271626 )
( 22 , 3.276597 )
( 23 , 3.269047 )
( 24 , 3.270212 )
( 25 , 3.268808 )
( 26 , 3.270301 )
( 27 , 3.268319 )
( 28 , 3.265333 )
( 29 , 3.264436 )
( 30 , 3.259739 )
( 31 , 3.26153 )
( 32 , 3.273737 )
( 33 , 3.262569 )
( 34 , 3.257198 )
( 35 , 3.254073 )
( 36 , 3.258203 )
( 37 , 3.257099 )
( 38 , 3.273522 )
( 39 , 3.258458 )
( 40 , 3.256925 )
( 41 , 3.256511 )
( 42 , 3.254568 )
( 43 , 3.252688 )
( 44 , 3.254049 )
( 45 , 3.251386 )
( 46 , 3.253356 )
( 47 , 3.254114 )
( 48 , 3.252059 )
( 49 , 3.25647 )
( 50 , 3.248902 )
( 51 , 3.247057 )
( 52 , 3.250503 )
( 53 , 3.25212 )
( 54 , 3.250188 )
( 55 , 3.250462 )
( 56 , 3.249385 )
( 57 , 3.252262 )
( 58 , 3.248474 )
( 59 , 3.246871 )
( 60 , 3.245687 )
( 61 , 3.248359 )
( 62 , 3.246209 )
( 63 , 3.246162 )
( 64 , 3.246241 )
( 65 , 3.242001 )
( 66 , 3.243899 )
( 67 , 3.2436 )
( 68 , 3.24811 )
( 69 , 3.242657 )
( 70 , 3.242028 )
( 71 , 3.243752 )
( 72 , 3.242996 )
( 73 , 3.242522 )
( 74 , 3.244993 )
( 75 , 3.24042 )
( 76 , 3.242336 )
( 77 , 3.240738 )
( 78 , 3.240776 )
( 79 , 3.245044 )
( 80 , 3.245004 )
( 81 , 3.240481 )
( 82 , 3.240084 )
( 83 , 3.240043 )
( 84 , 3.239603 )
( 85 , 3.23932 )
( 86 , 3.241601 )
( 87 , 3.243924 )
( 88 , 3.247981 )
( 89 , 3.239087 )
( 90 , 3.240083 )
( 91 , 3.240161 )
( 92 , 3.242783 )
( 93 , 3.239969 )
( 94 , 3.239006 )
( 95 , 3.241327 )
( 96 , 3.241164 )
( 97 , 3.239936 )
( 98 , 3.242039 )
( 99 , 3.242018 )
( 100 , 3.240419 )
};
\addlegendentry{$C$}
\end{axis}
\begin{axis}[
  axis y line*=right,
  axis x line=none,
  ymin=0, ymax=1,
  ylabel=Volume fraction $V_{\Omega_\mathrm{m}} / V_\Omega$,
  ylabel style={xshift=0cm, yshift=-8.25cm},
  yticklabel pos=right,
width = 6cm,
  legend style={at={(0.6, 0.92)}, anchor=north west, draw=none, fill=none}
]

\addplot[smooth,black, dashed]
  coordinates{
( 0 , 0.6110979545714862 )
( 1 , 0.5621562465806342 )
( 2 , 0.4305083541259042 )
( 3 , 0.4578458496754617 )
( 4 , 0.4431245752116011 )
( 5 , 0.4436569000409036 )
( 6 , 0.4447334569334889 )
( 7 , 0.4474703892744281 )
( 8 , 0.44100779574788784 )
( 9 , 0.43171591617368527 )
( 10 , 0.43061671264348633 )
( 11 , 0.43416174235851546 )
( 12 , 0.44106049744122316 )
( 13 , 0.4460183809743423 )
( 14 , 0.44746950356658766 )
( 15 , 0.444647409539511 )
( 16 , 0.44685361417610237 )
( 17 , 0.45077001805910466 )
( 18 , 0.4520467046977475 )
( 19 , 0.45130055039816286 )
( 20 , 0.45143448078073584 )
( 21 , 0.45031588550569585 )
( 22 , 0.4467385183568582 )
( 23 , 0.4506121395037429 )
( 24 , 0.44718855084472664 )
( 25 , 0.4494460134787933 )
( 26 , 0.44571042231141744 )
( 27 , 0.44870775319356493 )
( 28 , 0.4472035199878958 )
( 29 , 0.44954739463764154 )
( 30 , 0.4487877771804338 )
( 31 , 0.4486194242600304 )
( 32 , 0.44238312554955517 )
( 33 , 0.4466645115130772 )
( 34 , 0.44921380611379647 )
( 35 , 0.44939085737130224 )
( 36 , 0.4486956481367535 )
( 37 , 0.4485467956823654 )
( 38 , 0.4418503706466609 )
( 39 , 0.44846881322949217 )
( 40 , 0.44901008713071966 )
( 41 , 0.4483691645462425 )
( 42 , 0.44898317011250033 )
( 43 , 0.4487744191734093 )
( 44 , 0.44850486961533903 )
( 45 , 0.44876654532909055 )
( 46 , 0.44827714915041256 )
( 47 , 0.4470920377890365 )
( 48 , 0.44933280377197415 )
( 49 , 0.4461469514707887 )
( 50 , 0.44904326110274173 )
( 51 , 0.4500988873538236 )
( 52 , 0.44883542704437895 )
( 53 , 0.44733305391345934 )
( 54 , 0.44894847112497216 )
( 55 , 0.4481359236904481 )
( 56 , 0.4486354190103833 )
( 57 , 0.4478701835470005 )
( 58 , 0.4489807861631394 )
( 59 , 0.44913854108625734 )
( 60 , 0.44978938508383765 )
( 61 , 0.44856218386961777 )
( 62 , 0.4501269765285036 )
( 63 , 0.449368872001242 )
( 64 , 0.44897566787691073 )
( 65 , 0.45159999891434865 )
( 66 , 0.4497020728932057 )
( 67 , 0.4501954046826145 )
( 68 , 0.44808202818982235 )
( 69 , 0.4501752240454942 )
( 70 , 0.4503488847806911 )
( 71 , 0.44941168539545806 )
( 72 , 0.4490982751314733 )
( 73 , 0.45008030715950725 )
( 74 , 0.44810313010731806 )
( 75 , 0.4504765455365691 )
( 76 , 0.4492174640453977 )
( 77 , 0.44980000229836115 )
( 78 , 0.44973224658229854 )
( 79 , 0.44853415242750866 )
( 80 , 0.4488644520910577 )
( 81 , 0.4502279171349406 )
( 82 , 0.45030370793435476 )
( 83 , 0.45045849811275496 )
( 84 , 0.45050169877617713 )
( 85 , 0.45022833310816185 )
( 86 , 0.450417542163272 )
( 87 , 0.44927047784886803 )
( 88 , 0.4463823531653971 )
( 89 , 0.45091195610264595 )
( 90 , 0.44967037432457974 )
( 91 , 0.4506071098673251 )
( 92 , 0.44920642440715086 )
( 93 , 0.45022440279646775 )
( 94 , 0.4510303688011761 )
( 95 , 0.44925662707922115 )
( 96 , 0.449019924682844 )
( 97 , 0.45090734961206663 )
( 98 , 0.44945229296038386 )
( 99 , 0.45029037904676167 )
( 100 , 0.44978386038271645 )
}; \label{plot_vol_one}
\addlegendentry{$V_{\Omega_\mathrm{m}} / V_\Omega$}
\end{axis}
\end{tikzpicture}}
\caption{\SJVDB{Results of the heat sink problem: (a) shows the optimized design of the heat sink problem. Observe that narrow features are created by placing the zero-contour of the levelset near the edges of the original mesh element. In the convergence plot shown in (b), initially some small oscillations are observed in both the objective and volume convergence, which can be prevented by the use of a smaller move limit.}}\label{fig:heat_sink}
\end{figure*}
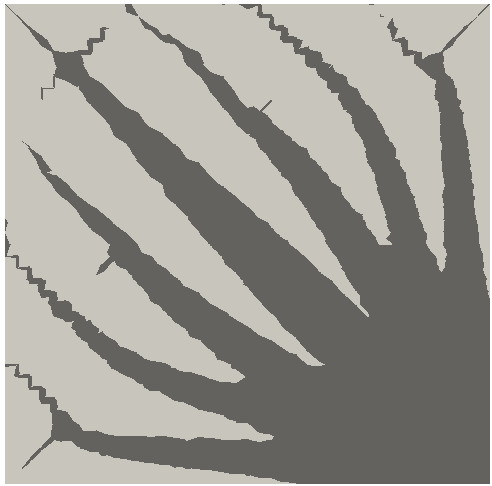
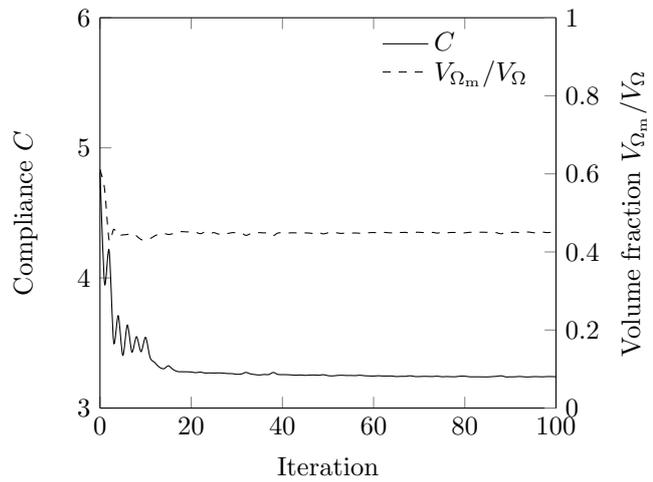

The results of this optimization problem are shown in Figure~\ref{fig:heat_sink}. In the optimized design, narrow features can be distinguished that follow the edges of original elements in the background mesh. This is an effect \SJVDB{caused by} how the intersections are detected, and is investigated in more detail in \S\ref{sec:disc}. 
The convergence plot shows that, \SJVDB{although} there are \SJVDB{initially} some oscillations in both the objective and constraint (also investigated further in \S\ref{sec:disc}), they converge in the end.

\section{Discussion}
\label{sec:discussion}

\subsection{Oscillations: the levelset discretization}
\label{sec:disc}

Oscillations in the objective functions are visible in the convergence of the heat sink problem in Figure~\ref{fig:heat_sink}, and in the coarsest RBF mesh of the MBB beam in Figure~\ref{fig:MBB_convergence}.  As these oscillations might point to inaccurate modeling or sensitivities, the phenomenon is \SJVDB{discussed} here in more detail.

\begin{figure}
\centering
\def\svgwidth{0.7\columnwidth}
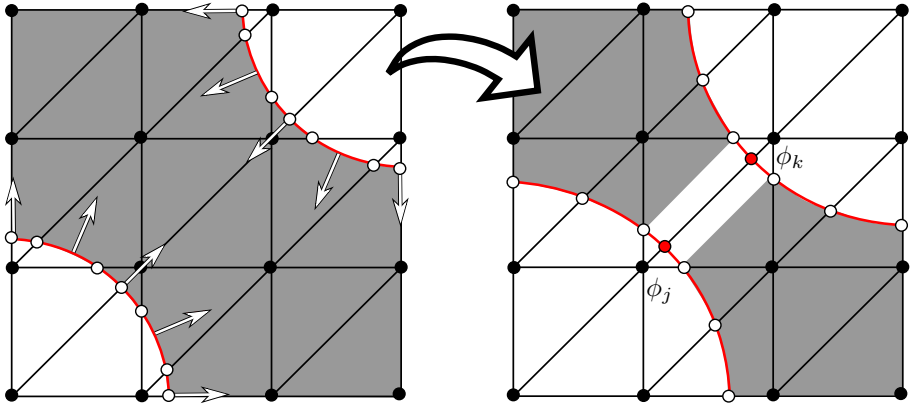
\caption{Illustration of the structure disconnecting due to the levelset discretization. On the left the zero-contour of the levelset, shown in red, defines the design shown in gray. The white arrows indicate the update of the levelset in the next iteration. On the right, the next iteration is shown, where the narrowest part of the zero-contour lies within a single element, and the nodal levelset values $\phi_j$ and $\phi_k$ have the same sign. Therefore, the two intersections shown as \wrongDOF{} are not found, and the structure disconnects, as shown \SJVDB{by the new gray design}. }

\label{fig:oscillations}
\end{figure}

Recall that intersections between the zero contour of the levelset function and element edges are found using a linear interpolation of nodal levelset values. Because the levelset function is discretized, no intersections can be found if two adjacent nodes have the same sign. This effect is illustrated in Figure~\ref{fig:oscillations}. On the left, the zero-contour of a levelset function \SJVDB{is shown} in red, which defines a design shown in white/gray. The white arrows indicate the movement of the material boundary during the next design update. On the right, the updated levelset contour is shown in red. As the two adjacent original nodes $\phi_j$ and $\phi_k$ now have the same sign, the two intersections between them, shown as \wrongDOF{} cannot be found.

The sudden disconnection of the structure due to the levelset discretization is a discontinuous event that cannot be captured by the sensitivity information. Therefore, as soon as such discontinuous event occurs, the sensitivities and the modeling deviate, and oscillations may occur.

This problem can be alleviated by using a smaller move limit, as was done in the 3D MBB example. Another approach that could mitigate this issue is to evaluate the parametrized levelset function on a finer grid, so that multiple intersections are found on an element edge. However, the procedure that creates integration elements would also need to allow for these more complex intersections. It should be noted that neither of these methods completely eliminates the problem of discontinuous events. Rather, the methods alleviate the problem by limiting their chance of occurrence. 

\begin{figure}
\centering
\includegraphics[width=0.7\columnwidth]{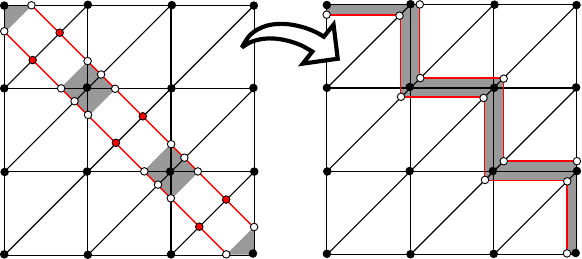}
\caption{Illustration of the zigzagged pattern that appears in Figure~\ref{fig:heat_sink}. When a narrow diagonal line is desired in the opposite direction of the diagonal lines of the mesh, the problem illustrated in Figure~\ref{fig:oscillations} results in a disconnected line, as shown on the left. Instead, the optimizer will create narrow features along element edges, as illustrated on the right.}

\label{fig:morezigzags}
\end{figure}

A related observation \SJVDB{can be made} in the zigzagged \SJVDB{features} in the heat sink design of Figure~\ref{fig:heat_sink}. As illustrated in Figure~\ref{fig:morezigzags}, this pattern occurs when the optimizer tries to make a narrow diagonal feature in the opposite direction of the mesh diagonals. The red intersections cannot be detected, and therefore the structure is disconnected. As a result, the optimizer can only create diagonal narrow features by zigzagging them along element edges, as illustrated in Figure~\ref{fig:morezigzags} on the right. 

\subsection{Zigzagging: approximation error}
\label{sec:zigzag}

\begin{figure}
\centering
\def\svgwidth{0.6\columnwidth}
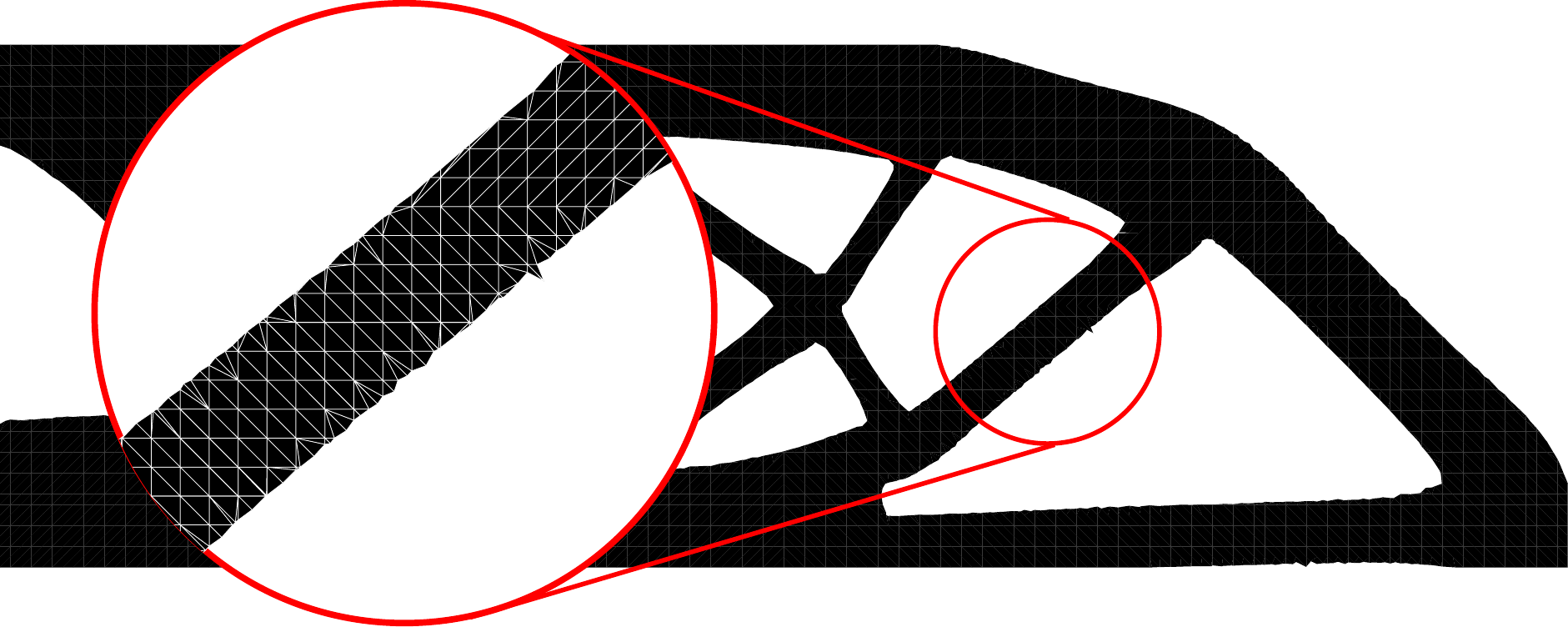
\caption{Detail of zigzagging that might occur when the design space is not reduced with respect to the FE mesh.}
\label{fig:wiggleinset}
\end{figure}

\begin{figure}
\centering
\def\svgwidth{180pt}
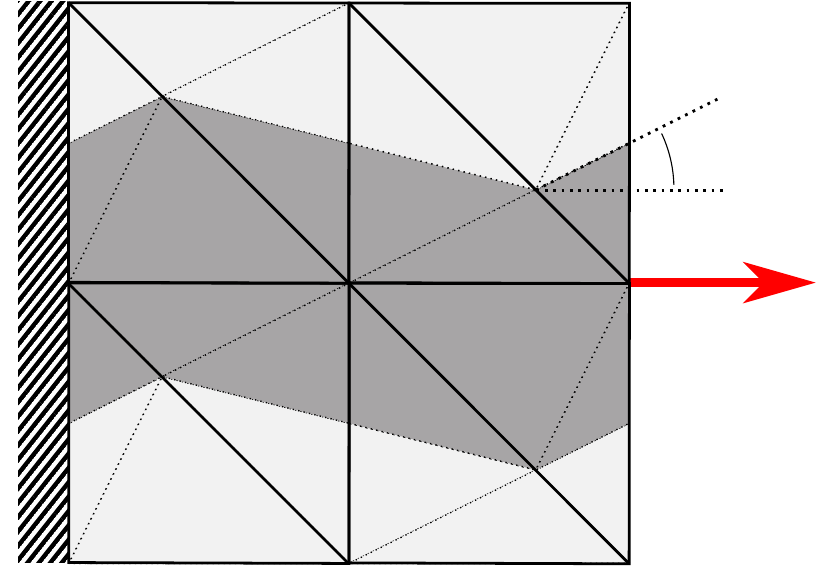
\caption{\SJVDB{Schematic} for the zigzagging approximation error. A beam with zigzagging angle $\alpha$ is clamped on the left, while a concentrated axial loading is applied on the right. \SJVDB{The} angle $\alpha$ is varied without changing the material volume, and the compliance is evaluated.}
\label{fig:wiggles}
\end{figure}

In the final designs of some of the numerical examples, zigzagging of the edges occurred where the zero contour of the levelset function is not perfectly smooth, as detailed in Figure~\ref{fig:wiggleinset}. To investigate the cause of this artifact, the test problem of a clamped beam loaded axially shown in Figure~\ref{fig:wiggles} was investigated.
The compliance was computed for a varying zigzagging angle $\alpha$ while keeping the material volume constant. 

\begin{figure}
\centering
\def\svgwidth{180pt}
\includegraphics[scale=1]{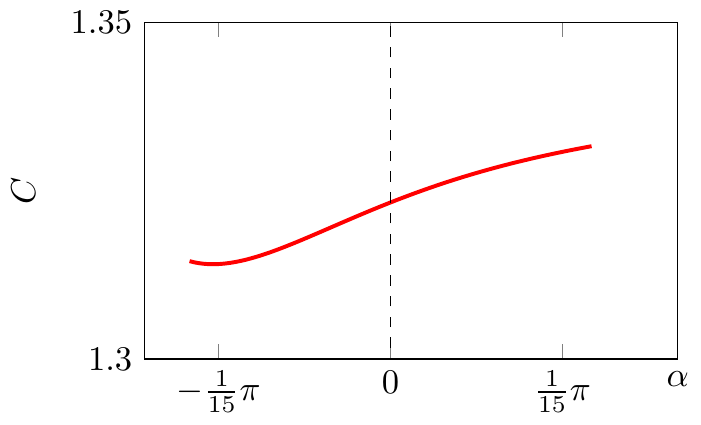}
\caption{The compliance of the test case, illustrated in Figure~\ref{fig:wiggles}, as a function of the zigzagging angle $\alpha$. The compliance for this coarse test case is non-symmetric with respect to $0$.  }
\label{fig:wiggles_comp}
\end{figure}

The results in Figure~\ref{fig:wiggles_comp} show that the minimum compliance is not found at $\alpha = 0$, as one would expect, but instead it is found at a negative value  of $\alpha$. Furthermore, the compliance is not symmetric with respect to $\alpha = 0$ due to the asymmetry of the analysis mesh. The cause of this zigzagging is an approximation error, as the mesh is too coarse to accurately describe the deformations and stresses in the structure, similarly to the effect described for nodal design variables in~\cite{Braibant:1984}. This effect can be resolved by reducing the design space with respect to the analysis mesh, for example with the use of RBFs, or by increasing the element order. Furthermore, as the non-smoothness is confined to a single layer of background elements, mesh-refinement makes the issue less pronounced.  

\section{Summary and Conclusions} 

In this work we introduced a new enriched topology optimization approach based on the Interface-enriched Generalized Finite Element Method (IGFEM). The technique yields non-pixelized black and white designs, that do not require any postprocessing. We have derived an analytic expression for the sensitivities, and have shown that they can be computed with relatively low computational effort. Furthermore, the method was compared to a number of open source topology optimization codes, based on SIMP, the Ersatz approach, and discrete levelsets. The influence of decoupling the design discretization from the analysis mesh was investigated using the classical MBB beam optimization problem. A 3-D cantilever beam and a heat sink problem were also demonstrated. The convergence behavior was provided for each numerical example. Any numerical artifacts, such as approximation errors and discretization errors of the levelset, as discussed in~\S\ref{sec:discussion}, can be mitigated by means of suitable move limits and radial basis functions, where the latter serve as a sort of filter because they can control the design complexity. 

A number of conclusion\SJVDB{s can} be drawn from this work:
\begin{itemize}
\item The combination of IGFEM with the levelset topology optimization based on RBFs results in crisp boundaries in both the design representation and the modeling. Because the RBF mesh and analysis mesh are completely decoupled, the resolution of the design and the modeling can be chosen independently, as is the case in any parametrized levelset optimization. In addition, the radial basis functions help in reducing numerical artifacts, as they act like a black-and-white filter. Lastly, as the RBFs may extend over multiple elements, they allow the boundary to move further and the optimizer to converge faster;
\item As only one intersection can be detected per element edge, due to the mapping of the levelset to the original mesh nodes, features smaller than a single element might not be described correctly.
As discussed in \S\ref{sec:disc}, this may lead to oscillations in the convergence. Using a finer grid for evaluating the levelsets, more intersection may be found, allowing for narrower features. However, this will require a more involved procedure for creating integration elements. Similarly, the method may be extended to be used on quadrilateral elements, which also requires more involved integration element procedures. Furthermore, for quadrilateral elements, higher order enrichment functions are needed~\cite{Aragon:2020};
\item Due to approximation error, numerical artifacts may occur which may be exploited by the optimizer when the RBF mesh is too fine with respect to the analysis mesh. Another known issue in IGFEM and other enriched methods, that may be exploited by the optimizer, is \SJVDB{the fact that} the computation of stresses near material interfaces \SJVDB{may yield inaccurate results}~\cite{Soghrati:2017,Nagarajan:2018};
\item In this work, \SJVDB{we chose} to model the void along with the material domain for a number of reasons, including the ease of implementation, and the \SJVDB{ease of comparing} to other methods. However, \SJVDB{we could have} chosen to completely remove the void from the analysis~\cite{vandenBoom:2018a}, which would reduce computation times and \SJVDB{eliminate the} artificial stiffness in the void. 
\end{itemize}

The benefits of using an enriched formulation are expected to be more pronounced for problems that rely heavily on an accurate boundary description, such as fluid-structure interaction and wave scattering. In fact, the optimization of the latter is the subject of an incoming publication. Furthermore, by having an enriched formulation, the technique has the potential to easily handle problems where boundary conditions have to be prescribed on evolving boundaries. In particular, and unlike other immersed methods, with IGFEM it is even possible to prescribe \textit{strongly} non-homogeneous essential (Dirichlet) boundary conditions~\cite{Aragon:2017,vandenBoom:2018a}. 



\section*{Acknowledgements}
The authors would like to thank Krister Svanberg for providing us with the MMA implementation.

\bibliographystyle{unsrt}      
\bibliography{bibliography}   

\appendix
\section{Derivatives of the Jacobian inverse and determinant} \label{appendix}
In the sensitivity computation discussed in \SJVDB{\S\ref{sec:sens}}, the derivative of the Jacobian inverse and determinant are required. 
According to Jacobi's formula~\cite{Magnus:2007}, the derivative of the determinant of a matrix can be computed as the trace of the adjugate of the matrix, multiplied by the derivative of the matrix. For the Jacobian determinant $j_e$, the derivative can thus be computed as:
\begin{equation} \label{eq:djdx}
\frac{\partial j_e}{\partial \bs x_n} = \Tr \left( \adj \left( \bs J_e \right) \frac{\partial \bs J_e}{\partial \bs x_n}\right) ,
\end{equation} 

The sensitivity of the Jacobian inverse can be computed by realizing that $\bs J_e \bs J_e^{-1} = \bs I$:
\begin{equation} \label{eq:dJ-1dx-pre}
\frac{\partial \bs J_e \bs J_e^{-1}}{\partial \bs x_n} = \frac{\partial \bs J_e}{\partial \bs x_n} \bs J_e^{-1} + \bs J_e \frac{\partial \bs J_e^{-1}}{\partial \bs x_n}  = \frac{\partial \bs I}{\partial \bs x_n} = 0, 
\end{equation}
and solving for $\partial \bs J_e^{-1} /\partial \bs x_n$:
\begin{equation} \label{eq:dJ-1dx}
\frac{\partial \bs J_e^{-1}}{\partial \bs x_n} = - \bs J_e^{-1} \frac{\partial \bs J_e}{\partial \bs x_n} \bs J_e^{-1} .
\end{equation} For both~\eqref{eq:djdx} and~\eqref{eq:dJ-1dx}, the sensitivity of the Jacobian is required; as the Jacobian \SJVDB{of the integration element} is computed as \SJVDB{$\bs J_e = \bs x_e \bs \Delta \bs \psi_e$} it can be computed as \SJVDB{
\begin{equation} \label{eq:dJdx}
\frac{\partial \bs J_e}{\partial \bs x_n} = \frac{\partial \bs x_e}{\partial \bs x_n}  \bs \Delta \bs \psi_e,
\end{equation} where }$\partial \bs x_e / \partial \bs x_n$ is simply a selection matrix consisting of zeros except for a one on the coordinates of interest for enriched node $n$.

\end{document}

%% file: figures/topopt_potato.pdf_tex
\begingroup%
  \makeatletter%
  \providecommand\color[2][]{%
    \errmessage{(Inkscape) Color is used for the text in Inkscape, but the package 'color.sty' is not loaded}%
    \renewcommand\color[2][]{}%
  }%
  \providecommand\transparent[1]{%
    \errmessage{(Inkscape) Transparency is used (non-zero) for the text in Inkscape, but the package 'transparent.sty' is not loaded}%
    \renewcommand\transparent[1]{}%
  }%
  \providecommand\rotatebox[2]{#2}%
  \ifx\svgwidth\undefined%
    \setlength{\unitlength}{157.3999939bp}%
    \ifx\svgscale\undefined%
      \relax%
    \else%
      \setlength{\unitlength}{\unitlength * \real{\svgscale}}%
    \fi%
  \else%
    \setlength{\unitlength}{\svgwidth}%
  \fi%
  \global\let\svgwidth\undefined%
  \global\let\svgscale\undefined%
  \makeatother%
  \begin{picture}(1,0.76568963)%
    \put(0,0){\includegraphics[width=\unitlength,page=1]{topopt_potato.pdf}}%
    \put(0.44980942,0.76869768){\color[rgb]{0,0,0}\makebox(0,0)[lt]{\begin{minipage}{0.10165185\unitlength}\raggedright $\bs n_m$\end{minipage}}}%
    \put(0.53113089,0.19563538){\color[rgb]{0,0,0}\makebox(0,0)[lb]{\smash{$\bs n$ }}}%
    \put(0.11181703,0.01901531){\color[rgb]{0,0,0}\makebox(0,0)[lt]{\begin{minipage}{0.07115629\unitlength}\raggedright $\bs e_1$\end{minipage}}}%
    \put(0,0.1219378){\color[rgb]{0,0,0}\makebox(0,0)[lt]{\begin{minipage}{0.068615\unitlength}\raggedright $\bs e_2$\end{minipage}}}%
    \put(0.25794157,0.62765576){\color[rgb]{0,0,0}\makebox(0,0)[lb]{\smash{$\Gamma_t$}}}%
    \put(0.0978399,0.70516527){\color[rgb]{0,0,0}\makebox(0,0)[lb]{\smash{$\bs{\bar{t}}$}}}%
    \put(0.17916138,0.47644862){\color[rgb]{0,0,0}\makebox(0,0)[lb]{\smash{$\Gamma_m$}}}%
    \put(0.50571793,0.60986666){\color[rgb]{0,0,0}\makebox(0,0)[lb]{\smash{$\Omega_v$}}}%
    \put(0.35705212,0.3252415){\color[rgb]{0,0,0}\makebox(0,0)[lb]{\smash{$\Omega_m$}}}%
    \put(0,0){\includegraphics[width=\unitlength,page=2]{topopt_potato.pdf}}%
    \put(0.28481938,0.08437567){\color[rgb]{0,0,0}\makebox(0,0)[lt]{\begin{minipage}{0.10602108\unitlength}\raggedright $\Gamma_u$\end{minipage}}}%
    \put(0.73284628,0.25726182){\color[rgb]{0,0,0}\makebox(0,0)[lt]{\begin{minipage}{0.06734435\unitlength}\raggedright $\bs x_j$\end{minipage}}}%
    \put(0.86912329,0.36526691){\color[rgb]{0,0,0}\makebox(0,0)[lt]{\begin{minipage}{0.09529861\unitlength}\raggedright $\bs x_k $ \end{minipage}}}%
    \put(0.69417159,0.04001527){\color[rgb]{0,0,0}\makebox(0,0)[lb]{\smash{$\phi < 0$}}}%
    \put(0.85163499,0.04001527){\color[rgb]{0,0,0}\makebox(0,0)[lb]{\smash{$\phi > 0$}}}%

    \put(0,0){\includegraphics[width=\unitlength,page=3]{topopt_potato.pdf}}%
  \end{picture}%
\endgroup%

%% file: figures/enrichment.pdf_tex
\begingroup%
  \makeatletter%
  \providecommand\color[2][]{%
    \errmessage{(Inkscape) Color is used for the text in Inkscape, but the package 'color.sty' is not loaded}%
    \renewcommand\color[2][]{}%
  }%
  \providecommand\transparent[1]{%
    \errmessage{(Inkscape) Transparency is used (non-zero) for the text in Inkscape, but the package 'transparent.sty' is not loaded}%
    \renewcommand\transparent[1]{}%
  }%
  \providecommand\rotatebox[2]{#2}%
  \ifx\svgwidth\undefined%
    \setlength{\unitlength}{58.85902023bp}%
    \ifx\svgscale\undefined%
      \relax%
    \else%
      \setlength{\unitlength}{\unitlength * \real{\svgscale}}%
    \fi%
  \else%
    \setlength{\unitlength}{\svgwidth}%
  \fi%
  \global\let\svgwidth\undefined%
  \global\let\svgscale\undefined%
  \makeatother%
  \begin{picture}(1,0.44610779)%
    \put(0,0){\includegraphics[width=\unitlength,page=1]{enrichment.pdf}}%
    \put(0.64202193,0.40403673){\color[rgb]{0,0,0}\makebox(0,0)[lb]{\smash{$\Psi_i$}}}%
    \put(0.67571367,0.09575678){\color[rgb]{0,0,0}\makebox(0,0)[lb]{\smash{$\bs \alpha_i$}}}%
    \put(0,0){\includegraphics[width=\unitlength,page=2]{enrichment.pdf}}%
  \end{picture}%
\endgroup%

%% file: figures/Cantilever_comparison.pdf_tex
\begingroup%
  \makeatletter%
  \providecommand\color[2][]{%
    \errmessage{(Inkscape) Color is used for the text in Inkscape, but the package 'color.sty' is not loaded}%
    \renewcommand\color[2][]{}%
  }%
  \providecommand\transparent[1]{%
    \errmessage{(Inkscape) Transparency is used (non-zero) for the text in Inkscape, but the package 'transparent.sty' is not loaded}%
    \renewcommand\transparent[1]{}%
  }%
  \providecommand\rotatebox[2]{#2}%
  \ifx\svgwidth\undefined%
    \setlength{\unitlength}{1637.28211725bp}%
    \ifx\svgscale\undefined%
      \relax%
    \else%
      \setlength{\unitlength}{\unitlength * \real{\svgscale}}%
    \fi%
  \else%
    \setlength{\unitlength}{\svgwidth}%
  \fi%
  \global\let\svgwidth\undefined%
  \global\let\svgscale\undefined%
  \makeatother%
  \begin{picture}(1,0.60213692)%
    \put(0,0){\includegraphics[width=\unitlength,page=1]{Cantilever_comparison.pdf}}%
    \put(0.10245969,0.60320348){\color[rgb]{0,0,0}\makebox(0,0)[lt]{\begin{minipage}{0.13839593\unitlength}\raggedright IGFEM\end{minipage}}}%
    \put(0.33139502,0.59091591){\color[rgb]{0,0,0}\makebox(0,0)[lb]{\smash{SIMP~\cite{sigmund:2001}}}}%
    \put(0.5680909,0.5928561){\color[rgb]{0,0,0}\makebox(0,0)[lb]{\smash{Density mapping~\cite{wei:2018}}}}%
    \put(0.79702627,0.59091591){\color[rgb]{0,0,0}\makebox(0,0)[lb]{\smash{Discrete levelset~\cite{challis:2010}}}}%
    \put(0.00286635,0.50833888){\color[rgb]{0,0,0}\makebox(0,0)[lb]{\smash{$21 \times 11$}}}%
    \put(0.00157294,0.40098476){\color[rgb]{0,0,0}\makebox(0,0)[lt]{\begin{minipage}{0.09829992\unitlength}\raggedright $41 \times 21 $\end{minipage}}}%
    \put(0.00286635,0.28457697){\color[rgb]{0,0,0}\makebox(0,0)[lt]{\begin{minipage}{0.10864728\unitlength}\raggedright $61 \times 31$\end{minipage}}}%
    \put(-0.00877442,0.1873689){\color[rgb]{0,0,0}\makebox(0,0)[lt]{\begin{minipage}{0.04785654\unitlength}\raggedright \end{minipage}}}%
    \put(0.00157294,0.17075601){\color[rgb]{0,0,0}\makebox(0,0)[lt]{\begin{minipage}{0.09312624\unitlength}\raggedright $81 \times 41$\end{minipage}}}%
    \put(-0.0010139,0.05176136){\color[rgb]{0,0,0}\makebox(0,0)[lt]{\begin{minipage}{0.11899464\unitlength}\raggedright $101 \times 51$\end{minipage}}}%
  \end{picture}%
\endgroup%

%% file: figures/3D_initial.pdf_tex
\begingroup%
  \makeatletter%
  \providecommand\color[2][]{%
    \errmessage{(Inkscape) Color is used for the text in Inkscape, but the package 'color.sty' is not loaded}%
    \renewcommand\color[2][]{}%
  }%
  \providecommand\transparent[1]{%
    \errmessage{(Inkscape) Transparency is used (non-zero) for the text in Inkscape, but the package 'transparent.sty' is not loaded}%
    \renewcommand\transparent[1]{}%
  }%
  \providecommand\rotatebox[2]{#2}%
  \ifx\svgwidth\undefined%
    \setlength{\unitlength}{1050.5086994bp}%
    \ifx\svgscale\undefined%
      \relax%
    \else%
      \setlength{\unitlength}{\unitlength * \real{\svgscale}}%
    \fi%
  \else%
    \setlength{\unitlength}{\svgwidth}%
  \fi%
  \global\let\svgwidth\undefined%
  \global\let\svgscale\undefined%
  \makeatother%
  \begin{picture}(1,0.79314431)%
    \put(0,0){\includegraphics[width=\unitlength,page=1]{3D_initial.pdf}}%
    \put(0.17881564,0.05820982){\color[rgb]{0,0,0}\makebox(0,0)[lt]{\begin{minipage}{0.08723491\unitlength}\raggedright $ \bar{\bs t}$\end{minipage}}}%
  \end{picture}%
\endgroup%

%% file: figures/schematic.pdf_tex
\begingroup%
  \makeatletter%
  \providecommand\color[2][]{%
    \errmessage{(Inkscape) Color is used for the text in Inkscape, but the package 'color.sty' is not loaded}%
    \renewcommand\color[2][]{}%
  }%
  \providecommand\transparent[1]{%
    \errmessage{(Inkscape) Transparency is used (non-zero) for the text in Inkscape, but the package 'transparent.sty' is not loaded}%
    \renewcommand\transparent[1]{}%
  }%
  \providecommand\rotatebox[2]{#2}%
  \ifx\svgwidth\undefined%
    \setlength{\unitlength}{167.2432459bp}%
    \ifx\svgscale\undefined%
      \relax%
    \else%
      \setlength{\unitlength}{\unitlength * \real{\svgscale}}%
    \fi%
  \else%
    \setlength{\unitlength}{\svgwidth}%
  \fi%
  \global\let\svgwidth\undefined%
  \global\let\svgscale\undefined%
  \makeatother%
  \begin{picture}(1,0.44602861)%
    \put(0,0){\includegraphics[width=\unitlength,page=1]{schematic.pdf}}%
    \put(0.85247547,0.26699576){\color[rgb]{0,0,0}\makebox(0,0)[lb]{\smash{$\phi_k$}}}%
    \put(0.70932701,0.11981492){\color[rgb]{0,0,0}\makebox(0,0)[lb]{\smash{$\phi_j$}}}%
  \end{picture}%
\endgroup%

%% file: figures/zigzaginset.pdf_tex
\begingroup%
  \makeatletter%
  \providecommand\color[2][]{%
    \errmessage{(Inkscape) Color is used for the text in Inkscape, but the package 'color.sty' is not loaded}%
    \renewcommand\color[2][]{}%
  }%
  \providecommand\transparent[1]{%
    \errmessage{(Inkscape) Transparency is used (non-zero) for the text in Inkscape, but the package 'transparent.sty' is not loaded}%
    \renewcommand\transparent[1]{}%
  }%
  \providecommand\rotatebox[2]{#2}%
  \ifx\svgwidth\undefined%
    \setlength{\unitlength}{541.61145872bp}%
    \ifx\svgscale\undefined%
      \relax%
    \else%
      \setlength{\unitlength}{\unitlength * \real{\svgscale}}%
    \fi%
  \else%
    \setlength{\unitlength}{\svgwidth}%
  \fi%
  \global\let\svgwidth\undefined%
  \global\let\svgscale\undefined%
  \makeatother%
  \begin{picture}(1,0.39941965)%
    \put(0,0){\includegraphics[width=\unitlength,page=1]{zigzaginset.pdf}}%
  \end{picture}%
\endgroup%

%% file: figures/zigzag.pdf_tex
\begingroup%
  \makeatletter%
  \providecommand\color[2][]{%
    \errmessage{(Inkscape) Color is used for the text in Inkscape, but the package 'color.sty' is not loaded}%
    \renewcommand\color[2][]{}%
  }%
  \providecommand\transparent[1]{%
    \errmessage{(Inkscape) Transparency is used (non-zero) for the text in Inkscape, but the package 'transparent.sty' is not loaded}%
    \renewcommand\transparent[1]{}%
  }%
  \providecommand\rotatebox[2]{#2}%
  \ifx\svgwidth\undefined%
    \setlength{\unitlength}{234.91774669bp}%
    \ifx\svgscale\undefined%
      \relax%
    \else%
      \setlength{\unitlength}{\unitlength * \real{\svgscale}}%
    \fi%
  \else%
    \setlength{\unitlength}{\svgwidth}%
  \fi%
  \global\let\svgwidth\undefined%
  \global\let\svgscale\undefined%
  \makeatother%
  \begin{picture}(1,0.69289105)%
    \put(0,0){\includegraphics[width=\unitlength,page=1]{zigzag.pdf}}%
    \put(0.82784872,0.49015288){\color[rgb]{0,0,0}\makebox(0,0)[lb]{\smash{$\alpha$}}}%
  \end{picture}%
\endgroup%